\documentclass[12pt,a4paper,epsf]{book}
\global\arraycolsep=2pt 
\input{epsf}
\input{psfig.sty}
\usepackage{graphicx}
\usepackage{cite}
\begin{document}
\makeatletter
\def\fmslash{\@ifnextchar[{\fmsl@sh}{\fmsl@sh[0mu]}}
\def\fmsl@sh[#1]#2{%
  \mathchoice
    {\@fmsl@sh\displaystyle{#1}{#2}}%
    {\@fmsl@sh\textstyle{#1}{#2}}%
    {\@fmsl@sh\scriptstyle{#1}{#2}}%
    {\@fmsl@sh\scriptscriptstyle{#1}{#2}}}
\def\@fmsl@sh#1#2#3{\m@th\ooalign{$\hfil#1\mkern#2/\hfil$\crcr$#1#3$}}
\makeatother
\thispagestyle{empty}
\begin{titlepage}
\begin{flushright}
\end{flushright}

%

\begin{center}
  \vspace{3cm} \LARGE{A Duality as a Theory  \\ for the Electroweak Interactions}\\ \vspace{1cm}
  \vspace{1cm} \Large{Dissertation der Fakult\"at f\"ur Physik der }\\ 
  \Large{Ludwig-Maximilians-Universit\"at M\"unchen}\\
  \vspace{1cm}
  { \large vorgelegt von} \\
  \vspace{1cm}
  {\LARGE Xavier Calmet}\\
  { \large aus Marignane} \\
  \vspace{2cm}
  \large{Januar 2002}
\end{center}
\vspace{2cm}
\begin{flushleft}
1. Gutachter: Prof.~Dr.~H.~Fritzsch\\
2. Gutachter: Prof.~Dr.~J.~Wess\\
Pr\"ufungstag: 8.05.2002
\end{flushleft}
\end{titlepage}
\mbox{}
\newpage
\chapter*{Deutsche Zusammenfassung}

Im Rahmen dieser Doktorarbeit wurde ein Modell f\"ur die
elektroschwache Wechselwirkung entwickelt. Das Modell basiert auf der
Tatsache, da\ss \ die sog. ``Confinement''-Phase und Higgs-Phase der
Theorie mit einem Higgs-Boson in der fundamentalen Darstellung der
Eichgruppe $SU(2)$ identisch sind. In der Higgs-Phase wird die
Eichsymmetrie durch den Higgsmechanismus gebrochen. Dies f\"uhrt zu
Massentermen f\"ur die Eichbosonen, und \"uber die Yukawa-Kopplungen
zu Massentermen f\"ur die Fermionen. In der ``Confinement''-Phase ist
die Eichsymmetrie ungebrochen. Nur $SU(2)$-Singuletts kann eine Masse
zugeordnet werden, d.h., physikalische Teilchen m\"ussen
$SU(2)$-Singuletts sein.  Man nimmt an, da{\ss} die rechtsh\"andigen
Quarks und Leptonen elementare Objekte sind, w\"ahrend die
linksh\"andigen Dupletts Bindungszust\"ande darstellen.

Es stellt sich heraus, da{\ss} das Modell in der ``Confinement''-Phase
dual zum Standard-Modell ist. Diese Dualit\"at erm\"oglicht eine
Berechnung des elektroschwachen Mi\-schungs\-win\-kels und der Masse
des Higgs-Bosons. Solange die Dualit\"at gilt, erwartet man keine neue
Physik.

Es ist aber vorstellbar, da{\ss} die Dualit\"at bei einer kritischen
Energie zusammenbricht. Diese Energieskala k\"onnte sogar relativ
niedrig sein. Insbesondere ist es m\"oglich, da\ss \ das
Standard-Modell im Yukawa-Sektor zusammenbricht. Falls die Natur durch
die ``Confinement''-Phase beschrieben wird, koennte man davon
ausgehen, da\ss \ die leichten Fermionmassen erzeugt werden, ohne
da\ss \ das Higgs-Boson an die Fermionen gekoppelt wird. Dann w\"urde
aber das Higgs-Boson anders als im Standard Modell zerfallen. Es ist
jedoch auch vorstellbar, da\ss \ die Verletzung der Dualit\"at erst
bei hohen Energien statt\-findet. Dann erwartet man neue Teilchen wie
Anregungen mit Spin 2 der elektroschwachen Bosonen. Ebenso vorstellbar
sind Fermionen-Substuktur-Effekte die beim anomal magnetischen
Moment des Muons sichtbar werden.

\newpage
\clearpage
\chapter*{}
\vspace{7.5cm}
 \begin{center}
 {\large Cette th\`ese est dedi\'ee \`a mes parents}
 \end{center}
\mbox{}
\clearpage
\chapter*{Acknowledgements}
It is a pleasure to thank Professor Harald Fritzsch for his scientific
guidance and for a fruitful collaboration. I have learnt a lot from
the numerous discussions we had during the completion of this work.

I would like to thank Dr.~Arnd Leike for the interesting discussions
we had and for trying to organize a bit of a social life in the
research group. I would also like to thank him for reading this work
thoroughly.

I am particularly happy to thank Professor Zhi-zhong Xing for the
discussions we had. These enlightening discussions were the sources of
my works on $B$ physics. I would like to thank him for his
encouragements to write down these ideas.

Professor Fritzsch's group was small during my stay in Munich, but it
was really stimulating due to the presence of Arnd and Zhi-zhong.

I would also like to thank Andrey Neronov and Michael Wohlgenannt for
the collaborations that emerged from inspiring discussions.

I am grateful to Dr.~E.~Seiler for enlightening discussions concerning
the complementarity principle.  I would like to thank Nicole Nesvadba
from Opal, Dr.~Philip Bambade, Jens Rehn and Marcel Stanitzki from
Delphi for discussions on searches at LEP for a Higgs boson that does
not couple to $b$-quarks.

Last but not least, I would like to thank my parents and my brother
Lionel for their love and for their moral support during the
completion of this work. It has always been a pleasure to discuss
physics with you Dad, and I am sure that you could have produced many
other interesting ideas in theoretical particle physics if you had
chosen to stick to physics instead of leaving for computer science.

\mbox{}
\clearpage
\mbox{}
\newpage
\tableofcontents
\chapter{Introduction}
During the past century, particle physics has undergone at least three
revolutions.

The first of these revolutions happened when it was discovered by de
Broglie \cite{debroglie} that particles have a dual character,
sometimes they behave like solid entities sometimes like waves. In
particular, it became clear that light is sometimes behaving like a
stream of particles but, on the other hand, an electron is sometimes
behaving like a wave. This led to the development of quantum
mechanics.

Even more surprising was the second revolution.  Particles can be
created and annihilated. A particle and its antiparticle can be
produced from the vacuum, and then they can annihilate.  This had some
profound consequences for quantum mechanics which had to be improved
to take this fact into account.  The mathematical tool which was
developed to describe this phenomenon is called quantum field theory.

The third revolution was that the particles which were discovered
could be classified according to simple schemes.  The standard example
is the eightfold way \cite{Gell-Mann:nj} proposed by Gell-Mann which
allows to classify, according to a $SU(3)$ symmetry, all particles
that interact strongly. Symmetries allow a much deeper understanding
of the microscopic world. It was a big step between sampling particles
and classifying them according to a symmetry. The $SU(3)$ symmetry
allowed to predict particles that were not yet discovered and also
allowed to understand that the strongly interacting particles that
were observed could not be fundamental, but had to be bound states of
some more fundamental fields, called quarks \cite{Gell-Mann:nj}.

Another symmetry, Lorentz invariance, forced Dirac to introduce an
antiparticle in his equation \cite{Dirac:1928hu}, and to posit the
positron which was discovered shortly after. Actually it turns out
that a relativistic quantum theory, for example Dirac's equation, is
inconsistent, and that the wave functions of relativistic quantum
mechanics have to be replaced by quantum operators. This process is
called the second-quantization, and it enables to describe processes
where particles are created or destroyed. In that sense these
revolutions are connected.

Symmetries in particle physics are symmetries of the action or in
other words of the S-matrix.  It became evident that any valid theory
of particle physics should be Lorentz invariant or at least Lorentz
invariant in a very good approximation. Thus all fields introduced in
the action must fulfil the Klein-Gordon equation. The concept of
Lorentz invariance introduces also the question of the discrete
symmetries which are the charge conjugation $C$, the space reflection
$P$ and the time reflection $T$. It turns out that if the fermions are
quantized using anticommutation relations and bosons using commutation
relations, then the S-matrix, or action, is invariant under the
combination $CPT$.

Another concept which was discovered later is that of global and local
gauge symmetries, i.e. the invariance of the action under certain
global symmetries and local symmetries. Local gauge transformations
are gauge transformations which are space-time dependent whereas
global gauge transformations are independent on space-time. A gauge
transformation is a transformation of the fields entering the action.
Using Noether's theorem, one can then deduce which quantities are
conserved. For example in Quantum Electrodynamics (QED), there is a
conserved quantity, the electric charge, corresponding to a $U(1)$
local gauge symmetry. The success of QED led Yang and Mills
\cite{Yang:ek} to consider more complex non-Abelian gauge symmetries
which eventually led to the standard model of particle physics.

Fundamental symmetries, like gauge symmetries or Lorentz symmetry,
must be distinguished from approximate symmetries.  For many technical
issues it is often useful to consider symmetries that are exact in
some limit, especially in Quantum Chromodynamics (QCD) where these
approximatively valid symmetries are crucial to find relations between
different non-perturbative quantities. An example of these symmetries
is for example the isospin symmetry which is approximatively exact at
low energy QCD.

After a century of great success applying symmetries in particle
physics, it is still unclear why symmetries are so important in
physics. We know that if we can identify one, it will have some very
deep consequences, but there is still no primary principle which
forces to require the action to be invariant under some given
symmetry. We can only postulate a set of symmetries of the action,
quantize and renormalize this action to obtain the Feynman rules and
compute some observables to test whether a given symmetry is present
or not in nature. There are two possibilities if a given symmetry is
not observed, it can either be broken or it must be ruled out as a
symmetry the theory.

In the present work we shall not try to understand why symmetries, and
in particular gauge symmetries, are so crucial to particle physics. We
shall take this as an given fact. Our main concern will rather be to
try to understand how to break gauge symmetries. As we shall describe
in this first chapter, the electroweak interactions are described by a
broken $SU(2) \times U(1)$ local gauge symmetry. The main result of
this work is that the electroweak interactions can be described as
successfully by a confining theory, i.e. a theory based on an unbroken
gauge symmetry, with a weak coupling constant. It turns out that this
confining theory is dual to the standard model. This duality allows to
find relations between some of the parameters of the standard
electroweak model that are otherwise not present in the normal
standard model with a broken electroweak symmetry. We shall first
review the standard electroweak model, some of its problems and some
of the solutions to these problems.

\section{The standard electroweak model}
In this section we shall discuss the standard model of the electroweak
interactions. The weak interaction was first considered to be a local
or point like interaction, the so-called Fermi interaction
\cite{Fermi:1934hr}, before it was realized by Glashow \cite{Glashow},
following the work of Schwinger \cite{Schwinger:em}, that a $SU(2)
\times U(1)$ local gauge symmetry could account for this phenomenon
and for Quantum Electrodynamics.  But, if the electroweak gauge bosons
were massless the electroweak interactions would be long range
interactions.  This is only partially the case, since QED is a long
range interaction, but the weak interactions are short range. This
implies that the gauge bosons are either confined and cannot propagate
as free particles or that they are massive. The standard approach is
to assume the latter. But, the $SU(2) \times U(1)$ gauge symmetry
prohibits a mass term for the gauge bosons in the action.  This led
Weinberg and Salam \cite{Weinberg:1967tq} to assume that this symmetry
is spontaneously broken and to apply the Higgs mechanism
\cite{Higgs:1964pj} to break this symmetry. It turns out that a theory
with a gauge symmetry which is spontaneously broken remains
renormalizable \cite{'tHooft:rn}, and that this theory is thus
consistent to any order in perturbation theory.

The standard model of the electroweak interactions is based on the
gauge group $SU(2)_L \times U(1)_Y$, where the index $L$ stands for
left and where $Y$ stands for hypercharge. In that model, parity is
broken explicitly, left-handed fermions
$\Psi^a_L=1/2(1-\gamma_5)\Psi^a$ are transforming according to the
fundamental representation of $SU(2)_L$ whereas right-handed fermions
$\Psi^a_R=1/2(1+\gamma_5)\Psi^a$ are singlets under this gauge group.
The gauge boson of the $U(1)_Y$ gauge group is denoted by ${\cal
  A}_\mu$, and the three gauge bosons of the $SU(2)_L$ gauge group are
called $B^a_\mu$, $a \in \{1,2,3\}$. The anti-symmetric tensors
$f_{\mu \nu}$ and $F^a_{\mu \nu}$ are the field strength tensors of
$U(1)_Y$ respectively $SU(2)_L$.

We start by writing down the Lagrangian of the standard electroweak
model, taking into account only the first family of leptons
($L_L,e_r$) and quarks ($Q_L,u_R,d_R$):
\begin{eqnarray} \label{SML}
  {\cal L}_{h}&=&-\frac{1}{4} F^{a}_{\mu \nu}  F^{a \mu \nu}
                 -\frac{1}{4} f_{\mu \nu}  f^{ \mu \nu}
                 + \bar L_L i \fmslash{D} L_L+ \bar Q_L i \fmslash{D} Q_L
                 +\bar e_R i \fmslash{D} e_R\\ \nonumber &&
                 +\bar u_R i \fmslash{D} u_R
                 + \bar d_R i \fmslash{D} d_R-G_e \bar e_R ( {\bar{\phi}} L_L)
                 -G_d \bar d_R ( {\bar{\phi}} Q_L) \\ \nonumber &&
                 -G_u \bar u_R ({\phi} Q_L)
                 +h.c.
                 +\frac{1}{2}(D_{\mu} \phi)^\dagger (D^{\mu} \phi)
                 -\frac{\mu^2}{2} \phi^\dagger \phi
                 -\frac{\lambda}{4} (\phi^\dagger \phi)^2.
\end{eqnarray}
The scalar doublet $\phi$ is the Higgs field and $\bar{\phi}=i
\sigma_2 \phi^*$. In the standard model this field enters the theory
in the fundamental representation of the $SU(2)$ gauge group which
has, as we shall see later, some nontrivial consequences.  The
quantum numbers of the fields entering the standard model Lagrangian
are summarized in table \ref{tab:tablefield}. The covariant derivative
is given by:
\begin{eqnarray}
        D_{\mu}=\partial_{\mu}-i \frac{g'}{2} Y {\cal A}_\mu -
        i \frac{g}{2} \tau^a B^a_{\mu}. 
\end{eqnarray}
The field strength tensors are as usual
\begin{eqnarray}
F^a_{\mu \nu}&=&\partial_\mu B^a_\nu-  \partial_\nu B^a_\mu
+g \epsilon^{abc} B^b_\mu B^c_\nu \\
f_{\mu \nu}&=&\partial_\mu {\cal A}_\nu-  \partial_\nu {\cal A}_\mu.
\end{eqnarray}
We have used the definitions:
\begin{eqnarray}
L_L=\left(\begin{array}{c} \nu_L \\ e_L \end{array}
\right ),  \ \ Q_L= \left(\begin{array}{c} u_L \\ d_L \end{array}
\right ), \ \ \phi=\left(\begin{array}{c} \phi^0 \\ \phi^- \end{array}
\right ) \ \ \mbox{and} \ \
\bar{\phi}=i \sigma_2 \phi^*=\left(\begin{array}{c} \phi^+  \\ -{\phi^0}^*
  \end{array}
\right ).
\end{eqnarray}
\begin{table}
\centering
  \begin{tabular}{|c|c|c|c|c|}
   \hline
    & $SU(3)_C$ & $SU(2)_L$ & $U(1)_Y$  & $U(1)_Q$ 
   \\
    \hline
   $ L_L=\left(\begin{array}{c} \nu_L \\ e_L \end{array} \right )$
   & ${\bf 1}$
   & ${\bf 2}$ 
   & $-1$
   & $\left(\begin{array}{c} 0 \\ -1 \end{array} \right )$  \\
  \hline
   $ e_R$
   & ${\bf 1}$
   & ${\bf 1}$ 
   & $-2$
   & $-1$  \\
  \hline
   $Q_L=\left(\begin{array}{c}  u_L \\ d_L \end{array} \right )$
   & ${\bf 3}$
   & ${\bf 2}$ 
   & $1/3$
   & $\left(\begin{array}{c} 2/3 \\ -1/3 \end{array} \right )$  \\
  \hline
   $u_R$
   & ${\bf 3}$
   & ${\bf 1}$ 
   & $4/3$
   & $2/3$  \\
  \hline
   $d_R$
   & ${\bf 3}$
   & ${\bf 1}$ 
   & $-2/3$
   & $-1/3$  \\
  \hline
  \hline
$\phi=\left(\begin{array}{c}  \phi^0 \\  \phi^- \end{array} \right )$
   & ${\bf 1}$
   & ${\bf 2}$ 
   & $-1$
   & $\left(\begin{array}{c} 0 \\ -1 \end{array} \right )$  \\
  \hline
  \hline
$B^i$
   & ${\bf 1}$
   & ${\bf 3}$ 
   & $0$
   & $(\pm 1, 0)$  \\
  \hline
$A$
   & ${\bf 1}$
   & ${\bf 1}$ 
   & $0$
   & $0$  \\
   \hline
   $G^a$
   & ${\bf 8}$
   & ${\bf 1}$ 
   & $0$
   & $0$  \\
   \hline
 \end{tabular}
 \caption{The standard model fields,
 as usual the electric charge is given by the Gell-Mann-Nishijima relation
 $Q=\frac{1}{2}\left(\tau_3+Y\right)$.
 The fields $B^i$ with $i\in \{+,-,3\}$
 denote the three electroweak gauge bosons and $A$ is the photon.
 The gluons $G^i$ are in the octet representation of $SU(3)_C$.
 \label{tab:tablefield}}
\end{table}
Obviously a mass term for the electroweak bosons of the form $m_W^2
W^i_\mu W^{i \mu}$ would violate the gauge symmetry. In other words,
the gauge invariance of the theory requires the gauge bosons to be
massless. If the gauge bosons were massless, the electroweak
interactions would be long range interactions. But, we know that the
weak interactions are short range whereas QED is a long
range interaction. Thus we have to break this symmetry partially.
\subsection{The Higgs mechanism}
The symmetry breaking scheme has already been introduced in the
standard electroweak model Lagrangian. The Higgs mechanism
\cite{Higgs:1964pj} breaks the $SU(2)\times U(1)$ gauge symmetry
spontaneously, which insures that the resulting theory is
renormalizable. The potential of the Higgs boson is given by
\begin{eqnarray}
V(\phi^\dagger \phi)= \frac{\mu^2}{2}  \phi^\dagger \phi +\frac{\lambda}{4}
(\phi^\dagger \phi)^2
\end{eqnarray}
The position of the minimum is dependent on the sign of the squared
mass $\mu^2$ of the Higgs doublet. If it is positive, i.e. if the
Higgs doublet squared mass has the right sign for the squared mass
term of a scalar field, then the gauge symmetry is unbroken, and the
minimum is at $\phi^\dagger \phi=0$. The Higgs mechanism postulates
that the doublet is a tachyon, and thus requires $\mu^2<0$. In that
case the potential has two extrema which are given by
\begin{eqnarray}
  \frac{d V(\rho)}{d\rho}= \mu^2  \rho +\lambda \rho^2
  =\left(\mu^2+\lambda \rho \right) \rho=0
  \end{eqnarray}
  with $\rho^2=\phi^\dagger \phi$. The extrema are then
\begin{eqnarray}
  \rho_1&=&0 \\ \rho_{2}&=&\frac{-\mu^2}{\lambda}\equiv v^2,
\end{eqnarray}
where $v$ is the so-called vacuum expectation value. The first
solution is unstable and thus not the true vacuum of the theory.
The standard procedure is to expand the Higgs field around its
vacuum expectation value. It is convenient to fix the gauge,
performing a $SU(2)$ rotation, at this stage. We shall choose the
unitarity gauge
\begin{eqnarray}
\phi &=& U_2 \left(\matrix{\phi^0\cr \phi^-}\right) = \left(
  \matrix{\eta + v \cr 0} \right)
\end{eqnarray}
which allows to ``rotate away'' the Goldstone bosons. The Goldstone bosons
are the three degrees of freedom which remain massless after
spontaneous symmetry breaking. They are absorbed in the longitudinal
degrees of freedom of the gauge bosons. The Higgs field is expanded
around its vacuum expectation value $v$. This is a semi-classical
approach. Of the four generators of $SU(2)\times U(1)$ three are broken
by the Higgs mechanism. Only the linear combination $Q=1/2(\tau^3+Y)$
is left unbroken and thus leaves the vacuum invariant. This implies
that a linear combination of the gauge fields of $SU(2)\times U(1)$
remains massless. It can be identified with the photon. Inserting
the expansion of the Higgs field in the Lagrangian (\ref{SML}), one
finds that the Higgs mechanism gives a mass to the electroweak bosons
$W^\pm_\mu$ and $Z_\mu$, whereas the photon $A_\mu$ remains massless.
The $Z_\mu$ and $A_\mu$ bosons are the mass eigenstates given by
\begin{eqnarray}
         A_\mu&=& \sin{\theta_W} W^3_\mu+ \cos{\theta_W}{\cal A}_\mu
         \nonumber \\
         Z_\mu&=& \cos{\theta_W} W^3_\mu-  \sin{\theta_W}{\cal A}_\mu,
\end{eqnarray}
with
\begin{eqnarray}
  \cos{\theta_W}=\frac{g}{\sqrt{g^2+g'^2}} \ \ \mbox{and} \ \
  \sin{\theta_W}=\frac{g'}{\sqrt{g^2+g'^2}}.
\end{eqnarray}
The masses of the electroweak bosons are given by $m_{W^\pm}=g v/2$,
$m_{Z}=\sqrt{g^2+{g'}^2} v/2$ and $m_A=0$. It is important to notice
that the mechanism responsible for the fermion mass generation is
not the Higgs mechanism but rather the Yukawa mechanism. The Yukawa
interactions generate a mass term for the fermions of the type
\begin{eqnarray}
  m_e=G_e v, \ \ \ m_u=G_u v, \ \ \mbox{and} \ \ \ m_d=G_d v.
  \end{eqnarray}
There are thus two distinct mass generating mechanisms in the standard model. 
\begin{figure}
\begin{center}
\leavevmode
\epsfxsize=8cm
\epsffile{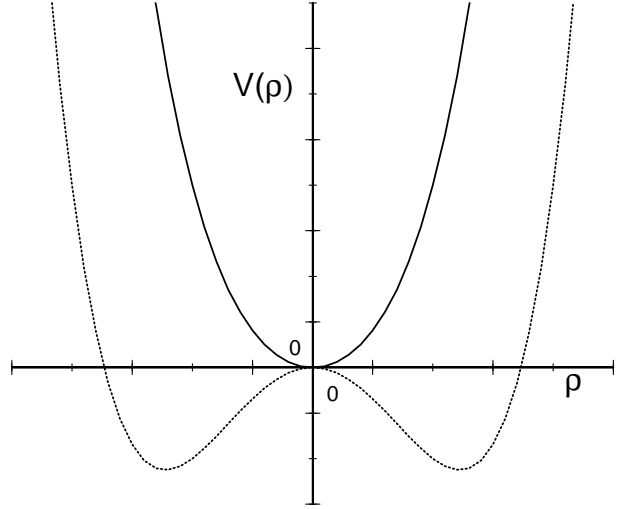}
\caption{Higgs potential before (continuous line)
  and after electroweak symmetry breaking (dotted line), the variable
  $\rho$ is defined by $\rho^2=\phi^\dagger \phi$.}
\label{Higgspot}
\end{center}
\end{figure}
\subsection{Naturalness and Hierarchy problem}
Albeit the standard model, which is the superposition of the standard
electroweak model, described by $SU(2)_L\times U(1)_Y$, and of Quantum
Chromodynamics, described by $SU(3)_C$, is extremely successful, it
might not be the final theory of particle physics. The major objection
is that it contains too many parameters that have to be measured and
cannot be calculated from first principles. This led to a quest for
the unification of the gauge interactions described by the three gauge
groups $SU(3)$, $SU(2)$ and $U(1)$.  There are two prime examples of
such unification groups: $SU(5)$ \cite{Georgi:sy} and $SO(10)$
\cite{Fritzsch:nn}. In that framework the standard model is embedded
in a larger gauge group whose gauge symmetry is broken at high energy
called the grand unification scale (GUT) scale. The running of the
coupling constant of the gauge group $SU(3) \times SU(2) \times U(1)$
suggests that the unification is taking place at a scale
$\Lambda_{GUT}\approx 10^{16}$ to $10^{17}$ GeV depending on whether
supersymmetry is present in Nature or not. The gauge hierarchy problem
states that it is unnatural for the electroweak breaking scale
$\Lambda_{EW}\approx 246$ GeV to be so small compared to the
fundamental scale $\Lambda_{GUT}$.

A second potential problem with the standard model is that the Higgs
boson is a scalar field. If a cutoff $\Lambda$ is used to renormalize
the theory, the Higgs boson mass receives quadratic ``corrections''
\begin{eqnarray}
  m_H^2&\approx& {m_H^{0}}^2+\frac{3 g^2 \Lambda^2}{32 \pi^2 m_W^2}
  \left(m^2_H+2 m^2_W+m^2_Z  - 4 \sum_f \left( \frac{n_f}{3} \right)
    m_f^2\right).
\end{eqnarray}
Nevertheless, this problem seems not very serious since the cutoff
would not be apparent in a different renormalization scheme and
secondly the standard model is a renormalizable theory. This means
that all divergencies can be absorbed in the renormalized coupling
constants and renormalized masses. Furthermore, it has been argued by
Bardeen \cite{Bardeen:1995kv} that there is an approximate scale
invariance symmetry of the perturbative expansion which protects the
Higgs boson mass. The Higgs boson mass can be viewed as a soft
breaking term for this symmetry. In that case fine tuning issues are
related to nonperturbative aspects of the theory or embeddings of the
standard model into a more complex theory.

The opinion of the author of this work is that none of these problems
is very serious. The main problem of the standard model is that the
symmetry breaking mechanism is implemented in a quite unnatural
fashion. The Higgs boson which is introduced in the standard model is
assumed to be a tachyon, i.e. its squared mass is adjusted to be
negative at tree level. This might be a sign that a mechanism is
required to trigger the Higgs mechanism.

There are many other motivations to extend the standard model. It is
not clear yet which of these is the right one. In the following, we
shall review a few typical models, which are all connected to
solutions of these problems.
\section{Extensions of the standard model}
\subsection{Composite models}
In this section we shall review a composite model proposed around
twenty years ago. The list of models proposed in the literature is
very long. Three popular models were those proposed by Greenberg and
Nelson \cite{Greenberg:1974qb}, Fritzsch and Mandelbaum
\cite{Fritzsch} and Abbott and Farhi \cite{class1}. For an
extensive list of citations see \cite{Lyons:1982jb} and
\cite{Chivukula:2000mb}.  The model Quantum Haplodynamics (QHD) we
have chosen to review has been proposed by Fritzsch and Mandelbaum
\cite{Fritzsch}.
 
This model is inspired by QCD.  In this approach, the weak
interactions are residual effects due to the substructure of leptons,
quarks and weak bosons.  The constituents are called haplons, their
quantum numbers are given in table \ref{tab:tableHaplons}.
\begin{table}
\centering
  \begin{tabular}{|c|c|c|c|c|}
   \hline
     & Color& Charge  & Spin & H 
   \\
    \hline
    $\alpha$
   &3
   &-1/2 
   & 1/2
   & +1 $n$ \\
  \hline
  $\beta$
   &3
   &+1/2 
   & 1/2
   & +1 $n$ \\
  \hline
   $x$
   &3
   &-1/6 
   & 0
   & -1 $\bar n$ \\
  \hline
   $y$
   &$\bar 3$
   &+1/2 
   & 0
   & -1 $\bar n$ \\
   \hline
 \end{tabular}
 \caption{Quantum numbers of the haplons.
 \label{tab:tableHaplons}}
\end{table}
The haplons are assumed to be bound together by a very strong confining
force, called hypercolor. The gauge group describing this interaction
could be a $SU(N)$ gauge group (e.g. $SU(4)$) or a $U(1)$ gauge group.
The spectrum of the model is as follows
\begin{eqnarray}
u&=(\bar \alpha \bar x)_{3}  \ \ \ \nu_e&=(\bar \alpha \bar y)_1 \\
\nonumber
d&=(\bar \beta \bar x)_{3}  \ \ \ e^-&=(\bar \beta \bar y)_1
\\
\nonumber
W^+&=(\bar \alpha \beta) \ \ \ W^-&=(\bar \beta \alpha)
\end{eqnarray}
and two neutral bosons
\begin{eqnarray} \nonumber
 W^3= \frac{1}{\sqrt{2}}(\bar \alpha \alpha -\bar \beta \beta) \ \ \
  W^0=\frac{1}{\sqrt{2}}(\bar \alpha \alpha +\bar \beta \beta).
\end{eqnarray}
The neutral boson $W^3$, which mixes with the $U(1)$-photon, is
identified with a component of the $Z$-boson. On the other side $W^0$
is assumed to be very heavy and not to contribute to the neutral
currents. This model had the very pleasant feature of solving the
gauge hierarchy problem and potentially the naturalness problem as in
that case the weak interactions are not a gauge theory, but an
effective theory with a cutoff at $\Lambda_{QHD}\approx 200$ GeV.
Unfortunately the simplest version of this model is nowadays ruled out
by experiments performed e.g. at LEP as are most of the composite
models proposed long ago.
\subsection{Technicolor}
A more elaborate approach is that of technicolor theories. Again the
literature is very rich, for reviews, see references
\cite{Chivukula:2000mb} and \cite{Lane:1993wz}.

We review the simplest possible (i.e. not extended) example of a
technicolor theories \cite{Weinberg:gm,Susskind:1978ms}.
Technicolor theories are models where the electroweak symmetry
breaking is due to dynamical effects.

Consider an $SU(N_{TC})$ gauge theory with fermions in the
fundamental representation of the gauge group
\begin{eqnarray}
\Psi_L=\left(
\begin{array}{c}
U\\D
\end{array}
\right) _L\,\,\,\,\,\,\,\,
U_R,D_R~.
\end{eqnarray}
The fermion kinetic energy terms
for this theory are
\begin{eqnarray}
{\cal L} &=& \bar{U}_L i\fmslash{D} U_L+
\bar{U}_R i\fmslash{D} U_R+ \bar{D}_L i\fmslash{D} D_L+
\bar{D}_R i\fmslash{D} D_R~,
\end{eqnarray}
and, like QCD in the $m_u$, $m_d \to 0$ limit, they exhibit a chiral
$SU(2)_L \times SU(2)_R$ symmetry.

As in QCD, exchange of technigluons in the spin zero, isospin zero
channel is attractive, causing the formation of a condensate
\begin{eqnarray}
{\lower35pt\hbox{
\epsfysize=1 truein \epsfbox{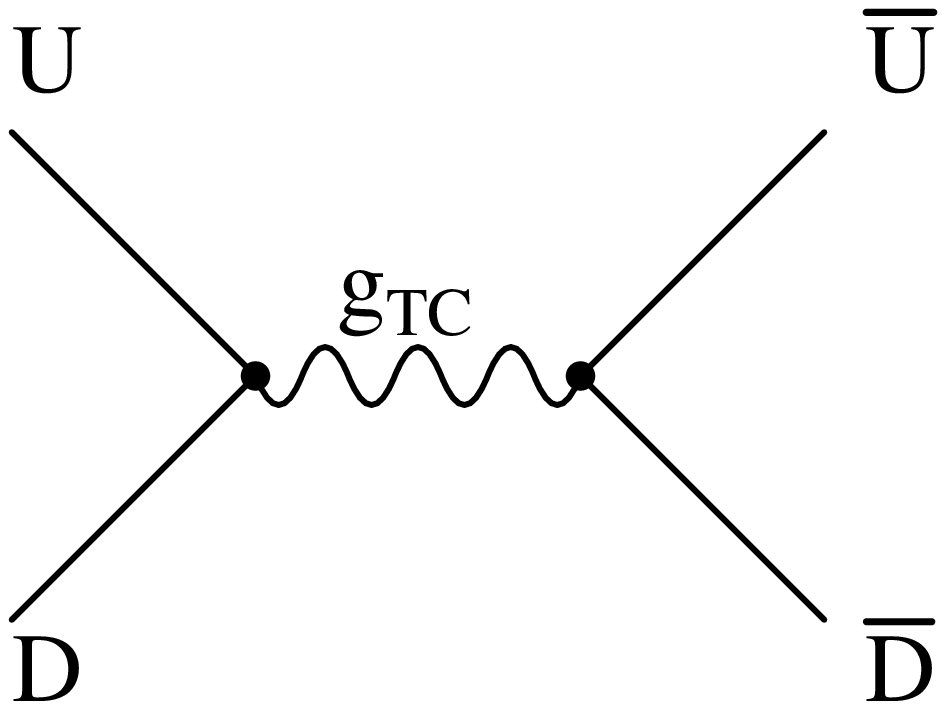}}
\ \ \rightarrow  \ \ \langle \bar U_LU_R\rangle
=\langle \bar D_LD_R\rangle \neq 0\, ,}
\end{eqnarray}
which dynamically breaks $SU(2)_L \times SU(2)_R \to SU(2)_V$.  These
broken chiral symmetries imply the existence of three massless Goldstone
bosons, the analog of the pions in QCD.

Now we consider gauging $SU(2)_W \times U(1)_Y$ with the left-handed
fermions transforming as weak doublets and the right-handed ones as
weak singlets. To avoid gauge anomalies, in this one-doublet
technicolor model, the left-handed technifermions are assumed to have
hypercharge zero and the right-handed up- and down-technifermions to
have hypercharge $\pm 1/2$.  The spontaneous breaking of the chiral
symmetry breaks the weak interactions down to electromagnetism. The
would-be Goldstone bosons become the longitudinal components of the
$W$ and $Z$
\begin{eqnarray}
\pi^\pm,\, \pi^0 \, \rightarrow\, W^\pm_L,\, Z_L~,
\end{eqnarray}
which
acquire a mass
\begin{eqnarray}
M_W = {g F_{TC} \over 2}~.
\end{eqnarray}
Here $F_{TC}$ is the analog of $f_\pi$ in QCD. In order to obtain the
experimentally observed masses, we must have $F_{TC} \approx 246$ GeV
and hence this model is essentially like QCD scaled up by a factor of
\begin{eqnarray}
{F_{TC}\over f_\pi} \approx 2500\, .
\end{eqnarray}

Since there are no fundamental scalars in the theory, there is not any
unnatural adjustment required to absorb quadratic divergencies of
scalar masses. The mass generation problem of the electroweak bosons
can thus be solved in a very elegant fashion. The gauge hierarchy
problem is solved in such a theory, because the scale of the
electroweak symmetry breaking is a dynamical quantity which could
eventually be calculated.  Nevertheless there is a potentially serious
problem with the mass generation of the fermions in such theories. The
model we have presented does not yet have a mechanism to generate
fermion masses.

The model has to be embedded in a more complex theory, so-called
extended Technicolor theories (ETC)
\cite{Eichten:1979ah,Dimopoulos:1979es}. In ETC models, technifermions
couple to ordinary fermions.  At energies low compared to the ETC
gauge-boson mass, $M_{ETC}$, these effects can be treated as local
four-fermion interactions
\begin{eqnarray}
{\lower35pt\hbox{\epsfysize=1 truein \epsfbox{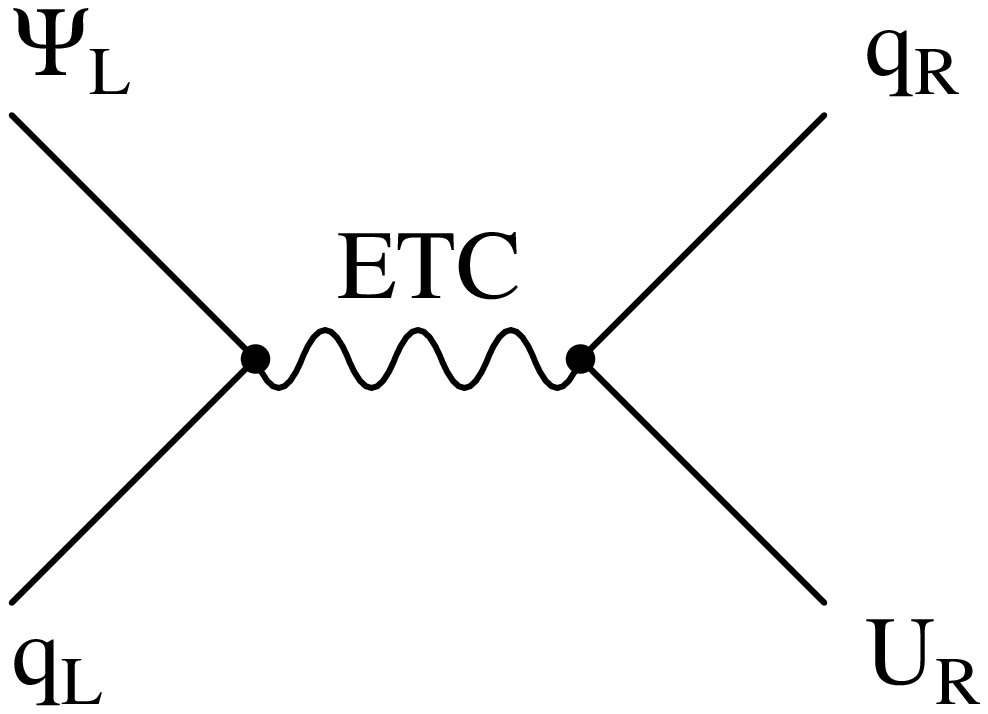}}}
\ \ \rightarrow\ \  {{g_{ETC}^2\over M^2_{ETC}}}(\overline{\Psi}_L U_R)
({\overline{q}_R q_L})~.
\label{etcint}
\end{eqnarray}
After technicolor chiral-symmetry breaking and the formation of a
$\langle \bar{U} U \rangle$ condensate, such an interaction gives rise
to a mass for an ordinary fermion
\begin{eqnarray}
m_q \approx {{g_{ETC}^2\over M^2_{ETC}}} \langle\overline{U} U\rangle_{ETC}~,
\label{fmassTC}
\end{eqnarray}
where $\langle \overline{U} U\rangle_{ETC}$ is the value of the
technifermion condensate evaluated at the ETC scale (of order
$M_{ETC}$). The condensate renormalized at the ETC scale in eq.
(\ref{fmassTC}) can be related to the condensate renormalized at the
technicolor scale as follows
\begin{eqnarray}\langle\overline{U} U\rangle_{ETC} =
\langle\overline{U} U\rangle_{TC}
\exp\left(\int_{\Lambda_{TC}}^{M_{ETC}} {d\mu \over \mu}
  \gamma_m(\mu)\right)~,
\end{eqnarray}
where $\gamma_m(\mu)$ is the anomalous dimension of the fermion mass
operator and $\Lambda_{TC}$ is the analog of $\Lambda_{QCD}$ for the
technicolor interactions. One finds
\begin{eqnarray}
\langle\overline{U} U\rangle_{ETC} \approx \langle\overline{U} U\rangle_{TC}
\approx 4\pi F^3_{TC}~,
\end{eqnarray}
using dimensional analysis.  In this case eq. (\ref{fmassTC}) implies
that
\begin{eqnarray}
\frac{M_{ETC}}{g_{ETC}} \approx 40 \mbox{TeV} 
\left(\frac{F_{TC}}{250 \ \mbox{GeV}}\right)^{3\over 2}
\left(\frac{100 \ \mbox{MeV}}{ m_q}\right)^{1\over 2}~.
\end{eqnarray}

It is not easy to build technicolor models that give a mass to
fermions while remaining simple. Besides this most of the ETC models
predict large deviations from the standard model predictions, and in
particular rare decays of the type $\mu \to e \gamma$ for which the
experimental limits are quite restrictive. It is also difficult to
understand how light fermion masses can be generated since
\begin{eqnarray}
  m_q \sim \frac{4 \pi g_{ETC}^2}{M_{ETC}^2} F^3_{TC}.
\end{eqnarray}
This requires $M_{ETC}$ to be in the range of $100$ TeV for the
$s$-quark or for the muon. But, $M_{ETC}=100$ TeV is too low to be
invisible, e.g. in $\bar K K$ mixing.

Nevertheless an interesting proposition has been made recently.  The
case of mass generation for fermions in a simple technicolor theory
has been reconsidered \cite{Tomboulis:2001uu}. If the fermion global
chiral symmetries are broken by the inclusion of four-fermion
interactions, it is found that the system can be nonperturbatively
unstable under fermion mass fluctuations driving the formation of an
effective coupling between the technigoldstone bosons and the ordinary
fermions.  A minimization of an effective action for the corresponding
composite operators leads to a dynamical generation of light fermion
masses $\sim M \exp(-k/g^2)$, where $M$ is some cutoff mass and where
$k$ is a parameter which depends on the coupling constants of the
four-fermion interactions.

Technicolor theories are still an acceptable alternative to the Higgs
mechanism. A better understanding of the non-perturbative aspects of
this theory might avoid to extend the plain technicolor models to ETC
models which are getting very complicated and are thus not very
elegant.

\subsection{Supersymmetry}
Low energy supersymmetry is a natural candidate to solve the
naturalness problem of the Higgs boson mass (see \cite{Louis:1998rx}
and \cite{Haber:1984rc} for reviews).  Supersymmetry \cite{BaggerWess}
is a symmetry between bosons and fermions, i.e. a symmetry between
states of different spin.  For example, a spin-0 particle is mapped to
a spin-$\frac{1}{2}$ particle under a supersymmetry transformation.
The particle states in a supersymmetric field theory form
representations (supermultiplets) of the supersymmetry algebra. There
is an equal number of bosonic degrees of freedom $n_B$ and fermionic
degrees of freedom $n_F$ in a supermultiplet
\begin{eqnarray}
n_B = n_F.
\end{eqnarray}
The masses of all states in a supermultiplet are degenerate.  In
particular the masses of bosons and fermions are equal
\begin{eqnarray}
m_B = m_F.
\end{eqnarray}

\begin{figure}
\begin{center}
\leavevmode
\epsfxsize=14cm
\epsffile{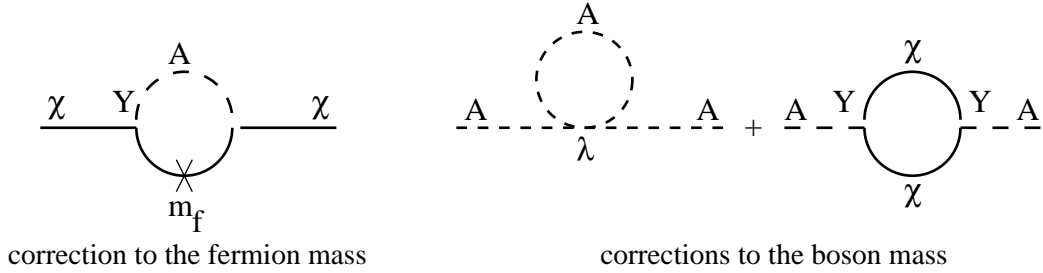}
\caption{One loop corrections to the fermion mass and to the boson mass.}
\label{corrmass}
\end{center}
\end{figure}

We shall illustrate how supersymmetry can solve the naturalness
problem. Consider the following (non-supersymmetric) Lagrangian of a
complex scalar $A$ and a Weyl fermion $\chi$
\begin{eqnarray}
{\cal L}  = & - & \partial_\mu\bar A \partial^\mu A
-i \bar \chi \bar \sigma^\mu
\partial_\mu \chi - \frac{1}{2} \,m_f\, (\chi\chi + \bar \chi \bar \chi)- 
 m^2_b\, \bar A A \nonumber \\
& - & \;Y\,(A\chi\chi + \bar A \bar \chi\bar \chi) \;-\;
\lambda \, (\bar A A)^2\ .
\label{toy model}
\end{eqnarray}
{}This Lagrangian is supersymmetric if $m_f=m_b$ and $Y^2=\lambda$, but
let us not consider this choice of parameters at first.  ${\cal L}$
has a chiral symmetry for $m_f=0$ given by
\begin{eqnarray}
A  \rightarrow  e^{-2 i \alpha}\, A\ , \qquad
\chi  \rightarrow e^{i \alpha}\, \chi \ .
\end{eqnarray}
This symmetry prohibits the generation of a fermion mass by quantum
corrections.  For $m_f\neq 0$ the fermion mass does receive radiative
corrections, but all possible diagrams have to contain a mass
insertion as can be seen from the one-loop diagram shown in figure
\ref{corrmass}. Since the propagator of the boson (upper dashed line
in the diagram) is $\sim \frac{1}{k^2} $ while the propagator of the
fermion (lower solid line) is $\sim \frac{1}{k} $ one obtains a mass
correction which is proportional to $m_f$

\begin{eqnarray}
\delta m_f \sim Y^2 m_f \ln {\frac{m_f^2}{\Lambda^2}}\ ,
\end{eqnarray}
where $\Lambda$ is the ultraviolet cutoff.  Hence the mass of a chiral
fermion does not receive large radiative corrections if the bare mass
is small.

The diagrams giving the one-loop corrections to $m_b$ are shown in
figure \ref{corrmass}.  Both diagrams are quadratically divergent but they
have an opposite sign because in the second diagram fermions are
running in the loop. One finds
 \begin{eqnarray}\label{bmasscorrection}
\delta m_b^2 \sim  (\lambda  -  ~Y^2)\, \Lambda^2\ .
\end{eqnarray}
Thus, in non-supersymmetric theories scalar fields receive large mass
corrections.  In supersymmetric theories the quadratic divergency in
(\ref{bmasscorrection}) exactly cancels due to the supersymmetric
relation $Y^2= \lambda$. The cancellation of quadratic divergencies is
a general feature of supersymmetric quantum field theories.  This
leads to the possibility of stabelizing the weak scale $M_Z$.

In that sense supersymmetry solves the naturalness problem. It allows
for a small and stable weak scale without fine-tuning. However,
supersymmetry does not solve the hierarchy problem in that it does not
explain why the weak scale is small in the first place. But, the main
problem of supersymmetric theories at low energy is to explain the
breaking of supersymmetry.

\subsection{New ideas and new dimensions}
It has recently been proposed that the gauge hierarchy problem could
be solved by lowering the scale of the unification of all forces and
in particular of the scale for gravity \cite{Arkani-Hamed:1998rs}.  In
this framework, the gravitational and gauge interactions become united
at the weak scale, which we take as the only fundamental short
distance scale in nature.  The observed weakness of gravity on
distances $\sim$ 1 mm is due to the existence of $n \geq 2$ new
compact spatial dimensions large compared to the weak scale.  The
Planck scale $M_{Pl} \sim G_N^{-1/2}$ is not a fundamental scale. Its
large value is simply a consequence of the large size of the new
dimensions.  While gravitons can freely propagate in the new
dimensions, at sub-weak energies the standard model fields must be
localized to a 4-dimensional manifold of weak scale ``thickness" in
the extra-dimensions.

A very simple idea is to suppose that there are $n$ extra compact
spatial dimensions of radius $\sim R$. The Planck scale $M_{Pl (4+n)}$
of this $(4+n)$ dimensional theory is taken to be $\sim m_{EW}$. Two
test masses of mass $m_1,m_2$ placed within a distance $r \ll R$ will
feel a gravitational potential dictated by Gauss's law in $(4+n)$
dimensions
\begin{equation}
V(r) \sim \frac{m_1 m_2}{M_{Pl (4+n)}^{n+2}} \frac{1}{r^{n+1}}, \, (r \ll R).
\end{equation}
On the other hand, if the masses are placed at distances $r \gg R$, their
gravitational flux lines can
not continue to penetrate in the extra-dimensions, and the usual $1/r$
potential is obtained,
\begin{equation}
V(r) \sim  \frac{m_1 m_2}{M_{Pl (4+n)}^{n+2} R^{n}} \frac{1}{r}, \, (r \gg R)
\end{equation}
so our effective 4 dimensional $M_{Pl}$ is
\begin{equation}
M_{Pl}^2 \sim M_{Pl (4+n)}^{2 + n} R^n. \label{fourplanck}
\end{equation}
Putting $M_{Pl (4+n)} \sim m_{EW}$ and demanding that $R$ be chosen to
reproduce
the observed $M_{Pl}$ yields
\begin{equation}
R \sim 10^{\frac{30}{n} - 17} \mbox{cm} \times \left(\frac{1
\mbox{TeV}}{m_{EW}}\right)^{1 + \frac{2}{n}}.  \label{radiuseq}
\end{equation}
For $n=1$, one finds $R \sim 10^{13}$ cm, implying deviations from
Newtonian gravity over solar system distances, so this case is
empirically excluded. For all $n \geq 2$, however, the modification of
gravity only becomes noticeable at distances smaller than those
currently probed by experiment. The case $n=2$ ($R \sim 100 \mu$m $-1$
mm) is particularly interesting, since it has not yet been ruled out
by experiments. Lowering the Planck scale to the TeV range solves the
gauge hierarchy problem.  Shortly after this observation was made, it
was proposed that these extra-dimensions might even be infinitely
large \cite{Randall:1998uk}. The main objection to these models with
extra-dimensions is that these quantum field theories are not
renormalizable.

A more exciting framework is that proposed in
\cite{Arkani-Hamed:2001ca} where extra-dimensions are created
dynamically. In that framework, which is essentially a reminiscence of
an old idea \cite{Kaplan:1983fs} using the language of
extra-dimensions, one considers the direct product of two groups
$SU(N) \times SU(M)$ in four dimensions, which are thus potentially
renormalizable. One of these is assumed to confine its charges at a
very high scale. The low energy effective action is a five dimensional
non-linear sigma model where the fifth dimension is discrete. In that
kind of models, the radiative corrections to the Higgs mass are finite
\cite{Arkani-Hamed:2001nc}, and the mass of this particle could thus
be calculated. This would solve the naturalness problem. But, these
extra-dimensions created dynamically at low energy are quite peculiar.
Indeed gravity would not propagate in these new dimensions.

\section{Discussion}
In this chapter we have presented the  standard electroweak model of
particle physics. We have discussed the so-called gauge hierarchy and
naturalness problem. These problems can at least be partially
addressed in different frameworks which are composite models,
technicolor models, supersymmetric models or models with
extra-dimensions. There are probably more frameworks were these
problems can be addressed.

Nevertheless all of these have in common the feature that they predict
a lot of new physics beyond the standard model, and while they are
able to address at least some of the these problems, they are unable
to reduce the number of free parameters introduced in the fundamental
theory. On the contrary they tend to increase them.  Besides this,
there are no signs of physics beyond the standard model.

We shall thus consider a different approach and reconsider the first
assumption we made, namely that the gauge theory describing the
electroweak interactions is broken. We shall argue that the
electroweak interactions can be described by a confining theory at
weak coupling which turns out to be dual to the standard model. This
duality allows in particular to calculate the electroweak mixing angle
and the Higgs boson's mass.

The remaining question is whether this duality is only a low energy
phenomenon or whether it is valid for all energies. This duality can
be tested by searching for deviations from the standard electroweak
model.

This work is organized in the following way. In chapter 2, we shall
establish the duality. We shall present the calculation of the
electroweak mixing angle and of the mass of the Higgs boson in chapter
3. A supersymmetric extension of the duality presented in chapter 2,
will be considered in chapter 4. In chapters 5 and 6, we shall present
some of the tests of this duality. We shall conclude in chapter 7.

\chapter{The dual phase of the standard model}
This chapter is dedicated to the description of the duality which is
the main achievement of this work. This duality is motivated by the
fact that the standard model action can be rewritten in terms of gauge
invariant fields and by the so-called complementarity principle. We
shall present both motivations in this chapter. The results were
published in \cite{Calmet:2000th}.
\section{The confinement phase}
In this work we will be constantly referring to the theory in the
Higgs phase and to the theory in the confinement phase. We shall adopt the
following definitions for the Higgs phase and for the confinement
phase:
\newtheorem{def1}{definition}
\newtheorem{def2}[def1]{definition}
\newtheorem{def3}[def1]{definition}
\begin{def1}[Higgs phase]
  By the theory in the Higgs phase we understand the standard model of
  particle physics with spontaneous electroweak symmetry breaking
  generated  at the classical level by the Higgs mechanism.
\end{def1}

\begin{def2}[confinement phase]
  By the theory in the confinement phase we understand the same theory
  as that of the standard model but with reversed sign of the Higgs
  boson squared mass, i.e. the $SU(2)_L$ gauge symmetry is unbroken at
  the classical level.  We do not make assumptions about the strength
  of the coupling constant of the theory.
\end{def2}

We shall consider a gauge theory with a gauge group which is the same
as that of the standard model, i.e. $SU(3)_C \times SU(2)_L \times
U(1)_Y$, but the gauge symmetry is assumed to be unbroken. The
parameters of the theory are, except for the Higgs potential and in
particular the sign of the Higgs doublet squared mass which has the
right sign for a scalar quantum field, i.e. the gauge symmetry is
unbroken, exactly the same as those of the standard model.  In
particular the coupling constant has its usual value and is thus weak.

We introduce the following fundamental left-handed dual-quark
doublets, which we denote as D-quarks (referring to duality):
\\
\begin{tabular}{lll}
leptonic D-quarks & $l_i=  \left(\begin{array}{c} l_1 \\ l_2 \end{array}

\right )$  &  (spin   $1/2$,  left-handed)  \\
& & \\
 hadronic  D-quarks   & $q_i= \left(\begin{array}{c}q_1 \\ q_2\end{array}
\right )$   &  (spin $1/2$, left-handed, $SU(3)_C$ triplet) \\
& & \\
scalar D-quarks  &  $h_i= \left(
  \begin{array}{c}
  h_1 \\ h_2
  \end{array}
\right )$ & (spin $0$), \\  \\
\end{tabular}
\\
taking into account only the first family of leptons and quarks.  The
right-handed particles are those of the standard model.  The
Lagrangian describing the electroweak interactions in the confinement
phase is
\begin{eqnarray} \label{conf}
  {\cal L}_{c}&=&-\frac{1}{4} F^{a}_{\mu \nu}  F^{a \mu \nu}
                 -\frac{1}{4} f_{\mu \nu}  f^{ \mu \nu}
                 + \bar l_L i \fmslash{D} l_L+ \bar q_L i \fmslash{D} q_L
                 +\bar e_R i \fmslash{D} e_R\\ \nonumber &&
                 +\bar u_R i \fmslash{D} u_R
                 + \bar d_R i \fmslash{D} d_R-G_e \bar e_R (\bar{h} l_L)
                 -G_d \bar d_R (\bar{h} q_L) \\ \nonumber &&
                 -G_u \bar u_R (h q_L)
                 +h.c.
                 +\frac{1}{2}(D_{\mu} h)^\dagger (D^{\mu} h)
                 -\frac{m_h^2}{2} h^\dagger h
                 -\frac{\lambda}{4} (h^\dagger h)^2,
\end{eqnarray}
with $\bar h=i \sigma_2 h^*$.  The covariant derivative is given by:
\begin{eqnarray}
        D_{\mu}=\partial_{\mu}-i \frac{g'}{2} Y {\cal A}_\mu -
        i \frac{g}{2} \tau^a B^a_{\mu}. 
\end{eqnarray}
The field strength tensors are as usual
\begin{eqnarray}
F^a_{\mu \nu}&=&\partial_\mu B^a_\nu-  \partial_\nu B^a_\mu
+g \epsilon^{abc} B^b_\mu B^c_\nu \\
f_{\mu \nu}&=&\partial_\mu {\cal A}_\nu-  \partial_\nu {\cal A}_\mu.
\end{eqnarray}

We are considering a theory in the weak coupling regime, i.e. the
strength of the $SU(2)_L$ interaction is that of the standard model,
but, we nevertheless assume that the confinement phenomenon can take
place at weak coupling. It has been conjectured by 't Hooft that
vortices which are classical solutions present in this theory can lead
to confinement of gauge charges at arbitrary weak coupling constant
\cite{'tHooft:1978hy}. Recently, a measurement of the vortex free
energy order parameter at weak coupling for $SU(2)$ has been performed
using so-called multi-histogram methods \cite{Kovacs:2000sy}. The
result shows that the excitation probability for a sufficiently thick
vortex in the vacuum tends to unity. It is claimed in
\cite{Kovacs:2000sy} that this rigorously provides a necessary and
sufficient condition for maintaining confinement at weak coupling in
$SU(N)$ gauge theories.

We thus have a consistent mechanism for the confinement of gauge
charges.  The mechanism for confinement might not be different from
that of QCD, but the basic difference between the weak interactions
and the strong interactions is, as stressed by 't Hooft
\cite{'tHooft:1998pk}, that in the weak interactions there is a large
parameter, the vacuum expectation value, which allows perturbation
theory whereas no such parameter is present in QCD, which explains why
QCD is nonperturbative. Nevertheless, in QCD the scale of the theory
coincides with the Landau pole of the theory, but obviously this
cannot be the case for a $SU(N)$ theory at weak coupling. This might
be the hint that QCD is a particular case of a more general class of
theories where confinement occurs. After these remarks on the
confinement mechanism, we study the spectrum of the theory.

The left-handed fermions are protected from developing a mass term by
the chiral symmetry, physical particles must thus be gauge singlets
under $SU(2)$ transformations.  The right-handed particles are those of
the standard model. We can identify the physical particles in the
following way:
\begin{eqnarray} \label{def1}
 &{\rm neutrino}: &  \nu_L \propto \bar h l    \\ 
 &{\rm electron}: &  e_L\propto h l  \\
 &{\rm up \ type \ quark}:&  u_L\propto \bar h q   \\
 &{\rm down\ type \ quark}:&  d_L\propto h q  \\
 &{\rm Higgs \ particle}:&
 \phi\propto \bar h h, \ \  \mbox{$s$-wave} \\
 &W^3\! \!- \!{\rm boson}: &   W^3\propto \bar h h,
 \ \ \mbox{$p$-wave} \\
 &W^-\! \!- \!{\rm boson}: &   W^- \propto h h, \ \ \mbox{$p$-wave} \\
 &W^+\! \!-\!{\rm boson}: & W^+ \propto (h h)^\dagger, \ \ \mbox{$p$-wave}. 
\end{eqnarray}

These bound states have to be normalized properly. We shall consider
this issue in the next section. Using a non-relativistic notation, we
can say that the scalar Higgs particle is a $\bar h h$-state in which
the two constituents are in a $s$-wave.  The $W^3$-boson is the
orbital excitation ($p$-wave). The $W^+$ ($W^-$)-bosons are $p$-waves
as well, composed of $(h h)$ $(\bar h \bar h)$ respectively. Due to
the $SU(2)$ structure of the wave function there are no $s$-wave
states of the type $( h h)$ or $(\bar h \bar h)$.

Notice that we have defined composite operators at the same space-time
point, i.e. $\Psi(x,x) =\bar \phi(x) \psi(x)$, where $\phi(x)$ and
$\psi(x)$ are the fields corresponding to the fundamental particles.
Those are not bound state wave functions which would be a function of
two space-time points, i.e. $ \Psi(x,y) = \bar \phi(x) \psi(y)$. The
space-time separation is taken to be vanishing.

\section{The duality}

As usual in a quantum field theory, the problem is to identify the
physical degrees of freedom. To do so we have to choose the gauge in
the appropriate way. The Higgs doublet can be used to fix the gauge.
Using the gauge freedom of the local $SU(2)$ group we perform a gauge
rotation such that the scalar doublet takes the form:

\begin{eqnarray}
   h_i=\left ( \begin{array}{c} F+h_{(1)} \\ 0 \end{array}\right),
\end{eqnarray}
where the parameter $F$ is a real number.  If $F$ is sufficiently
large we can perform an $1/F$ expansion for the fields defined above.
We have
\begin{eqnarray} \label{def2a}
 \nu_L&=&\frac{1}{F}(\bar h l)=l_1+\frac{1}{F} h_{(1)} l_1
=l_1+{\cal O}\left(\frac{1}{F}\right)
 \approx l_1 \\
  e_L&=&\frac{1}{F}(\epsilon^{ij} h_i l_j)=
  l_2+\frac{1}{F} h_{(1)} l_2
=l_2+{\cal O}\left(\frac{1}{F}\right)
  \approx l_2 \\
   u_L&=&\frac{1}{F}(\bar h q)=q_1+\frac{1}{F} h_{(1)} q_1
=q_1+{\cal O}\left(\frac{1}{F}\right)
   \approx q_1 \\
     d_L&=&\frac{1}{F}(\epsilon^{ij} h_i q_j)=q_2
     +\frac{1}{F} h_{(1)} q_2
=q_2+{\cal O}\left(\frac{1}{F}\right)
     \approx q_2 \\
     \phi&=&\frac{1}{2 F}(\bar h h)=h_{(1)}+\frac{F}{2} +\frac{1}{2 F} 
     h_{(1)} h_{(1)} \\ \nonumber 
&=&h_{(1)}+\frac{F}{2} +
     {\cal O}\left(\frac{1}{2 F}\right)
     \approx h_{(1)}+\frac{F}{2} \\
     W^3_{\mu}&=& \frac{2 i}{g F^2} ( \bar h D_{\mu} h) =
     \left ( 1 + 
     \frac{h_{(1)}}{F} \right)^2 B^3_{\mu} + \frac{2 i}{g F} \left (1+ 
     \frac{h_{(1)}}{F} \right) \partial_{\mu} h_{(1)}
    \\ \nonumber   
&=& B^3_{\mu} +{\cal O}\left(\frac{2}{F}
   \right)
   \approx B^3_{\mu}
     \\
   W^-_{\mu}&=& \frac{\sqrt{2} i}{g F^2} ( \epsilon^{ij} h_i D_{\mu} h_j) = \left ( 1 + 
     \frac{h_{(1)}}{F} \right)^2 B^-_{\mu} \\ \nonumber 
&=& B^-_{\mu}+{\cal O}\left(\frac{2}{F}
   \right)
   \approx B^-_{\mu}
   \\ \label{def2b}
   W^+_{\mu}&=& \left (\frac{\sqrt{2} i}{g F^2 }
     ( \epsilon^{ij} h_i D_{\mu} h_j)\right)^\dagger
 = \left ( 1 + 
     \frac{h_{(1)}}{F} \right)^2 B^+_{\mu} \\ \nonumber 
&=& B^+_{\mu}+{\cal O}\left(\frac{2}{F}
   \right)   \approx B^+_{\mu}.
\end{eqnarray}
The bound states have been normalized such that the expansion yields a
expression having the right mass dimension.

The parameter $g$ is the coupling constant of the gauge group
$SU(2)_{L}$ and $D_{\mu}$ is the corresponding covariant derivative.
As can be seen from (\ref{def2a}) to (\ref{def2b}), the physical
particles are those appearing in the standard model. We adopt the
usual notation $B^\pm_\mu=(B^1_\mu \mp i B^2_\mu)/\sqrt{2}$.  The
terms which are suppressed by the large scale $F$ are as irrelevant as
the terms which are neglected in the Higgs phase when the Higgs field
is expanded near its classical vacuum expectation value.  If we match
the expansion for the Higgs field $\phi=h_{(1)}+\frac{F}{2}$ to the
standard model, we see that $F=2 v=492$ GeV where $v$ is the vacuum
expectation value.  This parameter can be identified with a typical
scale for the theory in the confinement phase. The physical scale is
defined as $\Lambda=F/\sqrt{2}$, the $\sqrt{2}$ factor is included
here because the physical parameter is not $v$ but $v/\sqrt{2}$ as can
be seen from the Lagrangian (\ref{SML}). We see in the expansion for
$W^3_\mu$ that the suppressed irrelevant terms start at the order
$2/F$. We thus interpret the typical scale for the $W^3_\mu$ as
$\Lambda_W=\sqrt{2} F/4=173.9$ GeV. The scale corresponding to the
Higgs boson is deduced in a similar fashion. We find
$\Lambda_H=\sqrt{2} F=695.8$ GeV. The factor four between the scale of
the Higgs boson and that of the $W$ bosons is dictated by the
underlying algebraic structure of the gauge theory. In a similar
fashion, one could argue that the typical scale of the electroweak
interactions in the Higgs phase, is given by the scale around which
the Higgs field is expanded.

The bound states we are considering are point-like objects but with an
extension in momentum space corresponding to the typical scale of the
particle, which can thus be used a a cut-off in higher order
calculations.

At these stage, we shall like to stress that this model satisfies 't
Hooft criteria of anomaly matching \cite{'tHooft:1980xbis} which
states that chiral symmetry remains unbroken if the fundamental
fermions develop the same anomaly as the massless bound states
fermions.

\subsection{The gauge invariant standard model}
In this section we shall show that the standard model Lagrangian can
be rewritten using gauge invariant fields
\cite{'tHooft:1998pk,'tHooft:1980xb,Mack:1977xu}. Let us define the following $SU(2)$
gauge invariant fields
\begin{eqnarray}
  \Phi&=& \Omega^\dagger \phi,
  \\ \nonumber
\Psi^a_L&=&  \Omega^\dagger \psi^a_L
\\ \nonumber
W^i_\mu&=& \frac{i}{2g} \mbox{Tr} \ \Omega^\dagger \stackrel{\leftrightarrow}
{D}_\mu \Omega \tau^i
\\ \nonumber
({\cal D}_\mu)_{SU(2)}&=& \Omega^\dagger (D_\mu)_{SU(2)}
\Omega=\partial_\mu - i g W_\mu
\\ \nonumber
{\cal F}_{\mu \nu}&=&\frac{i}{g}[({\cal D}_\mu)_{SU(2)},({\cal D}_\nu)_{SU(2)}],
\end{eqnarray}
with $\phi^* \stackrel{\leftrightarrow}
{\partial}_\mu \phi= \phi^* \partial_\mu \phi -  \phi \partial_\mu \phi^*$ and where $\Omega$ is a gauge transformation given by
\begin{eqnarray}
\Omega=\frac{1}{\sqrt{\phi^\dagger
    \phi}}\left(\begin{array}{cc}  \phi_2^* & \phi_1 
    \\ -\phi_1^* & \phi_2
  \end{array}
\right ).
\end{eqnarray}
We start from the Lagrangian of the standard model
\begin{eqnarray}
{\cal L}&=&-\frac{1}{2} \mbox{Tr}F_{\mu \nu}F^{\mu \nu}-
\frac{1}{4} f_{\mu \nu}f^{\mu \nu}
+ i \bar \psi^a_L \left (
  (\fmslash D_\mu)_{SU(2)} - i\frac{1}{2} g' Y \fmslash {\cal A}
\right )  \psi^a_L
\\ \nonumber &&
+  i \bar \psi^a_R ( \fmslash \partial - i\frac{1}{2} g' Y \fmslash {\cal A} )
\psi^a_R
\\ \nonumber &&
+ \left( \left (  (D_\mu)_{SU(2)} - i\frac{1}{2} g' Y {\cal A}_\mu \right)
  \phi \right)^\dagger
\left( \left (  (D^\mu)_{SU(2)} - i\frac{1}{2} g' Y {\cal A}^\mu \right)
  \phi \right)
\\ \nonumber &&
+ V(\phi^\dagger \phi) - G_u ( \bar \psi^a_L \phi \psi^a_R
+ \bar \psi^a_R \phi^\dagger \psi^a_L )    - G_d ( \bar \psi^a_L \bar \phi \psi^a_R + \bar \psi^a_R \bar \phi^\dagger \psi^a_L )
\end{eqnarray}
where $\bar \phi=i \tau_2 \phi^\star$, the fermion $\psi^a$ is a
generic fermion field and the index $a$ runs over all the lepton and
quark flavors and the covariant derivative is given by $({\cal
  D}_\mu)_{SU(2)}=\partial_\mu -i g B_\mu$, with $B_\mu=1/2 \tau^a
B^a_\mu$. We denote the Yukawa couplings by $G_u$ and $G_d$.  The
gauge dependent fields can be replaced by their $SU(2)$ gauge
invariant counterparts.  One obtains
\begin{eqnarray}
{\cal L}&=&-\frac{1}{2} \mbox{Tr} {\cal F}_{\mu \nu} {\cal F}^{\mu \nu}-
\frac{1}{4} f_{\mu \nu}f^{\mu \nu}
+ i \bar \Psi^a_L \left (
  (\fmslash {\cal D}_\mu)_{SU(2)} - i\frac{1}{2} g' Y \fmslash {\cal A}
\right )  \Psi^a_L
\\ \nonumber &&
+  i \bar \psi^a_R ( \fmslash \partial - i\frac{1}{2} g' Y \fmslash {\cal A} )
\psi^a_R
\\ \nonumber &&
+ \left( \left (  ({\cal D}_\mu)_{SU(2)} - i\frac{1}{2} g' Y {\cal A}_\mu \right)
  \Phi \right)^\dagger
\left( \left (  ({\cal D}^\mu)_{SU(2)} - i\frac{1}{2} g' Y {\cal A}^\mu \right)
  \Phi \right)
\\ \nonumber &&
+ V(\Phi^\dagger \Phi)- G_u ( \bar \Psi^a_L \Phi \psi^a_R
+ \bar \psi^a_R \Phi^\dagger \Psi^a_L )    - G_d ( \bar \Psi^a_L \bar \Phi \psi^a_R + \bar \psi^a_R \bar \Phi^\dagger \Psi^a_L ).
\end{eqnarray}
The scalar field potential is taken of the form
\begin{eqnarray}
  V(\phi^\dagger \phi)=\frac{1}{2}\lambda \left( \phi^\dagger \phi -\frac{1}{2}v^2 \right)^2,
\end{eqnarray}
its $SU(2)$ gauge invariant counterpart is given by
\begin{eqnarray}
  V(\Phi^\dagger \Phi)=\frac{1}{2}\lambda \left( \Phi^\dagger \Phi -\frac{1}{2}v^2 \right)^2.
\end{eqnarray}
This potential can be minimized if the field $\Phi$ is forced to form
the gauge invariant condensate
\begin{eqnarray}
\langle
\Phi^\dagger \Phi \rangle=\langle
\phi^\dagger \phi \rangle=\frac{1}{2} v^2.
\end{eqnarray}
In that case we see that the gauge invariant charged vector bosons
receive a mass term of the form $m_W=g v/2$, the fermions receive
masses of the type $m_u=G_u v/\sqrt{2}$ for the up-type fermions and
$m_d=G_d v/\sqrt{2}$ for the down-type fermions. We also see that a term
\begin{eqnarray}
\frac{1}{2} g'g{\cal A}_\mu W^3_\mu \Phi^\dagger \Phi
\end{eqnarray}
appears, which gives rise to a mixing between the $U(1)$ generator and
the $W^3$ gauge invariant field.  After diagonalization according to
\begin{eqnarray}
         A_\mu&=& \sin{\theta_W} W^3_\mu+ \cos{\theta_W}{\cal A}_\mu
         \nonumber \\
         Z_\mu&=& \cos{\theta_W} W^3_\mu-  \sin{\theta_W}{\cal A}_\mu,
\end{eqnarray}
we find the correct property for the electromagnetic photon $A_\mu$
which couples with the right strength to the fermions and which is
massless. We also find the right property for the $Z_\mu$ boson whose
mass is shifted above that of other members of the triplet by the
electromagnetic interaction. Choosing the unitarity gauge, which
corresponds to the choice $\Omega=1$, one finds $\Phi\to \phi$, $\Psi
\to \psi$ and $W^i_\mu\to B^i_\mu$. This formulation is identical to
that presented in (\ref{def2a}-\ref{def2b}) if the higher dimensional
operators are neglected in (\ref{def2a}-\ref{def2b}). This is what is
done when one expands the Higgs fields around its vacuum expectation
value. Nevertheless for our purposes, the equations
(\ref{def2a}-\ref{def2b}) are more adequate as they describe
explicitly the relevant scale for each particle.

\section{The relation to lattice gauge theory}

Osterwalder and Seiler have shown that there is no fundamental
difference between the confinement phase and the Higgs phase of a
theory if there is a Higgs boson in the fundamental representation of
the gauge group \cite{Osterwalder:1978pc}. This is known as the
complementarity principle. 

\begin{figure}
\begin{center}
\epsfxsize=8cm
\epsffile{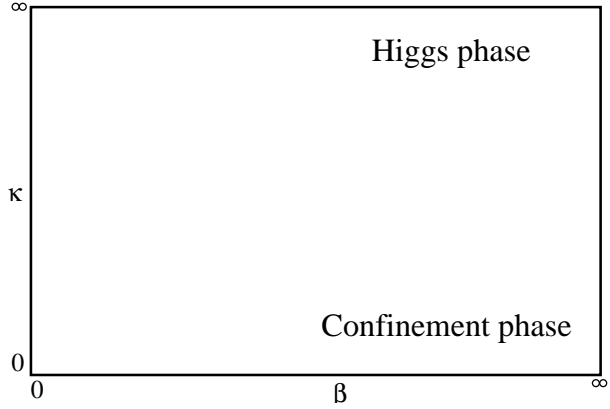}
\caption{Sketch of the phase diagram using the lattice method
  and frozen Higgs approximation. There is no phase transition between
  the Higgs phase and the confinement phase.}
\protect{\label{phase1}}
\end{center}
\end{figure}

\begin{def3}[Complementarity principle]
  If there is a Higgs boson in the fundamental representation of the
  gauge group then there is no phase transition between the Higgs and
  the confinement phase. 
\end{def3}
In this approach, the Higgs and confinement phase are defined at the
level of the effective action. It was shown by Fradkin and Shenker
\cite{Fradkin:1979dv} following the work of Osterwalder and Seiler
\cite{Osterwalder:1978pc} that in the lattice gauge theory there is no
phase transition between the the $SU(2)$ Yang-Mills-Higgs theory in
the confinement phase and in the Higgs phase (see figure \ref{phase1})
using the approximation of a frozen Higgs field and restricting
themselves to a $SU(2)$ gauge theory without fermions.

In order to understand this phenomenon we have to describe the lattice
Euclidean action. It reads:
\begin{eqnarray}
  S&=&S_g-\lambda \sum_x\left[ \phi^\dagger(x) \phi(x)-1\right]^2
  -\sum_{x,y,\mu} \phi^\dagger(x)Q_\mu(x,y) \phi(y),
  \end{eqnarray}
  where $S_g$ is the pure gauge piece
  \begin{eqnarray}
S_g&=&\frac{1}{g^2} \sum_{\rm plaq.}
     \left[ {\rm Tr}\, U_p +{\rm Tr}\, U_p^\dagger\right],
    \end{eqnarray}
    using the usual definition $\beta=4/g^2$.  The matrix $Q$ which
    couples the Higgs field to the link variables $U(x)$ reads
   \begin{eqnarray}
   Q_\mu(x,y)&=&\delta_{x,y} -\kappa \left [ \delta_{x,y-\mu} U_\mu(x)
     + \delta_{x,y+\mu} U_\mu^\dagger (x) \right ]
     \end{eqnarray}
     with a Higgs ``hopping parameter'' $\kappa$.
This action can be related to the Euclidean space-time continuum action
\begin{eqnarray}
  S_{cont.}&=&-\int d^4x\left [ |D_\mu \tilde \phi(x)|^2
    +m^2 |\tilde \phi(x)|^2 +\tilde \lambda |\tilde \phi(x)|^4 \right ]
  \end{eqnarray}
with $D_\mu=\partial_\mu +i g A_\mu(x)$ using the following relations
\begin{eqnarray}
  \tilde \phi(x) =\frac{\sqrt{\kappa}}{a} \phi(x), \
  \tilde \lambda=\frac{\lambda}{\kappa^2} \
  {\rm and} \
  m^2=\frac{1-2 \lambda - 8 \kappa}{\kappa a^2}.
\end{eqnarray}
Thus high values of $\kappa$ correspond to a negative mass for the
Higgs field and therefore to the Higgs phase whereas low values
correspond to a positive mass and therefore to the confinement phase.
This phase diagram was obtained making the assumption that no physical
information is lost when the Higgs field is frozen that is for
$\lambda=\infty$.
  \begin{figure}
\begin{center}
\epsfxsize=8cm
\epsffile{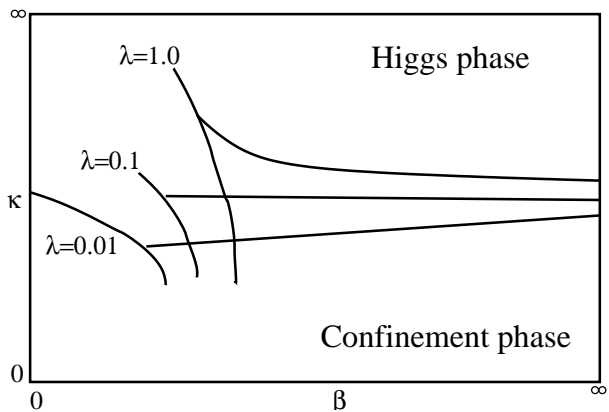}
\caption{Sketch of the phase diagram using the mean field techniques.} 
\protect{\label{phase2}}
\end{center}
\end{figure}
However some care has to be taken with the notion of complementarity
since it was shown by Damgaard and Heller \cite{Damgaard:1985nb} that
for certain small values of $\lambda$ a phase transition can appear
(see figure \ref{phase2}).  They performed an analysis of the phase
diagram of the $SU(2)$ gauge theory allowing the Higgs field to
fluctuate in magnitude using so-called mean field techniques.
Nevertheless the lattice method is more reliable than mean field
approximation techniques. The exact shape of phase diagram of the
theory is still an open question.

If there is no phase transition as conjectured by Osterwalder and
Seiler \cite{Osterwalder:1978pc} this implies that there is no
distinction between the two phases. This is analogous to the fact that
there is no distinction between the gaseous and liquid phases of
water. A continuous transition between the two phases is possible.

\begin{figure}
\begin{center}
\epsfxsize=8cm
\epsffile{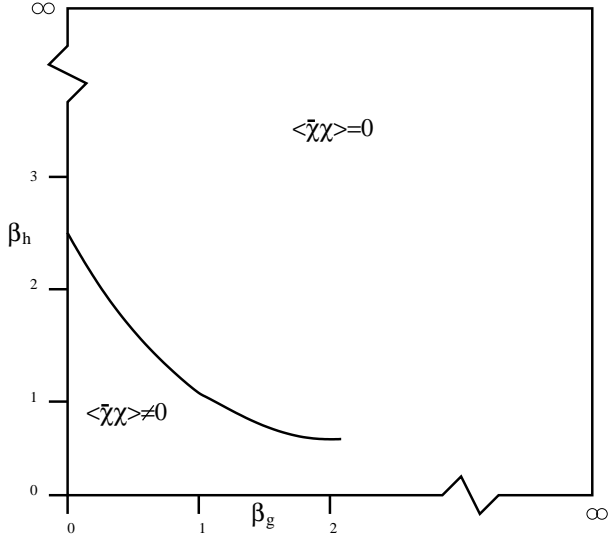}
\caption{Sketch of the chiral phase transition diagram.} 
\protect{\label{chiral}}
\end{center}
\end{figure}

Till this point, we were considering gauge theories that contain only
scalars. Nevertheless, if the complementarity is to be applied to the
standard model, fermions must be introduced in the theory. Therefore a
second phase diagram describing the chiral phase transition has to be
studied. This issue has been studied by Aoki, Lee and Shrock
\cite{Aoki:1988aw}. In order to overcome the well known difficulty of
placing chiral fermions on the lattice, they have rewritten the chiral
$SU(2)$ theory in a vectorlike form. However, this requires a very
specific form for the Yukawa couplings. Indeed the number of possible
Yukawa couplings has to be reduced and it is thus impossible to give
different masses to each of the fermion mass eigenstates. This is a
very serious limitation to their analysis as clearly the full standard
model with all its Yukawa couplings cannot be rewritten in a
vectorlike theory.  Aoki et al. have found that a phase transition
appears between the phase at weak gauge coupling and the phase at
large coupling (see figure \ref{chiral}). In their notation $\beta_h$
is proportional to the hopping parameter.
The standard model and the confining model at weak coupling we are
discussing are probably in the same phase in that phase diagram as the
chiral phase transition is dominantly determined by the strength of
the weak gauge coupling constant.  Nevertheless this analysis is a
constraint for models making use of the complementarity principle to
relate gauge theories at weak coupling and strong coupling constant.

All these analyses were performed a long time ago. It would be
important to study the phase diagram of the standard model using some
more modern techniques. The lack of phase transition has some very
deep consequences. If it is the case this implies that the mass
spectrum of both theories are really identical, there are the same
numbers of degrees of freedom and thus no new particle in the
confinement phase. Both theories are then identical.
\subsection{Discussion}

It had long been noted in the literature that the standard model can
be rewritten in terms of gauge invariant bound states, the so-called
confinement phase, but it has never been stressed that this represents
a new theory which is dual to the standard model. As we will see in
the next chapter, this duality allows to find relations between the
parameters of the standard model which are not apparent in the Higgs
phase and is therefore not trivial.

\begin{figure}
\begin{center}
\epsfxsize=10cm
\epsffile{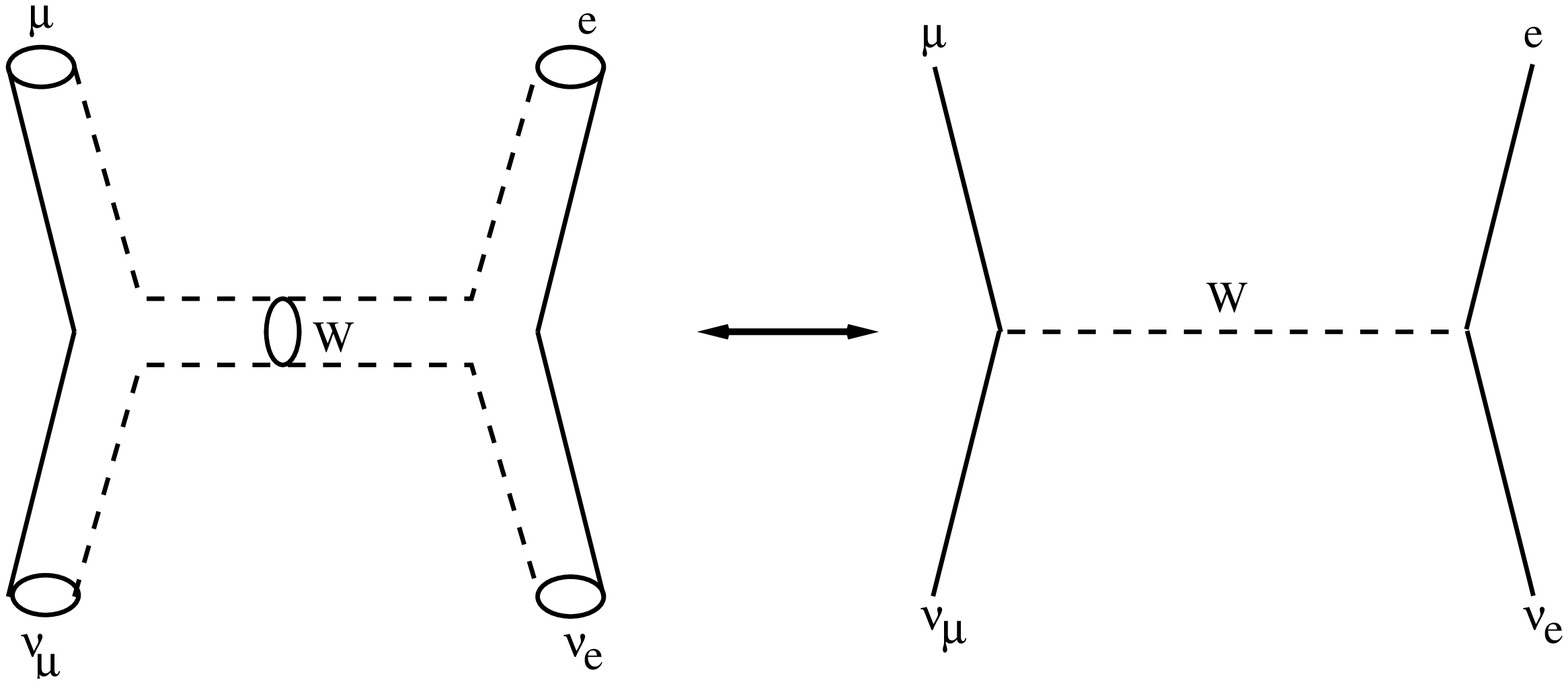}
\caption{Transition from the duality diagram to the Feynman graph.} 
\protect{\label{dualgraph}}
\end{center}
\end{figure}

We have presented above a duality between the Higgs phase of the
standard model Lagrangian and the confinement phase of the same
Lagrangian at weak coupling. We have shown that the fields of the
standard model can be rewritten in gauge invariant manner. This
implies that the duality diagram (diagrams in the confinement phase)
can be evaluated in the Higgs phase using perturbation theory. The
lines of the duality diagrams are shrinking together when moving from
the confinement phase to the Higgs phase (see graph \ref{dualgraph}).
This follows from the fact that the standard model can be rewritten in
terms of gauge invariant fields and that in a certain gauge, the
unitarity gauge, we obtain the usual standard model.  The idea that
the standard model in the Higgs phase and in the confinement phase are
dual if the confinement is caused by a weak coupling is supported by
the complementary principle.

This duality allows to identify relations between some of the
parameters of the standard model. In particular we shall see that the
electroweak mixing angle can be related to the typical scale of the
$W$-bosons which allows to compute this parameter. The mass of the
Higgs boson can be related to that of the $W$-bosons in the
confinement phase because the Higgs boson is the ground state of the
theory and the $W$-bosons are the excited states corresponding to this
ground state.

\section{A global SU(2) symmetry}

In the absence of the $U(1)$ gauge group the theory has a global
$SU(2)$ symmetry besides the local $SU(2)$ gauge symmetry. The scalar
fields and their complex conjugates can be written in terms of two
doublets arranged in the following matrix:
\begin{eqnarray}
M=\left(\begin{array}{cc} h_2^* & h_1 \\ -h_1^* & h_2 \end{array}
\right ).
\end{eqnarray}
 The potential of the scalar field $V(h h^*)$ depends solely on
\begin{eqnarray}
 h^* h&=& h_1^* h_1 +h_2^* h_2 \\ \nonumber 
 &=&(\mbox{Re}\, h_1)^2 +  (\mbox{Im\,} h_1)^2
 + (\mbox{Re}\, h_2)^2 +  (\mbox{Im}\, h_2)^2 = \mbox{det} M.
\end{eqnarray}
{}This sum is invariant under the group $SO(4)$, acting on the real
vector $H$
\begin{eqnarray}
H=(\mbox{Re} \, h_1,\mbox{Im} \, h_1,\mbox{Re} \, h_2,\mbox{Im}
\, h_2).
\end{eqnarray}
 This group is isomorphic to $SU(2) \times SU(2)$. One of
these groups can be identified with the confining gauge group
$SU(2)_L$, since $\mbox{det} M$ remains invariant under $SU(2)_L$:
\begin{eqnarray}
  \mbox{det} (UM)= \mbox{det} (M), \  U \in SU(2)_L.
\end{eqnarray}

Now the second $SU(2)$ factor can be identified by considering the matrix
$M^\top$

\begin{eqnarray}
M^\top=\left(\begin{array}{cc}  h_2^* & -h_1^* \\ h_1 & h_2\end{array}
\right ).
\end{eqnarray}
The determinant of $M^\top$, which is equal to $\mbox{det} (M)$,
remains invariant under a $SU(2)$ transformation acting on the
doublets $(h_2^*,h_1)$ and $(-h_1^*,h_2)$.

These transformations commute with the $SU(2)_L$ transformations. They
constitute the flavor group $SU(2)_F$, which is an exact symmetry as
long as no other gauge group besides $SU(2)_L$ is present. With
respect to $SU(2)_F$ the $W$-bosons form a triplet of states
$(W^+,W^-,W^3)$.  The left-handed fermions form $SU(2)_F$ doublets.
Both the triplet as well as the doublets are, of course, $SU(2)_L$
singlets. Once we fix the gauge in the $SU(2)_L$ space such that
$h_2=0$ and $\mbox{Im} \, h_1 =0$, the two $SU(2)$ groups are linked
together, and the $SU(2)_L$ doublets can be identified with the
$SU(2)_F$ doublets. The global and unbroken $SU(2)$ symmetry dictates
that the three $W$-bosons states, forming a $SU(2)_F$ triplet, have
the same mass. Once the Yukawa-type interactions of the fields $e_R$,
$u_R$ and $d_R$ with the corresponding left-handed bound systems are
introduced, the flavor group $SU(2)_F$ is in general explicitly
broken.  This symmetry is the analogon of the custodial symmetry,
present in the Higgs phase of the theory.

\section{Electromagnetism and mixing}

The next step is to include the electromagnetic interaction. The gauge
group is $SU(2)_L \times U(1)_Y$, where $Y$ stands for the
hypercharge. The covariant derivative is given by

 \begin{eqnarray} \label{covder}
        D_{\mu}=\partial_{\mu}-i \frac{g'}{2} Y {\cal A}_\mu -
        i \frac{g}{2} \tau^a B^a_{\mu}.
       \end{eqnarray}
 The assignment for $Y$ is as follows:
   \begin{eqnarray}
   Y \left( \begin{array}{c}
     l_1\\
   l_2
   \end{array} \right) & = & \left( \begin{array}{cc}
                 1 & 0\\
           0 &  1 \end{array} \right)
         \left( \begin{array}{c}
         l_1\\
         l_2 \end{array} \right) \nonumber \\
   Y \left( \begin{array}{c}
     q_1\\
   q_2
   \end{array} \right) & = & \left( \begin{array}{cc}
                  - \frac{1}{3}& 0\\
           0 & - \frac{1}{3} \end{array} \right)
         \left( \begin{array}{c}
             q_1\\
             q_2 \end{array} \right)\nonumber \\
    Y \left( \begin{array}{c}
     h_1\\
   h_2
   \end{array} \right) & = & \left( \begin{array}{cc}
                     1  & 0 \nonumber \\
              0 &  1 \end{array} \right)
         \left( \begin{array}{c}
             h_1\\
             h_2 \end{array} \right).
\end{eqnarray}
The complete Lagrangian of the model in the confinement phase is given by:
\begin{eqnarray}
  {\cal L}_{c}&=&-\frac{1}{4} G^{a}_{\mu \nu}  G^{a \mu \nu}
                 -\frac{1}{4} f_{\mu \nu}  f^{\mu \nu}
                 + \bar l_L i \fmslash{D} l_L+ \bar q_L i \fmslash{D} q_L
                 +\bar e_R i \fmslash{D} e_R \\ \nonumber &&
                 +\bar u_R i \fmslash{D} u_R
                 + \bar d_R i \fmslash{D} d_R-G_e \bar e_R (\bar{h} l_L)
                 -G_d \bar d_R (\bar {h} q_L)\\ \nonumber &&
                 -G_u \bar u_R (h q_L)
                 +h.c.
                 +\frac{1}{2}(D_{\mu} h)^\dagger
                 (D^{\mu} h)-\frac{m_h^2}{2} h h^\dagger
                 -\frac{\lambda}{4} (h h^\dagger)^2,
\end{eqnarray}
where $m_h^2>0$ and
\begin{eqnarray} \label{gmunu}
G^a_{\mu\nu}&=&\partial_\mu B^a_\nu-  \partial_\nu
B^a_\mu+g \epsilon^{abc}B^b_\mu B^c_\nu, \\ \nonumber
f_{\mu\nu}&=&\partial_\mu {\cal A}_\nu-  \partial_\nu
{\cal A}_\mu.
\end{eqnarray}

The $U(1)$ gauge group is an unbroken gauge group, like $SU(2)_L$. The
hypercharge of the $h$ field is $+1$, and that of the $h^*$ field is
$-1$, i.e.  the members of the flavor group $SU(2)_F$ have different
charge assignments. Thus the group $SU(2)_F$ is dynamically broken,
and a mass splitting between the charged and neutral vector bosons
arises.  The neutral electroweak boson $(\bar h D_\mu h)$, which is
not a gauge boson, mixes with the gauge boson ${\cal A}_\mu$.  As a
result these bosons are not mass eigenstates, but mixed states.  The
neutral electroweak boson $Z_\mu$ is a superposition of $(\bar h D_\mu
h)$ and of ${\cal A}_\mu$. The photon is the state orthogonal to the
neutral electroweak boson $Z_\mu$. The strength of this mixing depends
on the internal structure of the electroweak bosons.

We emphasize that the $B^a_\mu$ gauge bosons are as unphysical as the
gluons are in QCD. The hyperphoton ${\cal A}_\mu$ is not the physical
photon $A_\mu$ which is a mixture of ${\cal A}_\mu$ and of the bound
state $(\bar h D_\mu h)$. The fundamental D-quarks do not have an
electric charge but only a hypercharge. These hypercharges give a
global hypercharge to the bound states, and one can see easily that a
bound state like the electron has a global hypercharge and will thus
couple to the physical photon, whereas a neutrino has a vanishing
global hypercharge and thus will remain neutral with respect to the
physical photon. So we deduce that QED is not a property of the
microscopic world described by ${\cal L}_{c}$ but rather a property of
the bound states constructed out of these fundamental fields. The
theory in the confinement phase apparently makes no prediction
concerning the strength of the coupling between the bound states and
the electroweak bosons and the physical photon. This information can
only be gained in the Higgs phase.

The mixing between the two states can be studied at the macroscopic
scale, i.e. the theory of bound states, where one has
\begin{eqnarray}
         A_\mu&=& \sin{\theta_W} W^3_\mu+\cos{\theta_W}{\cal A}_\mu 
         \nonumber \\
         Z_\mu&=&\cos{\theta_W} W^3_\mu -
         \sin{\theta_W} {\cal A}_\mu.
\end{eqnarray}
Here $\theta_W$ denotes the electroweak mixing angle, and $A_\mu$
denotes the photon field.

\chapter{Making use of the duality}
In this chapter we shall make use of the duality to compute the weak
mixing angle and the the Higgs boson mass. The results of this chapter
were published in \cite{Calmet:2000th,Calmet:2001rp}
\section{Calculation of the weak mixing angle}
The electroweak mixing angle can be calculated using an effective
theory and a potential model to simulate the wave function of the
constituent.

In section chapter 2, we have matched the expansion for the Higgs
field to the standard model.  Using this point of view based on the
effective theory concept, we obtained a scale of $\Lambda_W=173.9$ GeV
for this boson.  Here we shall consider an effective Lagrangian to
simulate the effect of the $SU(2)_L$ confinement.

This Lagrangian was originally considered in an attempt to describe
the weak interactions without using a gauge theory \cite{Bjorken}.
The effective Lagrangian is given by
\begin{eqnarray} \label{effLag}
 {\cal L}_{eff}&=&-\frac{1}{4} F_{\mu \nu} F^{\mu \nu}
                            -\frac{1}{4} W^a_{\mu \nu} W^{a \mu \nu}
                            -\frac{1}{2} m^2_W W^a_\mu W^{a \mu}
                             \\ &&
                              -\frac{1}{4} \lambda \left
                              ( F_{\mu \nu} W^{3 \mu \nu}
                              + W^3_{\mu \nu} F^{\mu \nu} \right ) \nonumber
\end{eqnarray}
where we have
\begin{eqnarray}
           W^a_{\mu \nu}&=& \partial_\mu W^a_\nu -  \partial_\nu W^a_\mu,
          \nonumber \\ 
          F_{\mu \nu}&=& \partial_\mu {\cal A}_\nu - \partial_\nu {\cal A}_\mu.
\end{eqnarray}

\begin{figure}
\begin{center}
\epsffile{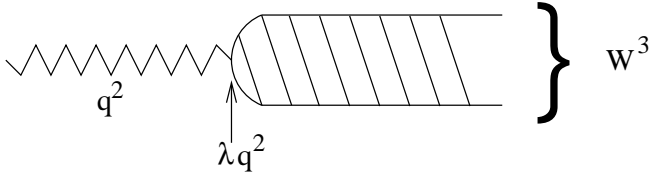}
\caption{Hyperphoton transition into a $W^3$} 
\protect{\label{vertex}}
\end{center}
\end{figure}

The first term in the effective Lagrangian (\ref{effLag}) describes
the field of the hyperphoton, the second term three spin one bosons
and the third term is a mass term which is identical for the three
spin one bosons.  In our case, the fourth term describes an effective
mixing between $W^3$-boson and the hyperphoton.

The effective mixing angle of the Lagrangian given in
equation (\ref{effLag}) reads
\begin{eqnarray}
  \sin^2\theta=\frac{e}{g} \lambda.
\end{eqnarray}
  
Using the duality, we deduce that the mixing angle of the theory in
the confinement phase has to be the weak mixing angle and therefore
$\lambda=\sin\theta_W$.
  
The diagram in figure \ref{vertex} enables us to relate the mixing
angle to a parameter of the standard model in the confinement phase,
the typical scale $\Lambda_W$ for the confinement of the $W^3$-boson.
For the annihilation of a $W^3$-boson into a hyperphoton we consider
the following relation
\begin{eqnarray}
    \langle 0 | J^\mu_{Y}(0)|W^3\rangle&=& \frac{\epsilon^\mu}{\sqrt{2 E_W}}
    \frac{m_W^2}{f_W}=\frac{\epsilon^\mu}{\sqrt{2 E_W}} m_W F_W,
\end{eqnarray}
where $J^\mu_{Y}$ is the hyper-current, $F_W=m_W/f_W$ is the decay
constant of the $W^3$-boson, and $\epsilon^\mu$ is its polarization. The
energy of the boson is $E_W$, and the decay constant is defined as follows:
\begin{eqnarray}
    \lambda=\frac{e}{f_W}.
\end{eqnarray}
On the other hand, this matrix element can be expressed using the wave
function of the $W^3$-boson which is a $p$-wave
\begin{eqnarray}
    \langle 0 | J^\mu_{Y}(0)|W^3\rangle&=& \frac{\epsilon^\mu}{\sqrt{2 E_W}}
    \sqrt{\frac{2}{m_W}} \partial_r \phi(0).
\end{eqnarray}
This leads to the following relation for the mixing angle
\begin{eqnarray}
      \sin^2\theta_{W}&=&\frac{8 \pi \alpha}{m_W^5}
      \left(\partial_r \phi(0)\right)^2,
\end{eqnarray}
where $\alpha\approx 1/128$ is the fine structure constant, normalized
at $m_W$.

We shall now consider two different models for the wave function:
\begin{itemize}
     \item[a)] Coulombic model.\\
       We adopt the following
       ansatz for the radial wave function
         \begin{eqnarray} 
           \phi(r)&=& \frac{1}{\sqrt{3}} \left(\frac{1}{2 r_B}\right)^{3/2}
           \frac{r}{r_B} \exp\left(-\frac{r}{2 r_B}\right),
           \end{eqnarray}
           where $r_B$ is the Bohr radius. Thus we obtain
    \begin{eqnarray}
      r_B^{-1}&=&m_W\left(\frac{\pi \alpha}{3 \sin^2\theta_W} \
      \right)^{-1/5}.
 \end{eqnarray}
 If we define the typical scale for confinement as $\Lambda_W=
 r_B^{-1}$, we obtain $\Lambda_H= 157$ GeV.
    \item[b)] Three-dimensional harmonic oscillator.
      \\
      The radial part of the wave function is defined as follows:
    \begin{eqnarray}
           \phi({r})&=& \sqrt{\frac{8}{3}} \frac{\beta^{3/2}}{\pi^{1/4}}
           \beta r \exp\left(-\frac{\beta^2 r^2}{2}\right),
\end{eqnarray}
where $\beta=\sqrt{m_W \omega}$, $\omega$ being the frequency of the
oscillator. We identify the typical confinement scale $\Lambda_W$ with
the energy $E=\left(n+\frac{3}{2}\right)\omega$ corresponding to the
quantum number of a $p$-wave i.e. $n=1$, and we obtain
\begin{eqnarray}
            \omega&=&m_W
            \left(\frac{3 \sin^2\theta_W}{64 \alpha \pi^{1/2}}\right)^{2/5}
\end{eqnarray}
and $\Lambda_W=\frac{5}{2} \omega=182$ GeV.
\end{itemize}
Although we have performed a non-relativistic calculation, we see that
the values we find for the typical composite scale are in good
agreement with our expectation based on the concept of an effective
theory.

In order to estimate the value of $\sin^2 \theta_W$, we had to rely on
the simple models, discussed above. However we should like to point
out that $\sin^2 \theta_W$ is not a free parameter in our approach but
fixed by the confinement dynamics. Thus the mixing angle can in
principle be calculated taking e.g. the three dimensional harmonic
oscillator:
\begin{eqnarray} \label{wein}
      \sin^2\theta_{W}&=&\frac{256}{375} \sqrt{10 \pi} \alpha
      \left(\frac{\Lambda_W}{m_W}\right)^{5/2}.
\end{eqnarray}
We can insert the value for $\Lambda_W$ obtained from the effective
theory point of view in equation (\ref{wein}) and we obtain
$\sin^2\theta_W=0.21$ which has to be compared to the experimental
value $(\sin^2\theta_W)_{exp}=0.23$.

\section{Calculation of the Higgs boson mass}

In the confinement phase the Higgs boson is the $s$-wave of the
$SU(2)$ theory, whereas the $W$-bosons are the corresponding
$p$-waves. Thus one naively expects the Higgs boson to be lighter than
the $W$-bosons. But, as we shall show, a dynamical effect shifts the
Higgs boson mass above that of the $W$-bosons mass. The reason for
this phenomenon is the large Higgs boson scale compared to that of the
$W$-bosons.

\begin{figure}[h]
\begin{minipage}[t]{0.3\linewidth}\centering
\includegraphics[width=\linewidth]{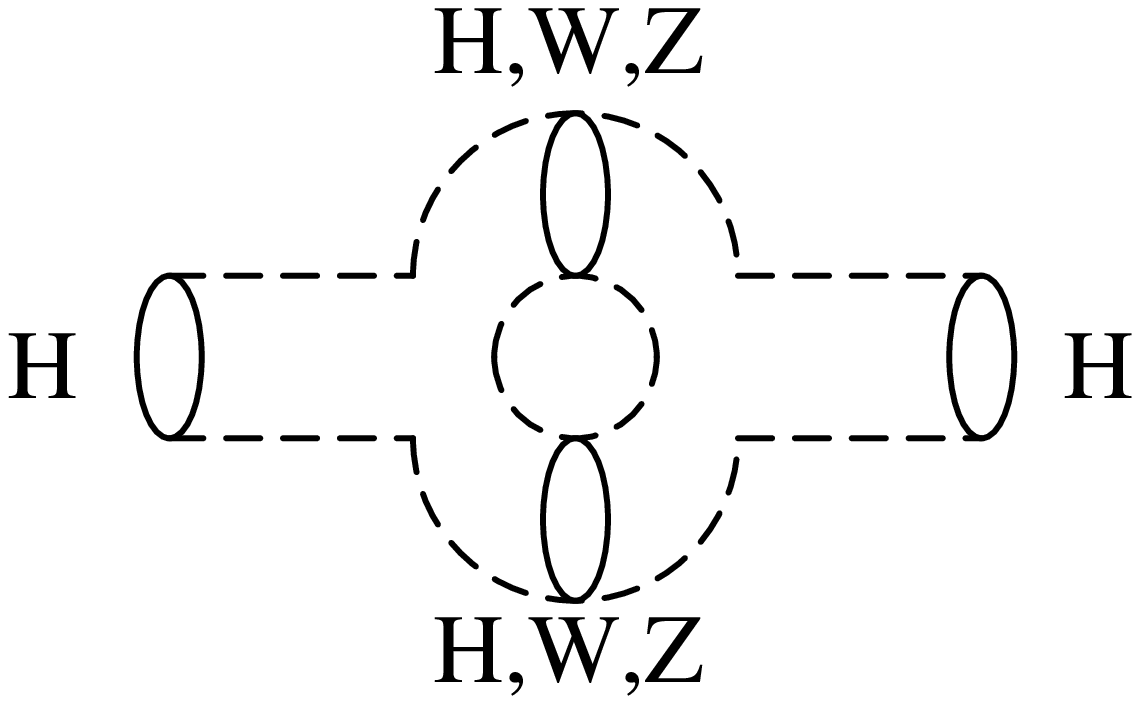}
\begin{minipage}{0.9\linewidth}
\caption{dual diagram: one loop 1PI contribution to $m_H$ \label{fig1HM}}
\end{minipage}
\end{minipage}
\begin{minipage}[t]{0.3\linewidth}\centering
\includegraphics[width=\linewidth]{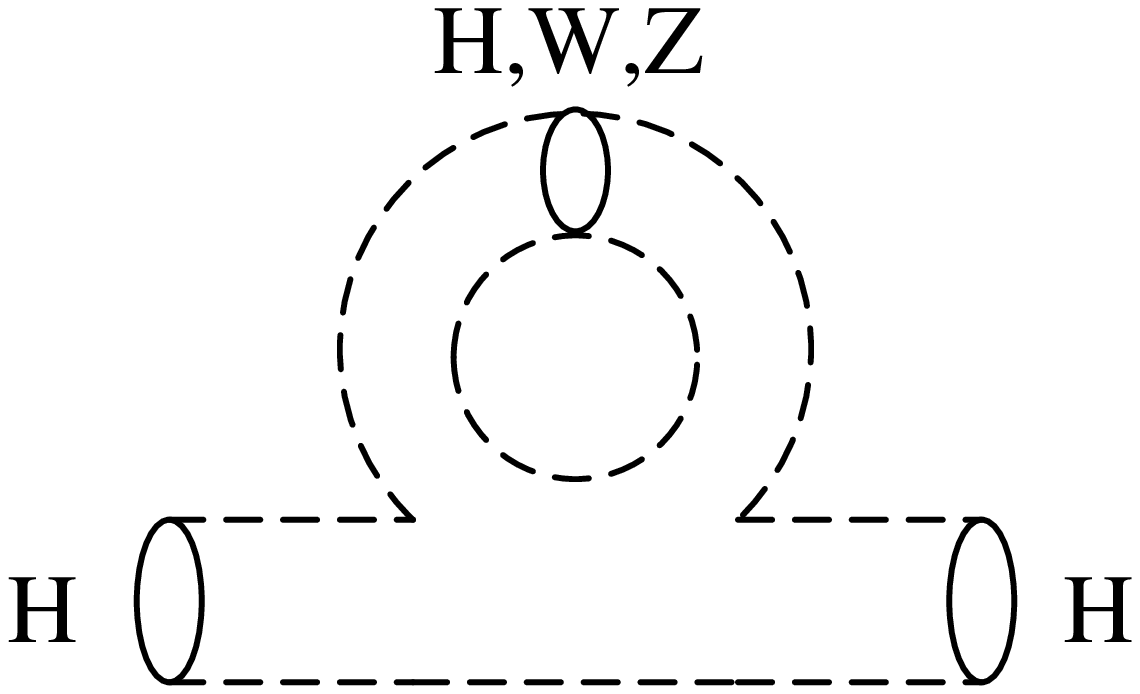}
\begin{minipage}{0.9\linewidth}
\caption{dual diagram: one loop 1PI contribution to $m_H$\label{fig2HM}}
\end{minipage}
\end{minipage}
\begin{minipage}[t]{0.3\linewidth}\centering
\includegraphics[width=\linewidth]{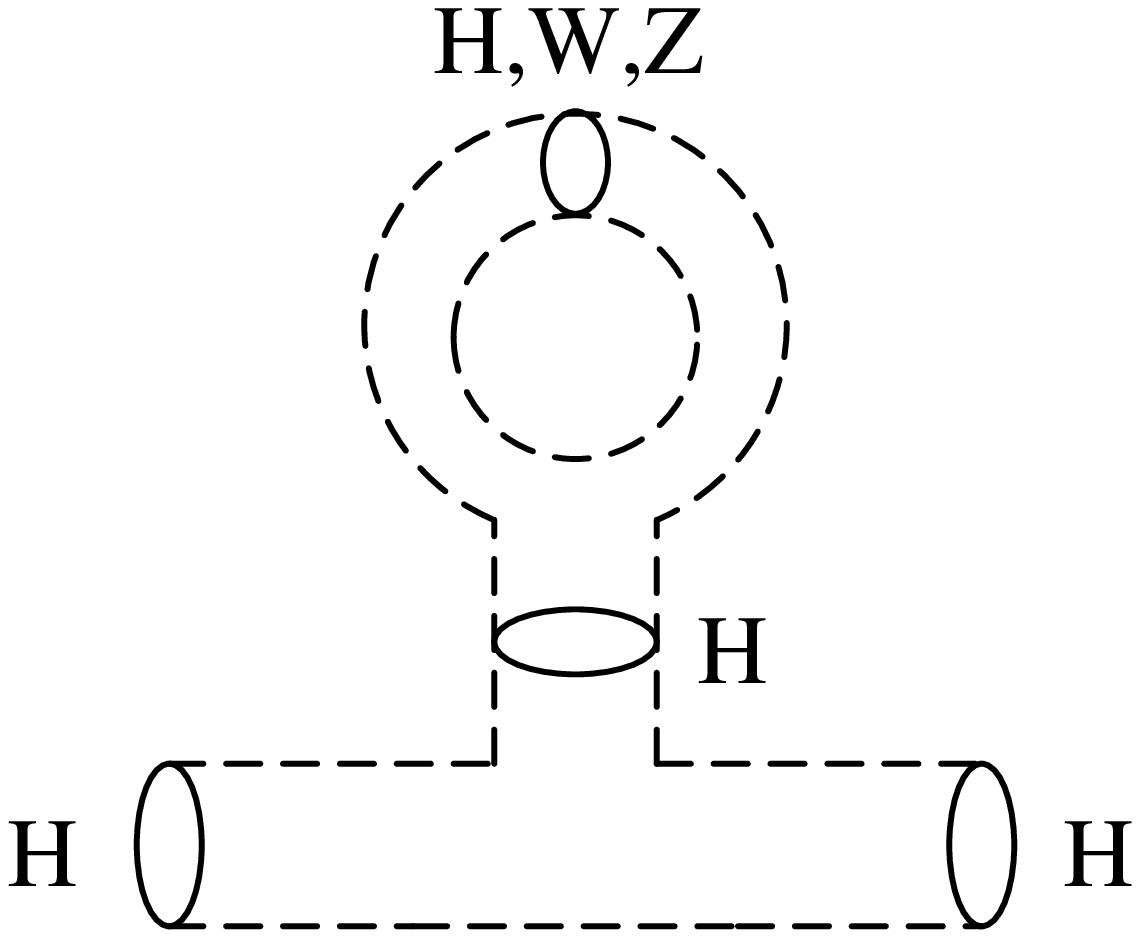}
\begin{minipage}{0.9\linewidth}
\caption{dual diagram: one loop 1PR contribution to $m_H$ \label{fig3HM}}
\end{minipage}
\end{minipage}
\end{figure}

The masses of the physical Higgs and W-bosons, being bound states
consist of a constituent mass $m_H^0=m^0_W=2 m_h$, where $m_h$ is the
mass of the scalar $D$-quark and of dynamical contributions. We have
to consider two types of diagrams: the one-particle reducible diagrams
(1PR) and the one-particle irreducible diagrams (1PI). For the Higgs
boson mass, we have to take the self-interaction and the contribution
of the $Z$ and $W^\pm$-bosons into account (see figures \ref{fig1HM},
\ref{fig2HM} and \ref{fig3HM}). The fermions couple via Yukawa coupling to
the Higgs boson, and as this interaction is not confining, fermions
cannot contribute to the dynamical mass of the Higgs boson.

The first task is to extract the constituent mass from the
experimentally measured $W$-bosons mass. The fermions contribute to
the dynamical mass of the $W$-bosons as they couple via $SU(2)$
couplings to the electroweak bosons but the divergence is only
logarithmic \cite{Veltman:1981mj} and we shall only keep the quadratic
divergences.
\begin{figure}[h]
\begin{minipage}[t]{0.3\linewidth}\centering
\includegraphics[width=\linewidth]{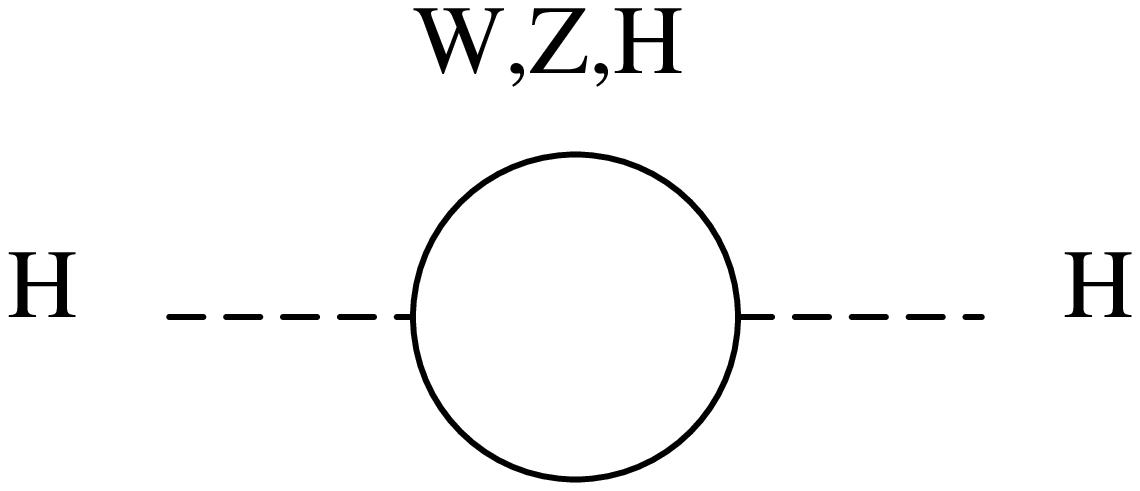}
\begin{minipage}{0.9\linewidth}
\caption{Feynman diagram: one loop 1PI contribution to $m_H$ \label{fig4HM}}
\end{minipage}
\end{minipage}
\begin{minipage}[t]{0.3\linewidth}\centering
\includegraphics[width=\linewidth]{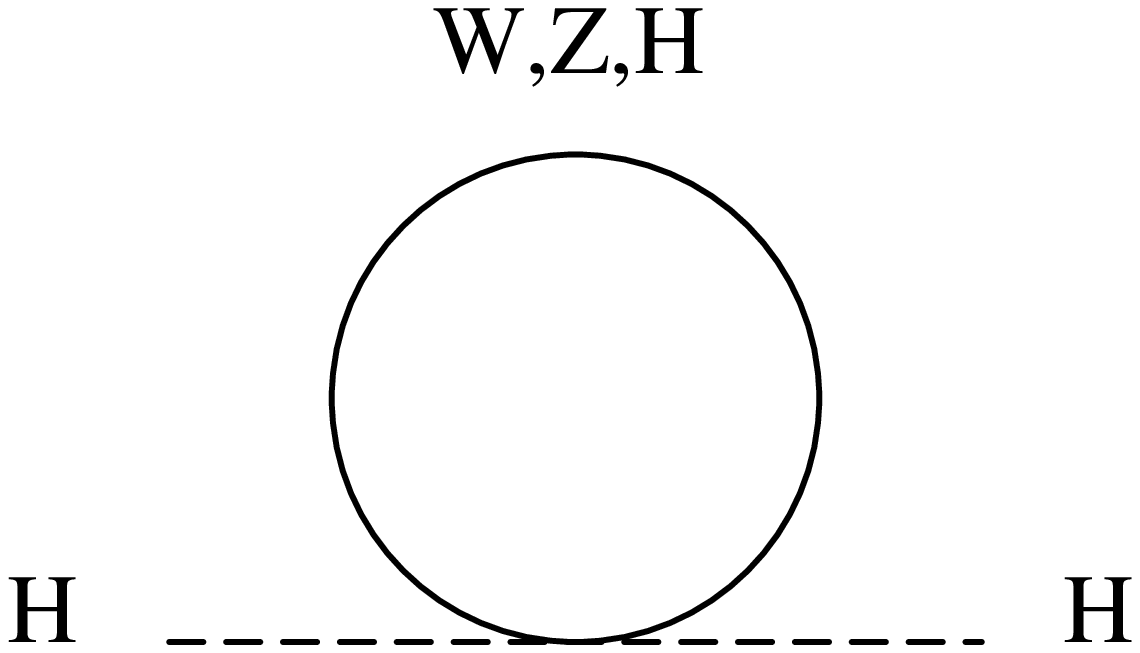}
\begin{minipage}{0.9\linewidth}
\caption{Feynman diagram: one loop 1PI contribution to $m_H$\label{fig5HM}}
\end{minipage}
\end{minipage}
\begin{minipage}[t]{0.3\linewidth}\centering
\includegraphics[width=\linewidth]{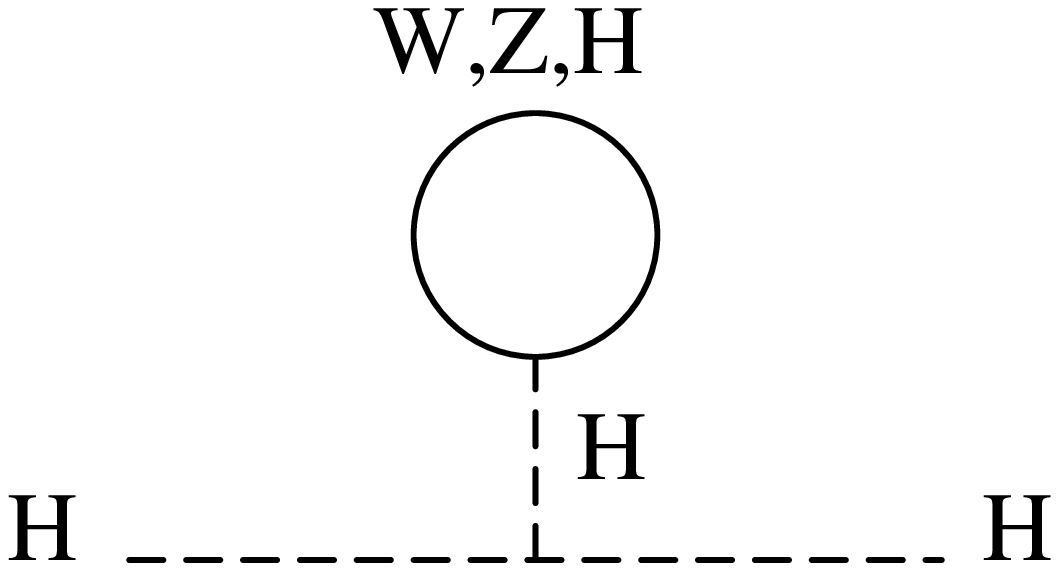}
\begin{minipage}{0.9\linewidth}
\caption{Feynman diagram: one loop 1PR contribution to $m_H$ \label{fig6HM}}
\end{minipage}
\end{minipage}
\end{figure}
We have considered the tadpoles and the one-particle-irreducible
contributions at the one loop order (the diagrams contributing to the
$W$-bosons mass are similar to those contributing to the Higgs-boson
mass).  Using the duality described in chapter 2, these
duality diagrams can be related to the Feynman graphs of figures
\ref{fig4HM}, \ref{fig5HM} and \ref{fig6HM}. The Feynman graphs have been
evaluated in ref.  \cite{Veltman:1981mj} as a function of a cutoff
parameter and we will only keep the dominant contribution which is
quadratically divergent.  We obtain:
\begin{eqnarray}
  m_W^2&=&{m_W^{0}}^2+\frac{3 g^2 \Lambda_W^2}{32 \pi^2 m_H^2}
  \left(m^2_H+2 m^2_W+m^2_Z \right).
\end{eqnarray}
This equation can be solved for $m_W^{0}$:
\begin{eqnarray}
 {m_W^{0}}^2&=&{m_H^{0}}^2= m_W^2-\frac{3 g^2 \Lambda_W^2}{32 \pi^2 m_H^2}
 \left(m^2_H+2 m^2_W+m^2_Z \right).
\end{eqnarray}

We can now compute the dynamical contribution to the Higgs boson mass.
The exact one loop, gauge invariant counterterm has been calculated in
refs. \cite{Veltman:1981mj}, \cite{Fleischer:1981ub} and
\cite{Ma:1993bt}. Using the results of ref. \cite{Ma:1993bt}, where
this counterterm was calculated as a function of a cut-off, we obtain:
\begin{eqnarray}
  m_H^2&=&{m_H^{0}}^2(m_H^2)+\frac{3 g^2 \Lambda_H^2}{32 \pi^2 m_W^2}
  \left(m^2_H+2 m^2_W+m^2_Z \right) \\ \nonumber &&
  + \frac{3 g^2 m_H^2}{64 \pi^2 m_W^2}
  \left(m_H^2 \ln \frac{\Lambda_H^2}{m_H^2} - 2 m_W^2 \ln \frac{\Lambda_H^2}{m_W^2} - m_Z^2 \ln \frac{\Lambda_H^2}{m_Z^2}\right).
\end{eqnarray}
The unknown of this equation is the Higgs boson's mass $m_H$. This
equation can be solved numerically. We obtain two positive solutions:
${m_H}_1$=14.1 GeV and ${m_H}_2$=129.6 GeV. The first solution yields
an imaginary constituent mass and is thus discarded. The second
solution is the physical Higgs boson mass. We obtain ${m_H}$=129.6 GeV
in the one loop approximation.  The constituent mass is then
$m_W^0$=78.8 GeV.

As expected the dynamical contribution to the $W$-bosons masses is
small and the Higgs boson mass is shifted above that of the $W$-bosons
mass because of the large intrinsic Higgs boson scale.

Note that our prediction ${m_H}$=129.6 GeV is in good agreement with
the requirement of vacuum stability in the standard model which
requires the mass of the Higgs boson to be in the range 130 GeV to 180
GeV if the standard model is to be valid up to a high energy scale
\cite{Sher:1989mj}. We can thus deduce that the duality we have
described in chapter 2 must also be valid up to some high energy
scale.  Our result is also in good agreement with the expectation
$m_H=98^{+58}_{-38}$ GeV based on electroweak fits
\cite{Kawamoto:2001ia}.

\chapter{Supersymmetry and Confinement}
In this chapter we shall consider a supersymmetric extension of the
ideas developed in chapter 2. These results were published in
\cite{Calmet:2000ws}.
\section{Supersymmetry and the confinement phase}
In this chapter we will present a supersymmetric extension of the
duality proposed in chapter 2.  If the confinement phase can describe
the electroweak interactions, all phenomena in particle physics are
described by exact gauge theories.  If Nature is such that its
fundamental Lagrangian has the maximal number of allowed symmetries,
it is natural to assume that supersymmetry could also be an exact
symmetry of this Lagrangian.  Supersymmetry is a crucial aspect of
particle physics. It is a desirable feature of many high energy
theories like some variants of grand unified theories. It is the
missing link between some theories at very high energies and low
energy particle physics.

It is thus meaningful to design mechanisms that explain why
supersymmetry is unobserved. A possibility is that supersymmetry is
broken. This leads to models such as the minimal supersymmetric
standard model (MSSM). We propose an alternative point of view. If the
electroweak interactions are described by a confining theory, the
microscopic theory can be supersymmetric but this symmetry is then
hidden at the macroscopic scale of fermions and electroweak bosons.
In other words we will break supersymmetry at the macroscopic scale
without breaking it at the scale of fundamental particles thus
providing a link between some theories at very high and low energy
particle physics.

In composite models, supersymmetry is not necessary to solve the
hierarchy problem because the Higgs boson is not a fundamental
particle but it remains important to have a supersymmetric theory to
reach the unification of the coupling constants at the unification
scale.

We then consider a supersymmetric extension of the model for the
electroweak interactions proposed in chapter 2 with broken
supersymmetry at the fundamental level.

\section{Hidden supersymmetry}

We shall consider a illustrative model with the gauge group $SU(2)_L$ and
unbroken $\!N\!=\!1\!$ supersymmetry.  The situation in a gauge theory
with unbroken supersymmetry is very similar to that of the confinement
phase in a non-supersymmetric theory. We assume that there is a
$SU(2)_L$ confinement: all physical particles are $SU(2)_L$ singlets.
We have the following particle spectrum: the right-handed fermions
$e_R$, $u_R$, $d_R$ and their superpartners $\tilde{e}_R$,
$\tilde{u}_R$, $\tilde{d}_R$. The right-handed particles are the usual
right-handed leptons and quarks of the standard model and their
superpartners, whereas the left-handed doublets are bound states of
some more elementary particles. The fundamental $SU(2)_L$ fields
(D-quarks) are:
\\
\begin{tabular}{lll}
leptonic D-quarks & $l_i=  \left(\begin{array}{c} l_1 \\ l_2 \end{array}
\right )$  &  (fermions)  \\
& & \\
 hadronic  D-quarks   & $q_i= \left(\begin{array}{c}q_1 \\ q_2\end{array}
\right )$   &  (fermions, $SU(3)_c$ triplets) \\ 
& & \\
scalar D-quarks  &  $h_i= \left(
  \begin{array}{c}
  h_1 \\ h_2
  \end{array}
\right )$ & (bosons).
\end{tabular}
\\
Notice that in order to cancel the anomalies we would have to
introduce a second scalar doublet. We discard this problem as our aim
is only to present a toy model to emphasize our idea. We then have the
superpartners
\\
\begin{tabular}{lll}
leptonic D-squarks & $\tilde{l}_i=  \left(\begin{array}{c}
    \tilde{l}_1 \\ \tilde{l}_2 \end{array}
\right )$  &  (bosons)  \\
& & \\
hadronic  D-squarks   & $\tilde{q}_i= \left(\begin{array}{c}
    \tilde{q}_1 \\ \tilde{q}_2\end{array}
\right )$   &  (bosons, $SU(3)_c$ triplets) \\ 
& & \\
scalar D-squarks  &  $\tilde{h}_i= \left(
  \begin{array}{c}
  \tilde{h}_1 \\ \tilde{h}_2
  \end{array}
\right )$ & (fermions).
\end{tabular}
\\
We shall refer to the theory involving the D-quarks and the
D-squarks as the microscopic theory. At the macroscopic level i.e,
the theory of bound states, a large number of $SU(2)_L$ invariant bound
states can be identified.  We see that bound states of different
particles can have the same quantum numbers. For example, the neutrino
can be identified with the bound state $\bar h l$ but also with the
bound state $\bar{\tilde{h}} \tilde{l}$.  It will thus be a
superposition of both bound states. This can be
applied to the rest of the known particles. The left-handed
fermions, normalized in the appropriate way, are defined as
follows. We have the leptons

\begin{eqnarray} \label{eq1CH}
\mbox{left-handed neutrino} \ \nu_L &=& \frac{1}{F}
\left ( (\bar h l) +  (\bar{\tilde{h}} \tilde{l}) \right ) \\
\mbox{left-handed electron} \ e_L &=& \frac{1}{F}
\left ( (\epsilon^{ij} h_i l_j)
+  (\epsilon^{ij} \tilde{h}_i \tilde{l}_j) \right ) \nonumber
\end{eqnarray}
where $F$ is a numerical, to be specified, normalization factor. The
quarks are also bound states

\begin{eqnarray} \label{eq2}
\mbox{left-handed up quark} \ u_L &=& \frac{1}{F}
\left ( (\bar h q) +  (\bar{\tilde{h}} \tilde{q}) \right ) \\
\mbox{left-handed down quark} \ d_L &=& \frac{1}{F}
\left ( (\epsilon^{ij} h_i q_j)
+  (\epsilon^{ij} \tilde{h}_i \tilde{q}_j) \right ). \nonumber
\end{eqnarray}

The Higgs and electroweak bosons are bound states of scalar D-quarks and
their superpartners:
\begin{eqnarray} \label{eq3}
\mbox{Higgs field} \ \phi &=& \frac{1}{2 F}
\left ( (\bar h h) + \beta  (\bar{\tilde{h}} \tilde{h}) \right) \\ \nonumber
\mbox{electroweak boson} \ W^3_\mu&=& \frac{2 i}{g F^2}
\left ( (\bar h D_\mu h) +  \beta (\bar{\tilde{h}} D_\mu \tilde{h}) \right ) \\
\mbox{electroweak boson} \ W^-_\mu&=& \frac{\sqrt{2} i}{g F^2}
\left ( (\epsilon^{ij} h_i D_\mu h_j) + \beta
(\epsilon^{ij} \tilde{h}_i D_\mu \tilde{h}_j) \right ), \nonumber
\end{eqnarray}
where $D_\mu$ is the covariant derivative of the gauge group $SU(2)_L$
involving the gauge bosons $B^a_\mu$ and $g$ is the gauge coupling of
this group. The second charged $W$ boson $W^+$ is defined as
$(W^-)^\dagger$. A simple dimensional analysis shows that a constant
$\beta$ with dimension $-1$ has to appear. This constant is a priori
unknown but the only scale of the theory being $F$, we could impose
$\beta=1/F$.  This apparently arbitrary choice is not a drawback for
the theory as we will see that only the terms containing a scalar
D-quark doublet will be relevant. 

The problem is to know whether a particle and its superparticle will
belong to the same supermultiplet, i.e, if they have the same mass. It
is a difficult question as dynamical effects can contribute to the
masses.  For example, the masses of the electroweak bosons are to a
large extent dominated by dynamical effects. Once we have introduced a
second Higgs doublet, we have the same gauge group and the same
particle content as in the MSSM, dynamical supersymmetry breaking is
thus possible.  There are two possibilities: either the masses of, for
example, an electroweak boson and of the corresponding superparticle
are identical and supersymmetry is unbroken at the macroscopic level
or they are different because of dynamical effects and supersymmetry
is dynamically broken. This possibility can't be excluded, but in the
sequel we assume that these particles indeed form a supermultiplet.
Thus, an electron is the superpartner of a selectron. Lattice
simulations could test the dynamical behavior of such a model.

All the particles we have identified up to this point are those
appearing in the standard model. We can also identify the bound states
corresponding to the macroscopic superparticles. For example, we have
\begin{eqnarray}
\mbox{selectron} \ 
\tilde e &=& \frac{1}{F}
\left ( (\epsilon^{ij} h_i \tilde{l}_j)
+ \beta (\epsilon^{ij} \tilde{h}_i l_j) \right ) \nonumber
\end{eqnarray}
for the left-handed selectron.

The complementarity principle was established in the framework of a
non-\-super\-symmetric theory with a single Higgs boson doublet. This
principle requires that the coupling constants between the bound
states and the electroweak bosons are the same in the Higgs phase and
in the confinement phase. 't Hooft proposed that the confinement
phenomenon is due to vortices \cite{'tHooft:1998pk,'tHooft:1978hy}.
This means that we have a confinement with a weak coupling constant
which avoids the problems due to chiral symmetry breaking
\cite{Aoki:1988aw}.

In a supersymmetric model the situation is more complex since the
theory is richer. Nevertheless the situation in such a theory is very
similar to that of the confinement phase in a non-\-super\-symmetric gauge
theory.  The question is whether our microscopic model which is
supersymmetric will have a supersymmetric macroscopic spectrum.  A
lattice study of the vacuum structure and of the dynamical behavior of
our model would be useful to answer this question. As long as this has
not been done, some place is left for speculation.

A discrete symmetry could explain why nature selects, at least at low
energy, only the particles.  We introduce a mechanism similar to the
so-called R-parity. We assign a new quantum number to the particles.
We call this new quantum number S-parity. The D-quarks are assigned
S-parity +1, whereas the D-squarks are assigned S-parity -1. We then
assume that the bound states appearing in nature have S-parity +1.

This selection rule shifts the masses of the superparticles to very
high energies. In other words we break supersymmetry at the
macroscopic level by imposing a discrete symmetry but it remains
intact at the microscopic level. It is thus clear that superparticles
corresponding to the left-handed particles, to the Higgs sector and to
the electroweak bosons will not be observable at least at low energy.
In that case, we expect that a confining theory describes the weak
interactions correctly.  Imposing this selection rule, which is
motivated by the apparent absence of superparticles in nature at low
energy, is not trivial as it would be in the case of the MSSM because
the fundamental D-squarks are confined in usual matter. It would not
be very surprising if this S-parity was broken in nature, as there are
already many examples of broken discrete symmetries. But, at this
stage it remains a speculation, which could be tested on the lattice.

That scenario is useful in the case of a grand unified theory.  If
there is a deconfinement phase at the scale of a few TeV,
supersymmetry is realized above that scale and the coupling constants
unification takes place at the unification scale, but supersymmetry
remains hidden at low energy under this deconfinement phase.  Two
scenarios are conceivable. The mass scale of the superparticles is
below the deconfinement scale, in which case one will observe
superparticles but the theory is not explicitly supersymmetric until
one reaches the deconfinement scale.  Another possibility is that the
mass scale for the superparticles is above the deconfinement scale in
which case the particle spectrum would suddenly become supersymmetric
above the deconfinement scale. This feature allows to test our idea.

Even if supersymmetry is broken by dynamical effects, it might still
be necessary, if the mass splitting is not sufficiently large, to
introduce the S-parity for phenomenological reasons.

\subsection{Back to known particles}

It remains to show that the definitions for the fields indeed describe
the observed particles. We use the unitary gauge for the scalar doublet
\begin{eqnarray}
   h_i=\left ( \begin{array}{c} F+h_{(1)} \\ 0 \end{array}\right).
\end{eqnarray}

The parameter $F$ is a real number.  If $F$ is sufficiently
large we can perform a $1/F$ expansion for the fields defined
previously. We then have

\begin{eqnarray} 
 \nu_L&=&l_1+\frac{1}{F} \left ( h_{(1)} l_1
   + \bar{\tilde{h}} \tilde{l} \right )\approx l_1  \\
  e_L&=&
  l_2+\frac{1}{F} \left ( h_{(1)} l_2 +
\epsilon^{ij} \tilde{h}_i \tilde{l}_j \right )
    \approx l_2 
  \nonumber \\
   u_L&=&q_1+\frac{1}{F} \left ( h_{(1)} q_1
 + \bar{\tilde{h}} \tilde{q} \right )
   \approx q_1 
   \nonumber \\
     d_L&=&q_2 
     +\frac{1}{F} \left ( h_{(1)} q_2
+
\epsilon^{ij} \tilde{h}_i \tilde{q}_j \right )
       \approx q_2
     \nonumber \\
     \phi&=&h_{(1)}+\frac{F}{2} +\frac{1}{2 F}  \left (
     h_{(1)} h_{(1)} + \beta
\bar{\tilde{h}} \tilde{h} \right)
     \approx h_{(1)}+\frac{F}{2}
     \nonumber \\ \nonumber
     W^3_{\mu}&=&\left ( 1 + 
     \frac{h_{(1)}}{F} \right)^2 B^3_{\mu} + \frac{2 i}{g F} \left (1+ 
     \frac{h_{(1)}}{F} \right) \partial_{\mu} h_{(1)}  \\ && \nonumber  
+\frac{2 i\beta}{g F^2} \left ( \bar{\tilde{h}} D_\mu \tilde{h} \right )
   \approx B^3_{\mu}
    \nonumber \\ \nonumber
   W^-_{\mu}&=& \left ( 1 +
     \frac{h_{(1)}}{F} \right)^2 B^-_{\mu} +
\frac{\sqrt{2} i\beta}{g F^2}
\left ( \epsilon^{ij} \tilde{h}_i D_\mu \tilde{h}_j \right )
   \approx B^-_{\mu}. \
\end{eqnarray}
As done in the non-supersymmetric case, we assume that the only
particles which are stable enough to be observable at presently
accessible energies are those containing the scalar doublet $h$, those
are the only fields who survive in the $1/F$ expansion. We consider
the terms suppressed by a factor $1/F$ as being irrelevant.  Therefore
the spectrum of this theory is, for the left-handed sector, identical
to the spectrum of the standard model. Nevertheless we are not able to
hide the superpartners of the right-handed particles at this stage.
Supersymmetry is apparently broken in the left-handed sector but in
fact it remains unbroken at the microscopic level of the theory.

We have considered a toy model with $SU(2)_L$ confinement and hidden
supersymmetry in the left-handed sector. Supersymmetry is broken at
the macroscopic level by a discrete symmetry. The first step towards a
realistic model is to include a second Higgs doublet. It can be done
without major difficulties as we shall show in the next section.

This model can be extended to a model with a $SU(3)_C \times SU(2)_R
\times SU(2)_L \times U(1)_Y$ gauge group with two Higgs doublets
for each $SU(2)$ sector.  Once this extension has been done, we can
hide supersymmetry completely at the microscopic level for the
$SU(2)_R \times SU(2)_L$ sector, assuming a $SU(2)_R \times SU(2)_L$
confinement.  Supersymmetry would have to be broken by usual means for
the two remaining gauge groups. The spectrum of the macroscopic theory
at low energy is then that of the standard model with ten Higgs
fields, i.e. five for each $SU(2)$ sector, 8 gluinos and a photino.

This model provides the missing link between low energy particle
physics and very high energy theories like grand unified theories.
Usual models with supersymmetry breaking are not able to explain a
small cosmological constant \cite{Witten:2000zk}. In our approach,
supersymmetry is not broken in the $SU(2)_L$ sector at the microscopic
level.  Thus the contribution of the energy of the fundamental vacuum
of that sector to the cosmological constant is vanishing. Our
mechanism could therefore help to explain a small or vanishing
cosmological constant.

Note that this model would nicely fit into a supersymmetric $SO(10)$
grand unified theory, which thus could be the fundamental theory of
D-quarks and D-squarks. It turns out that such a theory would be very
similar to the standard model if there is a confinement in the weak
interactions sector.

\section{The MSSM}

In this section, we assume that the complementarity principle remains
valid for supersymmetric theories once soft breaking terms have been
introduced. The model in the confinement phase corresponding to the
minimal supersymmetric standard model can easily be obtained by
requiring that supersymmetry is broken by usual means at the level of
the fundamental D-quarks and D-squarks. A second Higgs doublet $k$ and
the corresponding superparticle $\tilde k$ can be introduced without
any difficulty, and we basically have to replace $h$ and $\tilde h$ by
$s=h + i \sigma_2 k^*$ and $\tilde s= \tilde h + i \sigma_2 \tilde
k^*$ in the definitions of the fermions, superfermions, electroweak
bosons and of their superpartners. The gauge is fixed in such a way
that $s$ takes the form $s=(F+ h_{(1)} + k_{(1)},0)$, where
$F=F_1+F_2$, $F_1$ corresponding to the scalar doublet $h$ and $F_2$
to the scalar doublet $k$. We then have
\begin{eqnarray}
h= \left(\begin{array}{c}
    F_1 + h_{(1)} + i \chi^0 \\  - \phi^- \end{array}
\right ), \ \
k= \left(\begin{array}{c}
    - \phi^+ \\ F_2 + k_{(1)} + i \chi^0 \end{array}
\right ).
\end{eqnarray}
 
We can define the five Higgs bosons
\begin{eqnarray}
\mbox{$CP$ even Higgs boson} \
\phi_1&=& \frac{1}{2 F_1} \left ({\bar h h}
\right)
  = h_{(1)}+ \frac{F_1}{2} + {\cal O}\left(\frac{1}{ 2 F_1}\right)
\\ \nonumber
\mbox{$CP$ even Higgs boson} \
\phi_2&=& \frac{1}{2 F_2} \left ({\bar k k}
\right)
  = k_{(1)}+ \frac{F_2}{2} + {\cal O}\left(\frac{1}{ 2 F_2}\right)
\\\
\mbox{$CP$ odd Higgs boson} \
i \chi&=& \left (
\frac{1}{2 F} (\bar s h +  \epsilon^{ij} s_i k_j) 
-\frac{1}{2 F_1} (\bar h h)- \frac{1}{2 F_2} (\bar k k) \right)
\nonumber \\ &=& i \chi + {\cal O}\left(\frac{1}{ 2 F_1}\right)
+ {\cal O}\left(\frac{1}{ 2 F_2}\right)
\nonumber 
\\
\mbox{charged Higgs boson} \
\phi^+&=& \frac{-1}{F} \left ({\bar s k}
\right)
  = \phi^+ + {\cal O}\left(\frac{1}{ F}\right)
\nonumber
\\
\mbox{charged Higgs boson} \
\phi^-&=& \frac{-1}{F} \left ({\epsilon^{ij} s_i h_j}
\right)
  = \phi^- + {\cal O}\left(\frac{1}{ F}\right).
\nonumber 
\end{eqnarray}

The superpartners of these Higgs bosons can be obtained in a similar
way.  The duality presented in chapter 2 is thus compatible with a
supersymmetric extension provided that both $F_1$ and $F_2$ can be
chosen to be large. This model has the same vertices as the MSSM and
the same particle content. As in the case of the non-supersymmetric
model, we expect that radial and orbital excited versions of the known
particles will appear if the duality breaks down.

We thus have described a supersymmetric extension of the model proposed
in chapter 2 for the electroweak interactions with $SU(2)$
confinement. We have shown that this model is compatible with a
supersymmetric extension provided that the complementarity principle
remains valid for supersymmetric theories once soft breaking terms
have been introduced.

\chapter{Testing the duality}
The duality we have described in chapter 2 could break down at a
certain energy scale. In principle, effects of this breakdown could be
seen at relatively low energies. In that case we should assume that
the phase describing Nature correctly is the confinement phase. If the
duality breaks down and if Nature is indeed described by the
confinement phase, new particles are expected to appear. In this
chapter, we shall describe two possible scenarios which are
theoretically well motivated.  The first of these scenarios is a
failure of the duality in the Yukawa sector.  We shall assume that the
masses of the light fermions are of dynamical origin.  The Higgs boson
might thus not couple to the $b$-quark. This would modify the Higgs
decay modes in a fundamental fashion. The second scenario is a high
energy violation of the duality. In that case excitations of the
electroweak bosons, in particular the so-called electroweak $d$-waves,
could contribute in a sizable manner to the electroweak boson
scattering. The results presented in this chapter were published in
\cite{Calmet:2000vx,Calmet:2001yd}.

\section{The Higgs boson might not couple to $b$-quarks}

As far as the mass generation within the framework of the standard
electroweak model is concerned, one must differentiate between the
mass generation for the electroweak bosons $W$, $Z$, the mass
generation for the heavy $t$-quark, and the generation of mass for the
leptons and the five remaining, relatively light quarks. While there
exists no freedom in the choice of the interaction strengths of the
weak bosons with the scalar field, which is dictated by the gauge
invariance \cite{LlewellynSmith:1973ey}, there is such a freedom with
respect to the fermions. The masses of the fermions are given by the
various Yukawa coupling constants, which parametrize the interactions
of the leptons and quarks with the scalar field. The Yukawa coupling
constant of the $t$-quark field is of the same order as the gauge
coupling constant, while the other fermions couple much more weakly
($0.018$ for the $b$-quark, $0.005$ for the $c$-quark, etc.). The
origin of the light fermion masses is still mysterious, and
alternative views or slight variations of the standard electroweak
theory might indeed give a different view.  Taking into account the
observed flavor mixing phenomenon, one could speculate, for example,
that the masses of the light quarks and of the leptons are due to the
mixing. In the absence of the mixing the mass matrix of the quarks in
the $u$-sector would simply be proportional to a diagonal matrix with
the entries $(0, 0, 1)$, and there would only be a coupling of the
scalar field to the $t$-quark. Once the flavor mixing is switched on,
the mass eigenstates for the light quarks are not necessarily coupled
to the scalar field, with a strength given by the mass eigenvalues. In
particular these couplings could remain zero.

It is well-known that the renormalizability of the theory requires a
coupling of the fermions to the scalar field
\cite{LlewellynSmith:1973ey}. Otherwise the unitarity in the $s$
channel is violated at high energies for the reaction $f_L \, \bar f_L
\rightarrow W^+ W^-$. However, for all fermions except the $t$-quark
these problems appear only at extremely high energies. Modifications
of the electroweak theory, which involve an energy scale not orders of
magnitude above the typical electroweak scale of about $0.3$ TeV, e.g.
theories which do not rely on the Higgs mechanism, can take care of
this problem.

We have discussed an alternative description of the standard model,
based on the duality between confinement and Higgs phase in chapter 2.
We suppose that the electroweak interactions are described by the
confinement phase and that the duality breaks down in the Yukawa
sector.  This provides an alternative view of the electroweak bosons,
which are not the basic gauge bosons of the underlying gauge theory,
but ``bound states'' of an underlying scalar field, which in the Higgs
phase plays the role of the Higgs doublet.  Both the charged $W$
bosons and the neutral $Z$ boson are $J = 1$ bound systems of the type
$hh, (hh)^\dagger$ or $\bar h h$ respectively.  There is a
corresponding $J = 0$, $\bar h h$ system, which is to be identified
with the Higgs boson of the standard electroweak model.

We shall consider a deviation from our original model which would have
the same couplings as in the standard model.  It is conceivable that
in the confinement phase of the electroweak theory the coupling
strength of the fermions to the scalar boson are not proportional to
the light fermion masses, since these couplings depend strongly on the
dynamics of the model. In the simplest case only the fermion whose
mass is of the same order as the weak interaction energy scale, i.e.,
the $t$-quark, has such a coupling. Thus we proceed to calculate the
properties of the scalar boson, which couples only to the $t$-quark.
As far as the interaction of such a boson with the $W$ and $Z$ bosons
is concerned, there is no change in comparison to the standard
electroweak model.  However there is a substantial change of the decay
properties. Decay modes which were regarded as being strongly
suppressed become dominant.

\begin{figure}
  \begin{minipage}[t]{0.3\linewidth}\centering
    \includegraphics[width=\linewidth]{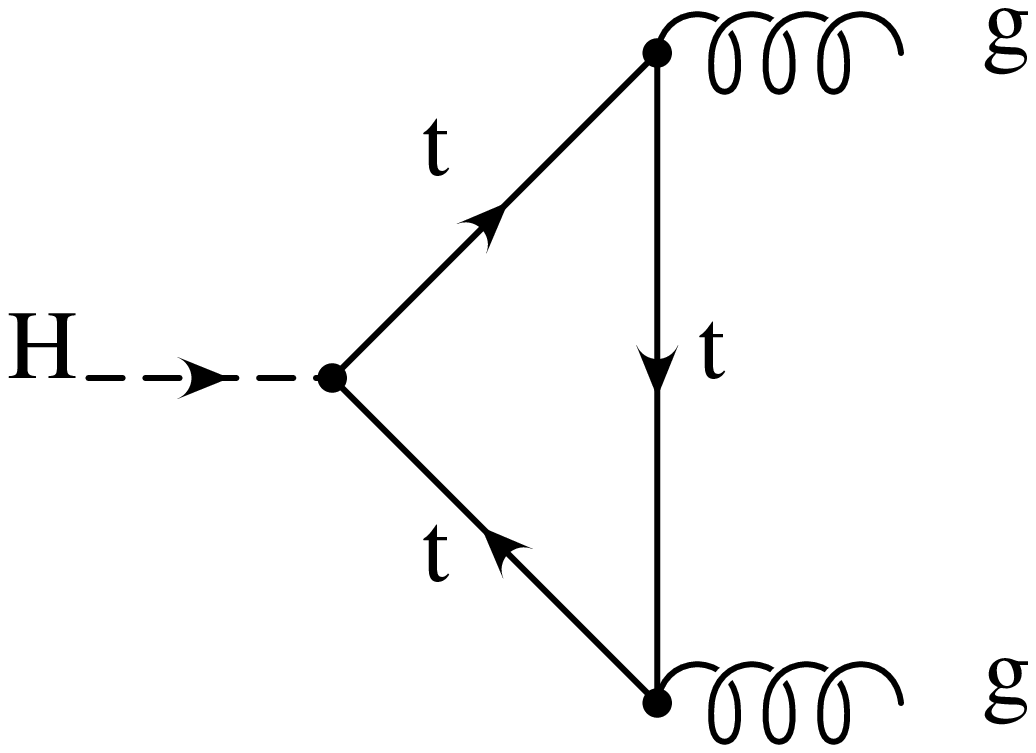}
    \begin{minipage}{1\linewidth}
      \caption{Top triangle.
        \label{graph1CH4}}
    \end{minipage}
  \end{minipage}
  \begin{minipage}[t]{0.3\linewidth}\centering
    \includegraphics[width=\linewidth]{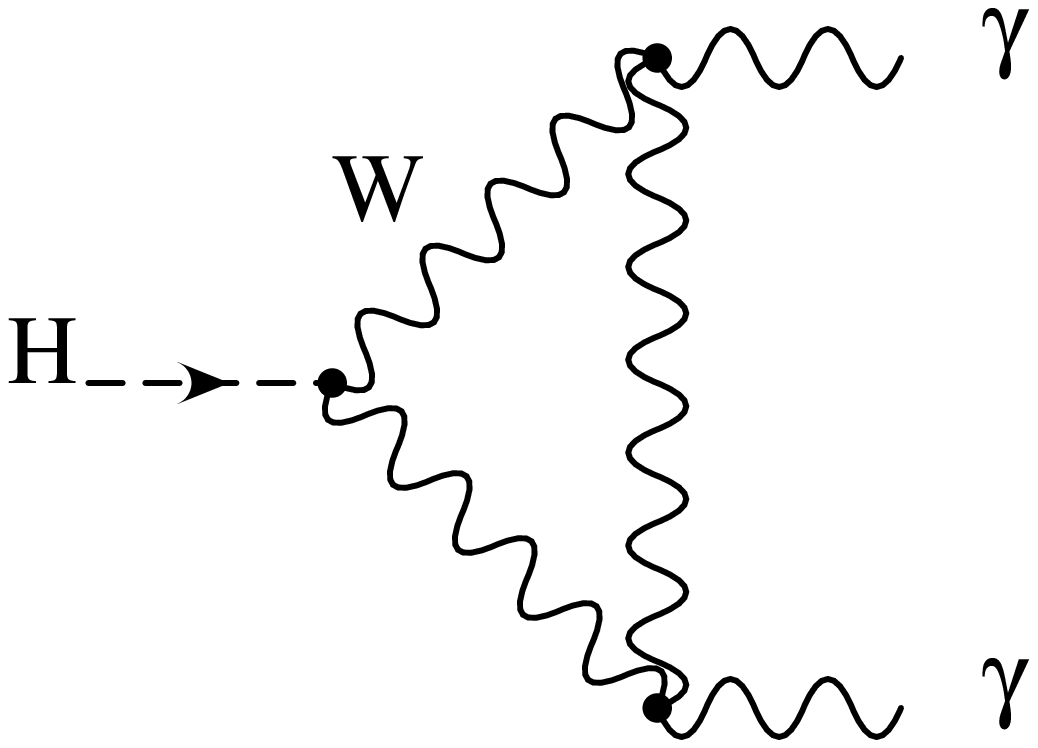}
    \begin{minipage}{1\linewidth}
      \caption{$W$ triangle.
        \label{graph2CH4}}
    \end{minipage}
  \end{minipage}
  \begin{minipage}[t]{0.3\linewidth}\centering
    \includegraphics[width=\linewidth]{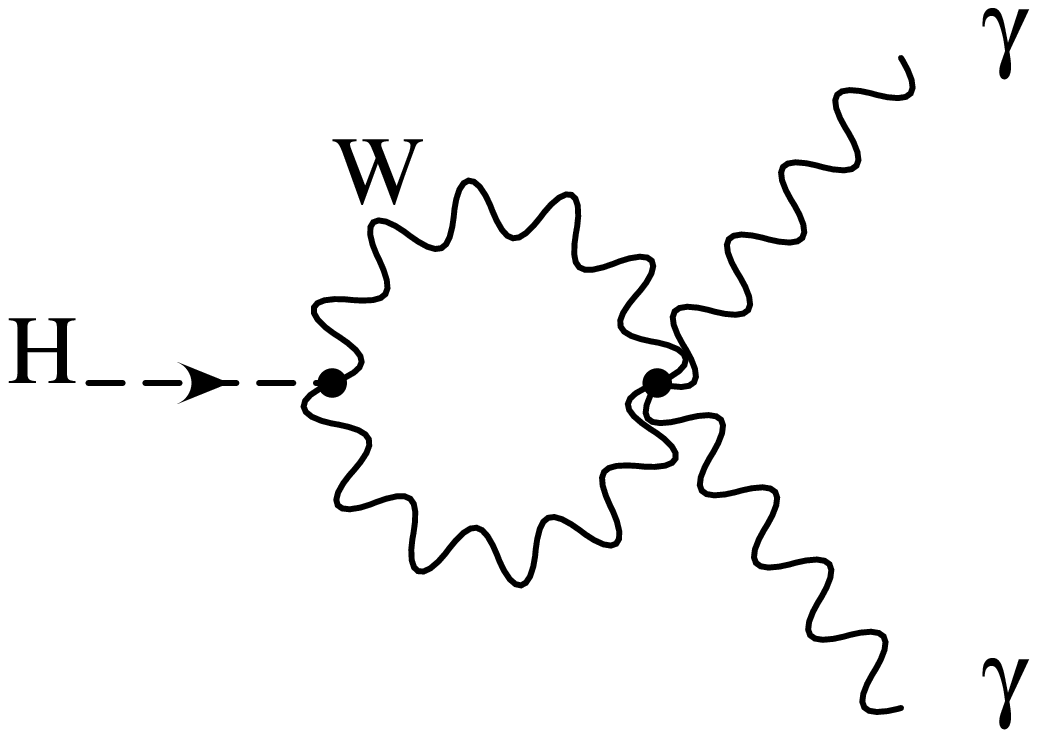}
    \begin{minipage}{1\linewidth}
      \caption{$W$ bubble.
        \label{graph3CH4}}
    \end{minipage}
  \end{minipage}
  \begin{minipage}[t]{0.3\linewidth}\centering
    \includegraphics[width=\linewidth]{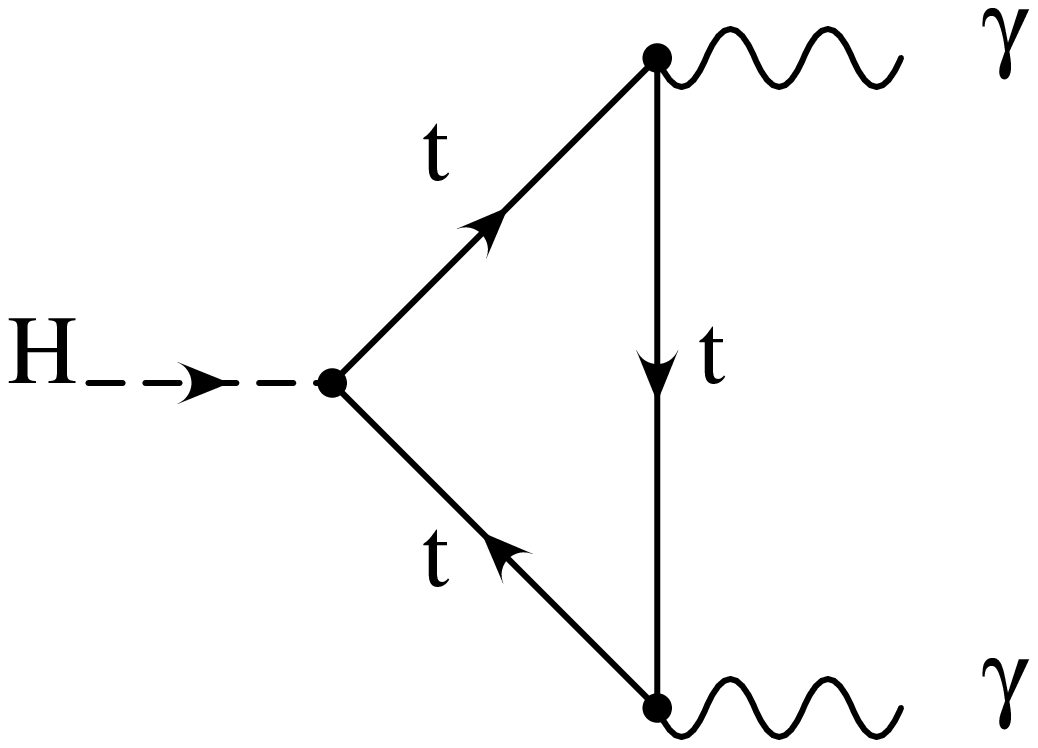}
    \begin{minipage}{1\linewidth}
      \caption{Top triangle.
        \label{graph4CH4}}
    \end{minipage}
  \end{minipage}
  \begin{minipage}[t]{0.3\linewidth}\centering
    \includegraphics[width=\linewidth]{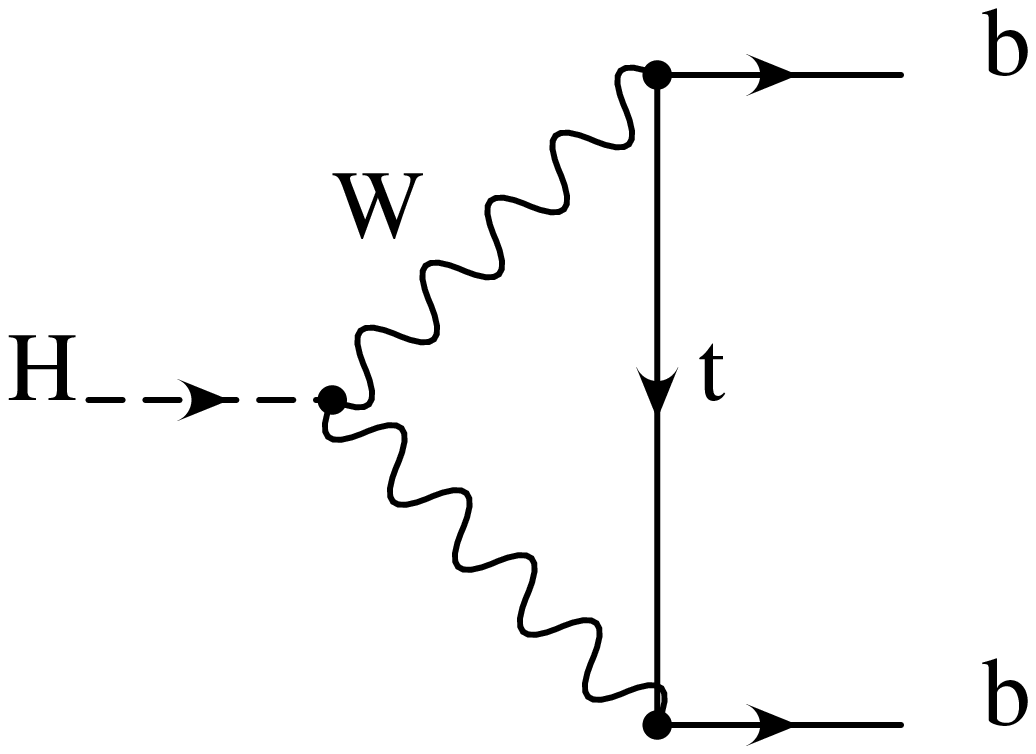}
    \begin{minipage}{1\linewidth}
      \caption{1st effective $b$-quark decay.
        \label{graph5CH4}}
    \end{minipage}
  \end{minipage}
  \begin{minipage}[t]{0.3\linewidth}\centering
    \includegraphics[width=\linewidth]{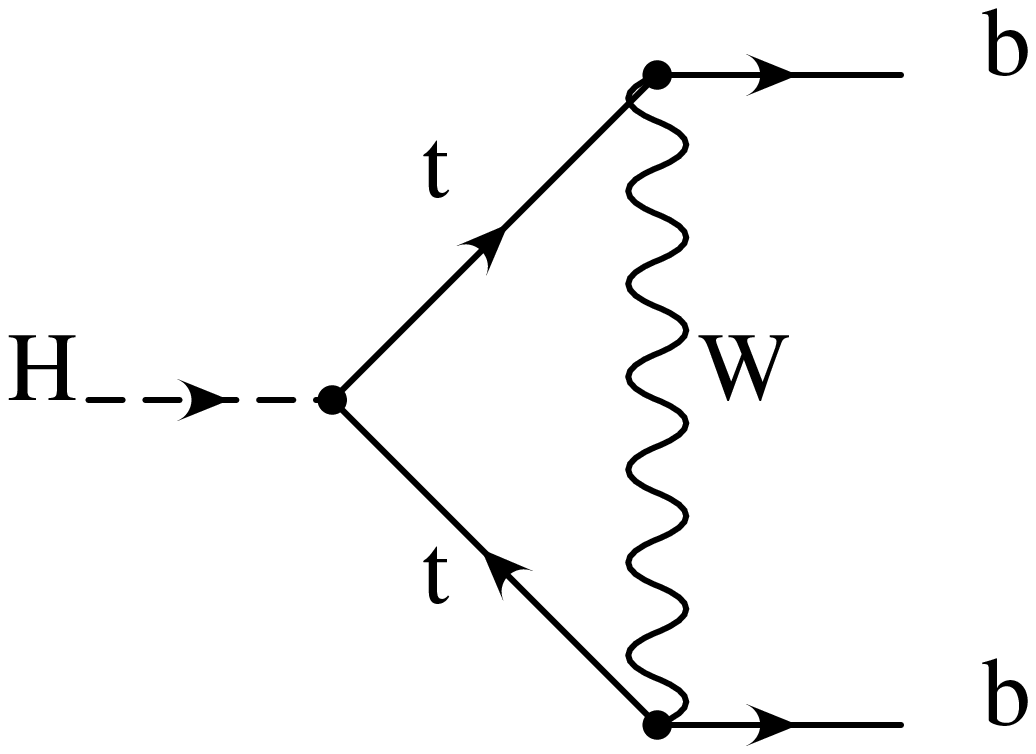}
    \begin{minipage}{1\linewidth}
      \caption{2nd effective $b$-quark decay.
        \label{graph6CH4}}
    \end{minipage}
  \end{minipage}
\end{figure}

We consider the following decay channels for the Higgs boson: $H \to g
g$ (see graph \ref{graph1CH4}) via a top quark triangle and $H \to
\gamma \gamma$ (see graphs \ref{graph2CH4}, \ref{graph3CH4} and
\ref{graph4CH4}) via a triangle involving top quarks and charged
electroweak bosons or a bubble diagram involving a neutral electroweak
boson.
For a two photon Higgs decay, ignoring radiative corrections,
one finds \cite{Ellis:1976ap,Gunion:1989we,Cahn:1989ru}
\begin{eqnarray}
\Gamma(H \to \gamma \gamma) &=& \frac{ \alpha^2 g^2}{1024 \pi^3} 
\frac{M_H^3}{M_W^2} \left | \sum_i e_i^2 N_{c\, i} F_i \right |^2 
= \frac{ \alpha^2 g^2}{1024 \pi^3} 
\frac{M_H^3}{M_W^2} \left | \frac{4}{3} F_{1/2} + F_{W} \right |^2 
\end{eqnarray}
where the functions $F_{1/2}$ and $F_{W}$ 
are given by
\begin{eqnarray} \label{eq1CH4}
F_{1/2}&=&-2 \tau [1+(1-\tau) f(\tau)]
\end{eqnarray}
and
\begin{eqnarray}
F_{W}&=& 2+3 \tau + 3 \tau (2-\tau) f(\tau)
\end{eqnarray}
where $\tau=4m_i^2/M_H^2$.  The first function corresponds to the
contribution of the top quark and the second to the contribution of
the charged $W$ bosons.  As we assume that the Higgs boson is light,
i.e., lighter than twice the mass of the $W$ bosons, the function
$f(\tau)$ reads
 \begin{eqnarray}
 f(\tau)&=& \left ( \arcsin{ \left( \sqrt{\frac{1}{\tau}}\right )} \right )^2.
 \end{eqnarray}  
 For the decay into two gluons one finds \cite{Ellis:1976ap,Gunion:1989we}
 \begin{eqnarray} 
\Gamma(H \to g g) &=& \frac{ \alpha_s^2 g^2}{512 \pi^3}
\frac{M_H^3}{M_W^2} \left |  F_{1/2} \right |^2, 
\end{eqnarray}
also neglecting the radiative corrections. The function $F_{1/2}$ was
given in equation (\ref{eq1CH4}).

Another possibility for the Higgs boson to decay are the electroweak
boson channels $H \to W W$ and $H \to Z Z$. The Higgs boson couples to
the electroweak bosons with the same strength as in the standard model.
The decay via two virtual electroweak bosons represents a
non-negligible contribution to the Higgs decay. For $m_W<m_H$ or
$m_Z<m_H$ one of the electroweak bosons is on-shell.  These decay
rates were evaluated using the program HDECAY \cite{Djouadi:1998yw}
and cross-checked using CompHEP \cite{Pukhov:1999gg}. The numerical
results are the sum of the decay over two electroweak bosons, for a
light Higgs both electroweak bosons are virtual, when allowed by the
kinematics, the contributions of on-shell electroweak bosons are also
taken into account.

\begin{table}
\centering
  \begin{tabular}{|l|l|l|l|l|l|}
   \hline
   channel & $\!m_H\!=\!60$ 
   & $\!m_H\!=\!70$ 
   & $\!m_H\!=\!80$ 
   & $\!m_H\!=\!90$ 
   & $\!m_H\!=\!100$
   \\
    \hline
   $\Gamma(H \! \to \! g  g)_{\alpha_s=0.119} \! \! 
   $ &$2.3\times 10^{-5}  $    &
  $3.7\times 10^{-5}$ & $5.5\times 10^{-5}$ 
  & $7.9\times 10^{-5}$  
  &$1.1\times 10^{-4}$    \\
 \hline
  $\Gamma(H \! \to \! g  g)_{\alpha_s=0.15} \! \! 
  $ &$3.6\times 10^{-5}$ & 
  $5.8\times 10^{-5}$ & $8.7\times 10^{-5}$ 
  & $1.3\times 10^{-4}$  
  &$1.7\times 10^{-4}$    \\
 \hline
 $\Gamma(H \!\to \! \gamma \, \gamma )$
 &$8.0\times 10^{-7}$ &
 $1.3\times 10^{-6}$&
 $2.0\times 10^{-6}$
 &$3.0\times 10^{-6}$ &
 $4.4\times 10^{-6}$
 \\
 \hline
 $\Gamma(H\! \to \!W \, W)\! \! 
 $ &$1.09\times 10^{-7}$ &$3.80\times 10^{-7}$ &
 $1.22\times 10^{-6} $
 &$4.49\times 10^{-6} $ &$2.66\times 10^{-5} $ \\
 \hline
  $\Gamma(H \! \to \! Z \, Z)\! \! 
  $ &$3.33\times 10^{-8}$ &
  $1.10\times 10^{-7} $ &
  $3.27\times 10^{-7} $
  &
  $9.12\times 10^{-7} $ &
 $2.72\times 10^{-6} $  \\
 \hline
\end{tabular}
 \caption{Higgs boson decay rates in GeV for different Higgs masses in GeV.
         \label{tab:table1CH4}}
 \end{table}
 
 The results of these calculations are given in table
 \ref{tab:table1CH4}.  The corresponding branching ratios are given in
 table \ref{tab:table2CH4}. We see that such a Higgs boson would decay in
 a fundamentally different way than the Higgs boson of the standard
 model. The results for the $H \to g g$ decay are strongly dependent
 of the value chosen for $\alpha_s$. Thus this decay channel has a
 considerable uncertainty.  We have done the calculations for two
 different values of the strong coupling constant $\alpha_s=0.119$ and
 $\alpha_s=0.15$. The fine-structure constant was taken to be
 $\alpha=1/128.9$.
 
 Even if the light fermions in particular the $b$-quark, do not couple
 directly to the Higgs boson, some $b$-quarks could be produced via
 the diagrams \ref{graph5CH4} and \ref{graph6CH4}. Their contributions
 is not easy to estimate but the electroweak corrections for a light
 Higgs boson are known to be very small \cite{Kniehl:1992ze},
 typically $0.3 \%$ of the tree level value.  Nevertheless they could
 still be of the same order of magnitude as the $\gamma \gamma$
 contribution.  Above $90$ GeV the decay channel $H \to Z \gamma$
 opens. For masses larger than $110$ GeV the Higgs boson mainly decays
 into two electroweak bosons.

 The present searches for the Higgs boson at LEP are mainly based on
 the assumption that the leading decay made in the mass region of
 about 100 GeV or less is the decay $H \rightarrow \bar b b$. The
 present experimental limit $m_H >113.3$ GeV \cite{Murray} is obtained
 on the basis of this assumption. In our model the decay is dominated
 by the decay $H \rightarrow g g$, i.e., the decay products do not show
 a specific flavor dependence. The lower limit on the mass of such a
 boson is much weaker and of the order of $70$ GeV \cite{Schaile}.
 
 The best way to detect the Higgs boson at LEP seems to us to search
 for the decay $H \rightarrow \gamma \gamma $. Since the invariant
 mass of the $2 \gamma $ system would be identical to the mass of the
 boson, the background coming from radiation effects could be
 substantially reduced. In our case this decay channel, having a small
 branching ratio, is not seriously constrained by fermiophobic Higgs
 studies \cite{Murray}.

 Typical fits of the Higgs boson mass indicate that the most likely
 mass of the boson is about $m_H=98^{+58}_{-38}$ GeV
 \cite{Kawamoto:2001ia}.  It might well be, that the mass of the Higgs
 boson is in the region $70$ to $110$ GeV, provided the decay proceeds
 via the mechanism discussed above. We note that in contrast to the
 standard expectation the Higgs particle is a relatively narrow object
 with a width of about $58.5$ KeV.
\begin{table}[ht]
\centering
  \begin{tabular}{|l|l|l|l|l|l|}
   \hline
   channel & $\!m_H\!=\!60$ 
   & $\!m_H\!=\!70$ 
   & $\!m_H\!=\!80$ 
   & $\!m_H\!=\!90$ 
   & $\!m_H\!=\!100$
   \\
    \hline
   ${\mbox Br}(H \! \to \! g  g)$
   & $96.06 \, \%$
   & $95.39 \, \%$
   & $93.94 \, \%$
   & $90.39 \, \%$ 
   & $76.54 \, \% $\\
 \hline
 ${\mbox Br}(H \!\to \! \gamma \, \gamma )$ &$3.34\, \%  $ &
 $3.35 \, \%$&
 $3.42 \, \%$
 &$3.43 \, \%$
 & $3.06\, \% $
 \\
 \hline
 ${\mbox Br}(H\! \to \!W \, W)$ &$0.46 \, \% $
 &$0.98 \, \% $
 & $2.08 \, \% $
 &$5.14 \, \% $
& $18.51\, \% $
 \\
 \hline
  ${\mbox Br}(H \! \to \! Z \, Z)$ &$0.14 \, \% $ &
  $0.28 \, \% $ &
  $0.56 \, \% $
  &
  $1.04 \, \% $
& $1.89 \, \% $
  \\
 \hline
\end{tabular}
 \caption{Branching ratios for different Higgs masses in GeV and
   for $\alpha_s=0.119$. \label{tab:table2CH4}}
 \end{table}

\section{Electroweak $d$-waves}

In the model considered in chapter 2, new particles corresponding to
exotic particles like leptoquarks can be introduced.  But, they do not
survive to the expansion in $1/F$, and therefore, the duality cannot
be applied to describe their properties. Leptoquarks are bound states
of two fermions. Forces between two fermions can be very much
different than those between a fermion and a scalar or between two
scalars. If leptoquarks do exist, their mass scale is presumably very
high.

Of particular interest are radially excited versions of the Higgs
boson $H^*$ and of the electroweak bosons $W^{3 *}$ and $W^{\pm*}$.
The most promising candidates for energies available at the LHC or at
future linear colliders are the excited states of the Higgs boson and
of the electroweak bosons.  Especially the orbital excitation, i.e.,
the spin 2 $d$-waves $D^3_{\mu \nu}$, $D^-_{\mu \nu}$ and $D^+_{\mu
  \nu}$, of the electroweak bosons have a well defined $1/F$ expansion
(we use the unitary gauge: $h=(h_{(1)}+F,0)$:
\begin{eqnarray} 
   D^3_{\mu \nu}&=& \frac{2}{g^2 F^2}
\left(   \left ( D_{\mu}h \right)^\dagger \left (D_\nu h\right) +
\left ( D_{\nu}h \right)^\dagger \left (D_\mu h\right)
\right)
   \approx B^3_\nu B^3_\mu + B^+_\nu B^-_\mu+  B^+_\mu B^-_\nu
    \\ \nonumber
   D^-_{\mu \nu}&=& \frac{ - \sqrt{2} }{g^2 F^2} \epsilon^{ij}
  \left(   \left ( D_{\mu}h \right)_i \left (D_\nu h\right)_j +
\left ( D_{\nu}h \right)_i \left (D_\mu h\right)_j
\right)
   \approx B^3_\mu B^-_\nu+ B^3_\nu B^-_\mu
  \\ \nonumber
   D^+_{\mu \nu}&=&\left(\frac{ - \sqrt{2} }{g^2 F^2} \epsilon^{ij}
  \left(   \left ( D_{\mu}h \right)_i \left (D_\nu h\right)_j +
\left ( D_{\nu}h \right)_i \left (D_\mu h\right)_j
\right) \right)^\dagger
   \approx  B^3_\mu B^+_\nu+ B^3_\nu B^+_\mu
\end{eqnarray}
where $D_{\mu}$ is the covariant derivative, $B_\mu^a$, $a=\{3,+,-\}$
are the gauge fields and $g$ is the coupling constant corresponding to
the gauge group $SU(2)_L$. Although the masses and the couplings of
these electroweak $d$-waves to other particles are fixed by the
dynamics of the model, it is difficult to determine these parameters.
In analogy to Quantum Chromodynamics, it is expected that the
$d$-waves couple with a reasonable strength to the corresponding
$p$-waves, the electroweak bosons. In the following, we assume in
accordance with the duality property, that the $d$-waves only couple
to the electroweak bosons and not to the photon, the Higgs boson or the
fermions.

\section{Production of the electroweak $d$-waves}

The cross-sections and decay width of $d$-waves predicted in a variety
of composite models were considered in \cite{Chiappetta:1987xc}.
Here we shall consider different effective couplings of our
electroweak $d$-waves that are more suitable for the model proposed in
chapter 2. If their masses are of the order of the scale of the
theory, they will be accessible at the LHC.  Of particular interest is
the neutral electroweak $d$-wave because it is expected to couple to
the $W^\pm$ electroweak bosons. This particle can thus be produced by
the fusion of two electroweak bosons at the LHC or at linear
colliders.

We  shall  use  the  formalism   developed  by  van  Dam  and  Veltman
\cite{vanDam:1970vg} for massive $d$-waves  to compute the decay width
of the $D^3_{\mu \nu}$ into $W^+ W^-$. We use the following relation:
\begin{eqnarray}
\sum^{5}_{i=1} e^i_{\mu \nu}(p) e^i_{\alpha \beta}(p) &=& \frac{1}{2}
\left (\delta_{\mu \alpha} \delta_{\nu \beta}
  + \delta_{\mu \beta}   \delta_{\nu \alpha}
  - \delta_{\mu \nu}\delta_{\alpha \beta}\right) \\ \nonumber
&&+\frac{1}{2} \left (\delta_{\mu \alpha} \frac{p_\nu p_\beta}{m_D^2}+
  \delta_{\nu \beta} \frac{p_\mu p_\alpha}{m_D^2}
  +  \delta_{\mu \beta} \frac{p_\nu p_\alpha}{m_D^2}
  + \delta_{\nu \alpha} \frac{p_\mu p_\beta}{m_D^2}
\right)
\\ \nonumber
&&+\frac{2}{3}
\left ( \frac{1}{2} \delta_{\mu \nu} -\frac{p_\mu p_\nu}{m_D^2}\right)
\left ( \frac{1}{2} \delta_{\alpha \beta} -\frac{p_\alpha p_\beta}{m_D^2}\right)
\end{eqnarray}
for the sum over the polarizations $e^i_{\mu \nu}$ of the $d$-wave.
In the notation of \cite{vanDam:1970vg} the sum over the
polarizations of the $W^\pm$ is given by
\begin{eqnarray}
  \sum^{3}_{i=1} e^i_{\mu}(p) e^i_{\nu}(p) &=& \delta_{\mu \nu}
  + \frac{p_\mu p_\nu}{m_W^2}
\end{eqnarray}
where $\delta_{\mu \nu}$ is the Euclidean metric. 
Averaging over the polarizations of the $d$-wave, we obtain
\begin{eqnarray}
  \Gamma(D^3\rightarrow W^+ W^-)&=&
  \frac{g^2_D}{1920 m_D \pi} \left(x_W-4\right)^2
  \sqrt{\left(1-\frac{4}{x_W}\right)}
\end{eqnarray}
with $x_W=(m_D/m_W)^2$, where $m_D$ is the mass of the $d$-wave and
$g_D$ is a dimensionfull coupling constant with $\mbox{dim}[
g_D]=\mbox{GeV}$. A dimensionless coupling constant is obtained by a
redefinition of the coupling constant $g_D \rightarrow m_D \bar{g}_D$.
We shall discuss plausible numerical inputs in the next section.
Assuming that the $Z$ boson couples with the same strength to the
$d$-wave as the $W$-bosons, we can estimate the decay width into
$Z$ bosons in the following way
\begin{eqnarray}
 \Gamma(D^3\rightarrow Z Z)&=&
  \frac{g^2_D}{3840 m_D \pi} \left(x_Z-4\right)^2
  \sqrt{\left(1-\frac{4}{x_Z}\right)} \\ \nonumber
   &\approx& \frac{1}{2} \Gamma(D^3\rightarrow W^+ W^-) 
\end{eqnarray}
with $x_Z=(m_D/m_Z)^2$. The Breit-Wigner resonance cross section for
the reaction $W^+ + W^- \rightarrow D^3$ thus reads (see e.g.
\cite{Donoghue:1992dd})
\begin{eqnarray}
\sigma^{(res)}_{W^+ + W^-\rightarrow D^3}&=& \frac{10 \pi}{ q^2} \frac{m_D^2 \Gamma^{\mbox{(tot)}}_D \Gamma(D^3\rightarrow W^+ W^-)}{\left(m_D^2-s\right)^2+m_D^2 {\Gamma^{\mbox{(tot)}}_D}^2}
\end{eqnarray}
where $q^2=(s-4 m_W^2)/4$ and $\Gamma^{(\mbox{tot})}_D\approx 3/2
\Gamma(D^3\rightarrow W^+ W^-)$ is the total decay width of the
neutral $d$-wave. Due to the background, the $W$ bosons might be
difficult to observe. But, if the electroweak $d$-waves states are
produced we also expect an excess of $Z$ bosons compared to the
standard model expectation. Note that the $Z$ bosons are easier to
observe.

As we shall see in the next section, the neutral $d$-waves give a sizable
contribution to the reaction $W^+ + W^- \to W^+ + W^-$.

\section{The reaction $W^+ + W^- \to W^+ + W^-$}

\begin{figure}
  \begin{minipage}[t]{0.32\linewidth}\centering
    \includegraphics[width=\linewidth]{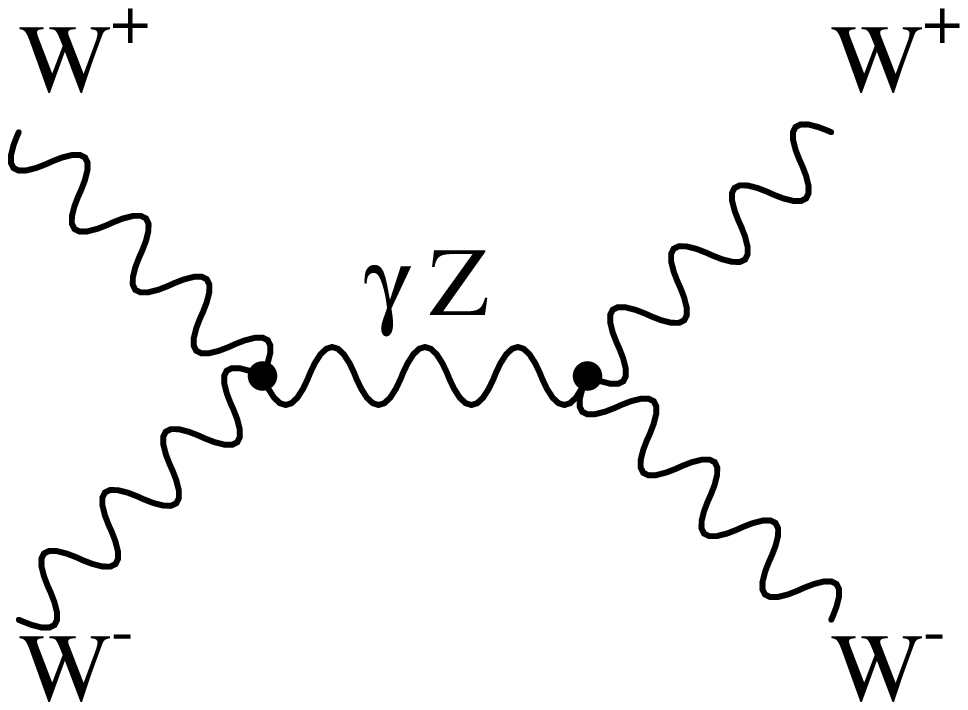}
    \begin{minipage}{1\linewidth}
      \caption{photon and Z boson in the s channel \label{fig1}}
    \end{minipage}
  \end{minipage}
  \begin{minipage}[t]{0.32\linewidth}\centering
    \includegraphics[width=\linewidth]{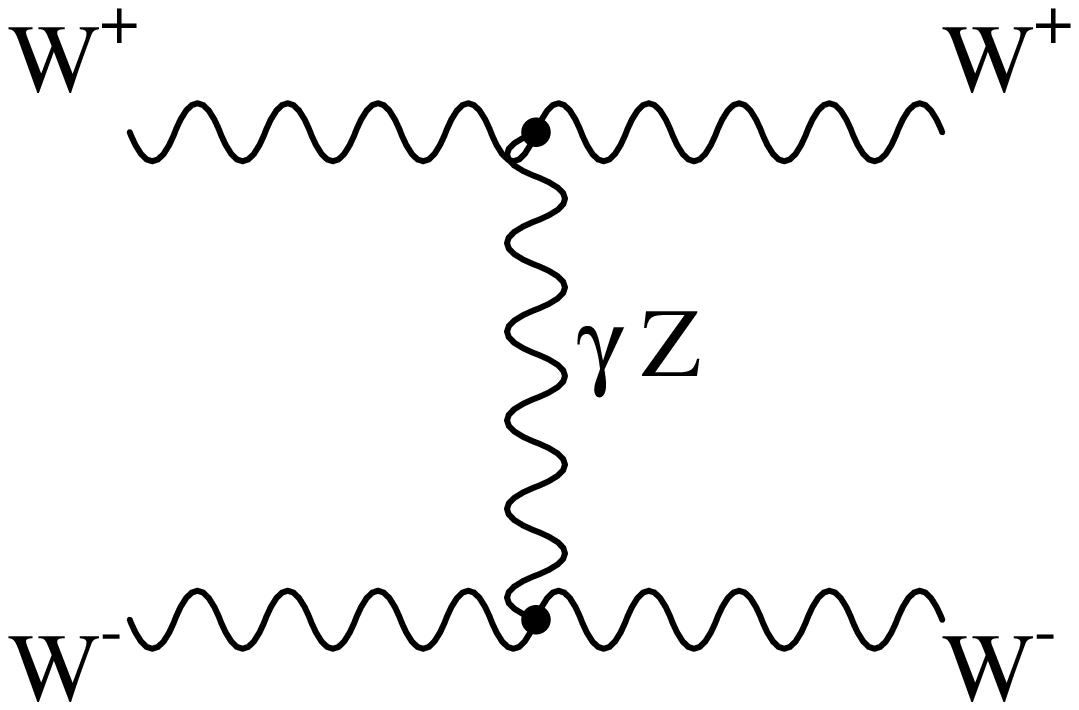}
    \begin{minipage}{1\linewidth}
      \caption{photon and Z boson in the t channel\label{fig2}}
    \end{minipage}
    \end{minipage}
  \begin{minipage}[t]{0.32\linewidth}\centering
    \includegraphics[width=\linewidth]{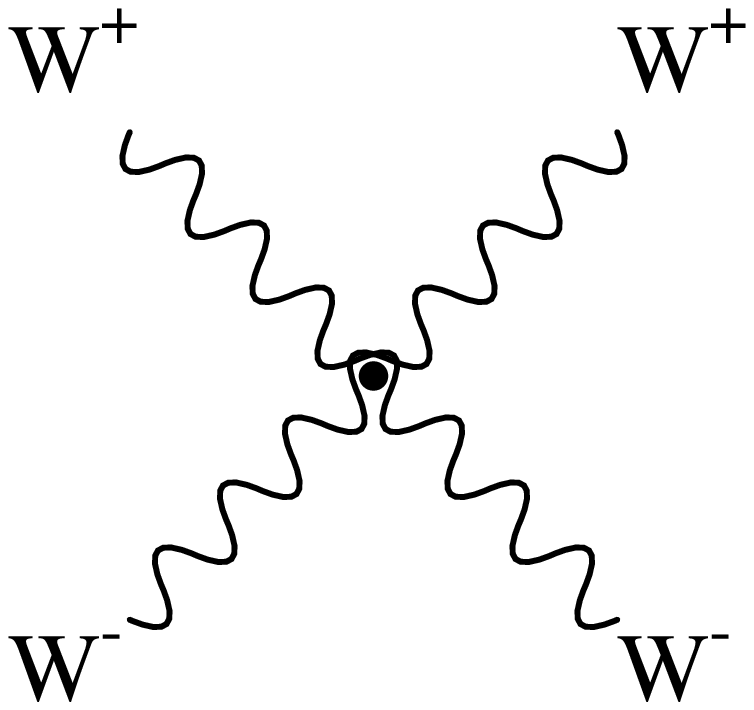}
    \begin{minipage}{1\linewidth}
      \caption{four W vertex \label{fig3}}
    \end{minipage}
  \end{minipage}
  \end{figure}

\begin{figure}
  \begin{minipage}[t]{0.45\linewidth}\centering
    \includegraphics[width=\linewidth]{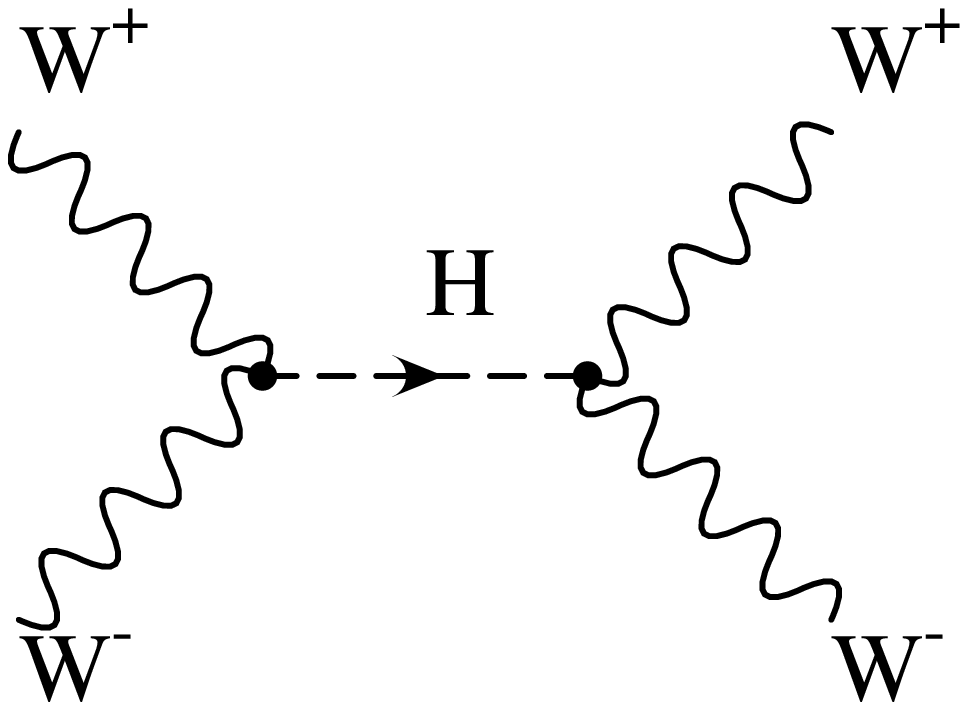}
    \begin{minipage}{0.9\linewidth}
      \caption{Higgs boson in the s channel\label{fig4}}
    \end{minipage}
    \end{minipage}
     \begin{minipage}[t]{0.45\linewidth}\centering
    \includegraphics[width=\linewidth]{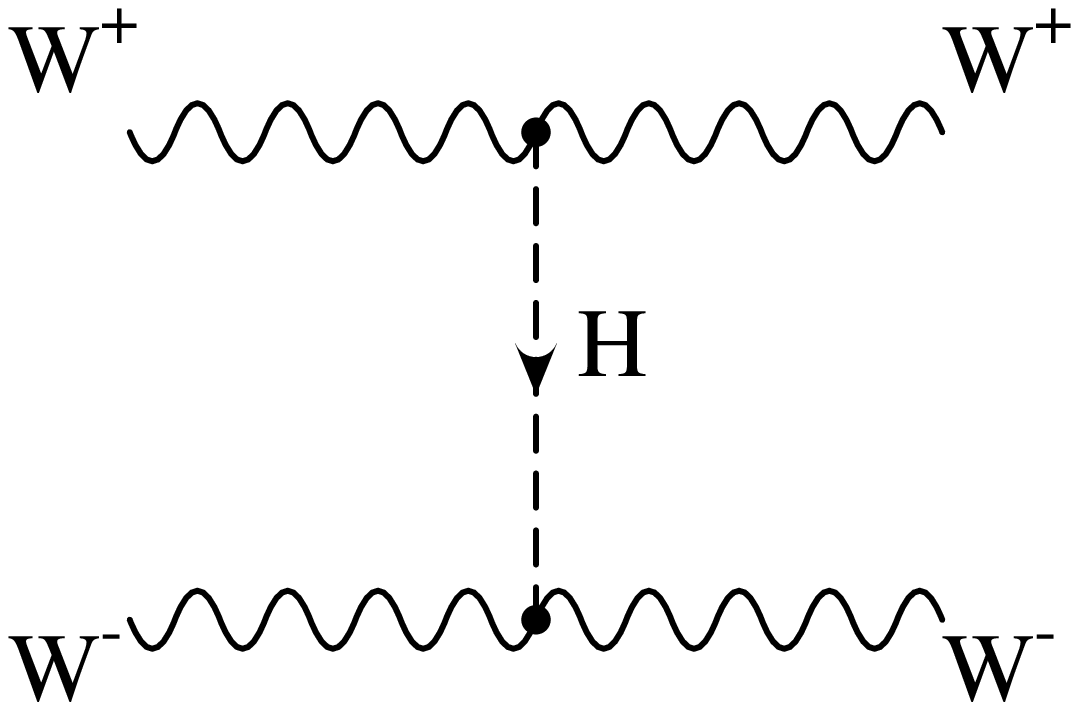}
    \begin{minipage}{0.9\linewidth}
      \caption{Higgs boson in the t channel\label{fig5}}
    \end{minipage}
  \end{minipage}
  \end{figure}

A considerable attention has been paid to the scattering of
electroweak bosons since this represents a stringent test of the gauge
structure of the standard model. In particular the reaction $W^+ + W^-
\to W^+ + W^-$ is of prime interest. If the Higgs boson is heavier
than 1 TeV, the electroweak bosons will start to interact strongly
\cite{Dicus:1973vj}. This reaction has been studied in the framework
of the standard model in \cite{Duncan:1986vj}. The one loop
corrections were considered in \cite{Denner:1998kq} and are known to
be sizable. For the sake of this work, the tree level diagrams are
sufficient to show that the contribution of the neutral electroweak
$d$-wave will be sizable and cannot be overlooked in forthcoming
experiments. As described in \cite{Gunion:1989we} (see also
\cite{Duncan:1986vj}) the $W$'s emitted by the beam particles are
dominantly longitudinally polarized if the following relations are
fulfilled: $m_W^2 \ll m^2_{WW} \ll s$ at an $e^+$ $e^-$ collider, and
$m_W^2 \ll m^2_{WW} \ll s_{q \bar q} \ll s$ at a hadron collider, and
we shall only consider the especially interesting reaction $W^+_L +
W^-_L \to W^+_L + W^-_L$ as described in \cite{Duncan:1986vj}. In the
standard model, this reaction is a test of the gauge structure of the
theory \cite{LlewellynSmith:1973ey}. The Feynman graphs contributing
in the standard model to this reaction can be found in figures
\ref{fig1}, \ref{fig2}, \ref{fig3}, \ref{fig4} and \ref{fig5}.
The amplitudes corresponding to these graphs are \cite{Duncan:1986vj}
\begin{eqnarray}
A_{s \gamma} &=& -\frac{1}{16}i g^2 x s^2 \beta^2 (3-\beta^2)^2 \cos{\theta},
\\ \nonumber 
 A_{s Z}&=& -\frac{1}{16}i g^2 (1-x) \frac{s^3}{s-\xi_Z} \beta^2
  (3-\beta^2)^2 \cos{\theta},
 \\ \nonumber  
  A_{t \gamma} &=& -\frac{1}{32 }i g^2 x \frac{s^3}{t}
                   [\beta^2(4- 2 \beta^2+\beta^4)   
                   +\beta^2(4- 10 \beta^2+\beta^4) \cos{\theta} \\
                    \nonumber &&
                   +(2-11 \beta^2+10\beta^4) \cos^2{\theta}
                   +\beta^2 \cos^3{\theta}],
\\ \nonumber 
A_{t Z} &=& -\frac{1}{32 }i
                 g^2 (1-x) \frac{s^3}{t - \xi_Z}
                   [\beta^2(4- 2 \beta^2+\beta^4)   
                   +\beta^2(4- 10 \beta^2+\beta^4) \cos{\theta}
                   \\
                    \nonumber &&
                   +(2-11 \beta^2+10\beta^4) \cos^2{\theta}
                   +\beta^2 \cos^3{\theta}],
\\ \nonumber 
   A_4&=& -\frac{1}{16} i g^2 s^2
   (1 + 2 \beta^2 -6 \beta^2 \cos{\theta}- \cos^2{\theta}),
\\ \nonumber 
     A_{s H}&=&-\frac{1}{16} i g^2 s^2
     \frac{(1+\beta^2)^2}{s-\xi_H+i\gamma_H}, 
\\ \nonumber 
     A_{t H}&=&-\frac{1}{16} i g^2 s^2 \frac{(\beta^2-\cos{\theta})^2}
     {t-\xi_H+i\gamma_H},
\end{eqnarray}
where $x=\sin^2{\theta_W}$, $\xi_Z=(1-x)^{-1}=m_Z^2/m^2_W$,
$\xi_H=m_H^2/m^2_W$, $\gamma_H=m_H \Gamma_H/m_W^2$ and
$\beta=\sqrt{1-4/s}$. The variables $s$ and $t$ are scaled with
respect to $m_W^2$. The scattering angle is $\theta$, $t=-1/2 s
\beta^2(1-\cos{\theta})$. These notations are the same as those
introduced in \cite{Duncan:1986vj}.  The standard model amplitude
is thus
\begin{eqnarray}
  A^{sum}_{SM}&=&A_{s \gamma}+A_{s Z}+A_{t \gamma}+A_{t Z}+A_{4}
  +A_{s H}+A_{t H}.
\end{eqnarray}
In the high energy limit, one observes the cancellation of the
leading powers in $s$ and finds \cite{Duncan:1986vj}
\begin{eqnarray}
  A^{sum}_{SM} \approx \frac{1}{2} i g^2 \left[ \xi_Z \left(1+ \frac{s}{t}
      + \frac{t}{s} \right) + \xi_H - i \gamma_H \right]
\end{eqnarray}
for the sum of these amplitudes.
The cross section with the angular cut $-z_0< \cos\theta<z_0$ is then
\begin{eqnarray}
\sigma&=& \frac{1}{16 \pi s^2 \beta^2} \int^{t_+}_{t_-} |A^{sum}|^2 \mbox{d}t
\end{eqnarray}
in dimensionless units, $t_\pm=(2-s/2)(1\mp z_0)$.

The excitations of the Higgs and electroweak bosons also contribute
via the $s$ and $t$ channel.  The amplitudes corresponding to the
contribution of a radially excited Higgs boson ($H^*$) of mass
$m_{H^*}$ and decay width $\Gamma_{H^*}$ to this reaction are
\begin{eqnarray}
     A_{s H^*}&=&-\frac{1}{16} i g^2_{H^*} s^2
     \frac{(1+\beta^2)^2}{s-\xi_{H^*}+i\gamma_{H^*}}, 
\\ \nonumber 
     A_{t H^*}&=&-\frac{1}{16} i g^2_{H^*} s^2 \frac{(\beta^2-\cos{\theta})^2}
     {t-\xi_{H^*}+i\gamma_{H^*}},
\end{eqnarray}
where $\xi_{H^*}=m_{H^*}^2/m^2_W$, $\gamma_{H^*}=m_{H^*} \Gamma_{H^*}/m_W^2$
and $g_{H^{*}}$ is the strength of the coupling between two $W$ bosons
and the $H^{*}$ scalar particle.

We shall now consider the contribution of the
radially $(W^{3 *})$ and orbitally $(D^{\mu \nu})$ excited neutral $Z$
boson. The amplitudes for the $W^{3 *}$ can be at once deduced from
those of the standard model contribution of the $Z$ boson
\begin{eqnarray}
  A_{s W^{3 *}}&=& -\frac{1}{16}i g^2_{W^{3 *}}
  \frac{s^3}
    {s-\xi_{W^{3 *}}+ i \gamma_{W^{3 *}}} \beta^2
  (3-\beta^2)^2 \cos{\theta},
\\
  \nonumber 
 A_{t W^{3 *}} &=& -\frac{1}{32 }i
                 g^2_{W^{3 *}} \frac{s^3}
                 {t - \xi_{W^{3 *}}+i\gamma_{W^{3 *}}}
                   [\beta^2(4- 2 \beta^2+\beta^4)
                   \\
                    \nonumber &&
                   +\beta^2(4- 10 \beta^2+\beta^4) \cos{\theta}
                   +(2-11 \beta^2+10\beta^4) \cos^2{\theta}
                   +\beta^2 \cos^3{\theta}],
\end{eqnarray}
where $\xi_{W^{3 *}}=m_{W^{3 *}}^2/m^2_W$, $\gamma_{W^{3 *}}=m_{W^{3 *}}
\Gamma_{W^{3 *}}/m_W^2$ and $g_{W^{3 *}}$ is the strength of the coupling
between two $W$ bosons and the $W^{3 *}$ boson.

The orbitally excited $Z$ boson $(D^{\mu \nu})$ is a $d$-wave, and its
propagation is thus described by a propagator corresponding to a
massive spin 2 particle. The propagator of a massive spin two particle
is as follows (see  \cite{vanDam:1970vg}):
\begin{eqnarray}
  \Gamma_{\mu \nu \rho \sigma} &=& \frac{1}{p^2-m_D^2}
   \frac{1}{2} ( g_{\mu \rho} g_{\nu \sigma} +
                 g_{\mu \sigma} g_{\nu \rho}- \frac{2}{3}
                 g_{\mu \nu} g_{\rho \sigma})
\end{eqnarray}
and we assume that the vertex $W^{+ \mu} W^{- \nu} D_{\mu \nu}$ is of
the form $i g_D$.  We obtain the following amplitudes for
the $s$ and $t$ channel exchange
\begin{eqnarray}
A_{s D}&=&\frac{-1}{48} i g_D^2 \frac{m_D^2}{m_W^2} 
\frac{s^2}{s-\xi_D+i \gamma_D}
\left(2\beta^4+3 \cos^2{\theta} - 2 \beta^2-1\right)
\end{eqnarray}
\begin{eqnarray}
\! \! \! \! \! \! \! \! \!
A_{t D}&=& \frac{-1}{96} i g_D^2  \frac{m_D^2}{m_W^2} 
\frac{s^2}{t-\xi_D+i \gamma_D}
\left(
  4\beta^4+6\beta^2+3+10\beta^2\cos{\theta}+1\cos{\theta}^2 \right).
\end{eqnarray}

Since there is a pole in the $t$ channel whose origin is the photon
exchange, one has to impose cuts on the cross sections. For the
numerical evaluation of the cross section, we impose a cut of
$10^\circ$, which is the cut chosen in   \cite{Denner:1998kq}. The
spin of the particle can be determined from the angular distribution
of the cross section. We have neglected the decay width of the $Z$
boson and that of the Higgs boson since we assume that the energy of the
process is such that no $Z$ boson or Higgs resonance appear. For
numerical estimates, we take $m_H=100$ GeV.

We have considered only the reaction involving longitudinally polarized 
$W$. The amplitudes for different polarizations for the standard model can 
be found in the literature \cite{Duncan:1986vj}. The amplitudes for a $H^*$ 
or a $W^{3*}$ can be deduced from the standard model calculations by 
replacing the masses, the decay widths and the coupling constants. Those 
for the neutral $d$-wave can be easily calculated using

\begin{eqnarray}
{\cal A}^{p_1 p_2 p_3 p_4}_s&=&-i g_D^2 \frac{m_D^2}{m_W^2} 
  \frac{1}{s-\xi_D+i\gamma_D} \\ \nonumber &&
   \frac{1}{2}\epsilon^\mu(p_1) \epsilon^\nu(p_2)
   ( g_{\mu \rho} g_{\nu \sigma} +
                 g_{\mu \sigma} g_{\nu \rho}- \frac{2}{3}
                 g_{\mu \nu} g_{\rho \sigma})
                 \epsilon^{* \rho}(p_3) \epsilon^{* \sigma}(p_4)
\end{eqnarray}
and
\begin{eqnarray}
{\cal A}^{p_1 p_2 p_3 p_4}_t&=&-i g_D^2  \frac{m_D^2}{m_W^2} 
  \frac{1}{t-\xi_D+i\gamma_D} \\ \nonumber &&
   \frac{1}{2}\epsilon^\mu(p_1) \epsilon^\rho(p_2)
   ( g_{\mu \rho} g_{\nu \sigma} +
                 g_{\mu \sigma} g_{\nu \rho}- \frac{2}{3}
                 g_{\mu \nu} g_{\rho \sigma})
                 \epsilon^{* \nu}(p_3) \epsilon^{* \sigma}(p_4)
\end{eqnarray}
where $p_i$ stands for the polarization and also using the following
relations
\begin{eqnarray}
  &\epsilon^\mu_1(0)=(-p,0,0,E)/m_W
  &\epsilon^\mu_1(\pm)=(0,-1,\pm i,0)/\sqrt{2}           \\
  \nonumber
   &\epsilon^\mu_2(0)=(-p,0,0,-E)/m_W
  &\epsilon^\mu_2(\pm)=(0,1,\pm i,0)/\sqrt{2} \\
  \nonumber
   &\epsilon^{*\mu}_{3}(0)=(p,-E \sin{\theta},0,-E\cos{\theta} )/m_W
  &\epsilon^{*\mu}_{3}(\pm)=(0,-\cos{\theta},\mp i,\sin{\theta})/\sqrt{2}           \\
  \nonumber
   &\epsilon^{*\mu}_4(0)=(p,E \sin{\theta},0,E\cos{\theta} )/m_W
  &\epsilon^{*\mu}_4(\pm)=(0,\cos{\theta},\mp i,-\sin{\theta})/\sqrt{2}
\end{eqnarray}
valid in the center of mass system where $E$ is the energy of the $W$
bosons, $p=\sqrt{E^2-m_W^2}$ is their momentum and $\theta$ is the
scattering angle.

\section{Discussion}

\begin{figure}
\begin{center}
\leavevmode
\epsfxsize=7.5cm
\epsffile{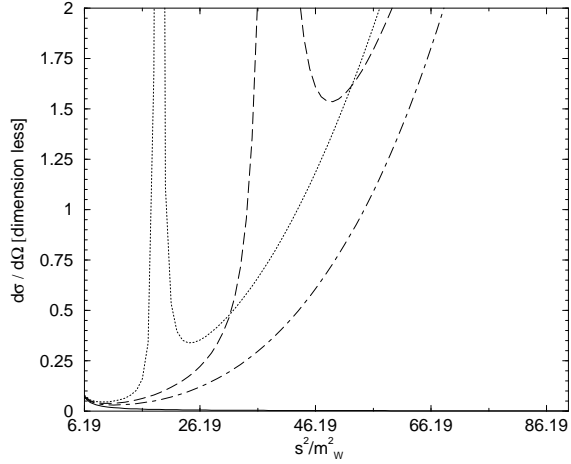}
\caption{Dimensionless cross section of the reaction $W_L^+ W_L^- \to W_L^+
  + W_L^-$ including the $d$-wave. The solid line is the standard
  model cross section, the dotted line corresponds to a $d$-wave of
  mass 350 GeV, with $\Gamma=4.38$ GeV and $\bar{g}_{W^{3 *}}=0.8 g$,
  the long dashed line to a $d$-wave of mass 500 GeV, with
  $\Gamma=27.49$ GeV and $\bar{g}_{W^{3*}}=0.7 g$ and the dot-dashed
  line to a $d$-wave of mass 800 GeV, with $\Gamma=251.03$ GeV and
  $\bar{g}_{W^{3 *}}=0.6 g$.}
\label{dwave}
\end{center}
\end{figure}

\begin{figure}
\begin{center}
\leavevmode
\epsfxsize=7.5cm
\epsffile{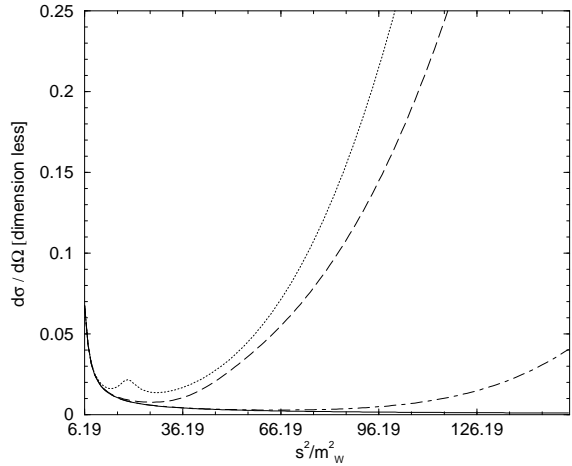}
\caption{Dimensionless cross section of the reaction
  $W_L^+ W_L^- \to W_L^+ + W_L^-$
  including the $W^{3*}$ boson. The solid line is the standard model
  cross section, the dotted line corresponds to a $W^{3*}$ boson of
  mass 350 GeV, with $\Gamma=66.2$ GeV and $\bar{g}_{W^{3 *}}=0.8
  \sin^2{\theta_W} g$, the long dashed line to a $W^{3*}$ boson of mass
  500 GeV, with $\Gamma=266.8$ GeV and $\bar{g}_{W^{3 *}}=0.7 \sin^2{\theta_W}
  g$ and the dot-dashed line to a $W^{3*}$ boson of mass 800 GeV, with
  $\Gamma=1795.5$ GeV and $\bar{g}_{W^{3 *}}=0.6 \sin^2{\theta_W} g$.}
\label{zprime}
\end{center}
\end{figure}

\begin{figure}
\begin{center}
\leavevmode
\epsfxsize=7.5cm
\epsffile{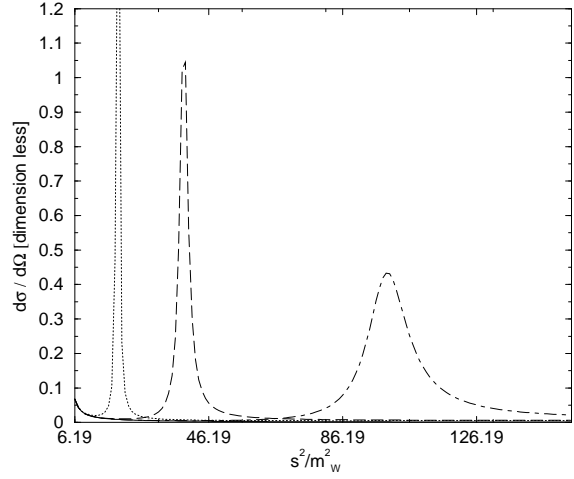}
\caption{Dimensionless cross section of the reaction
  $W_L^+ W_L^- \to W_L^+ + W_L^-$
  including the $H^{*}$ boson. The solid line is the standard model
  cross section, the dotted line corresponds to a $H^{*}$ boson of
  mass 350 GeV, with $\Gamma=6.72$ GeV and $\bar{g}_{H^{*}}=0.8 g$, the long
  dashed line to a $H^{*}$ boson of mass 500 GeV, with $\Gamma=17.6$
  GeV and $\bar{g}_{H^{*}}=0.7 g$ and the dot-dashed line to a $W^{3*}$
  boson of mass 800 GeV, with $\Gamma=58.25$ GeV and $\bar{g}_{H^{*}}=0.6 g$.}
\label{Hstar}
\end{center}
\end{figure}

\begin{figure}
\begin{center}
  \leavevmode \epsfxsize=7.5cm \epsffile{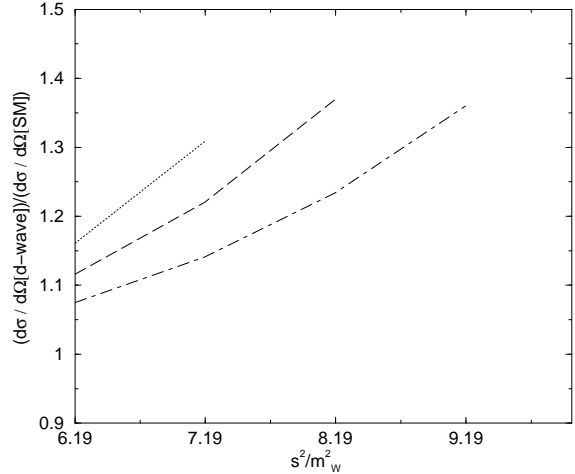}
\caption{Ratio of the cross-section for the of the reaction involving the $d$-wave to the standard model cross-section for different values of the $d$-wave mass and different coupling constants. The dotted line corresponds to a $d$-wave of mass 350 GeV, with $\Gamma=4.38$ GeV and $\bar{g}_{W^{3 *}}=0.8 g$, the long dashed  line to a $d$-wave of mass 500 GeV, with $\Gamma=27.49$ GeV and
  $\bar{g}_{W^{3 *}}=0.7 g$ and the dot-dashed line to a $d$-wave of
  mass 800 GeV, with $\Gamma=251.03$ GeV and $\bar{g}_{W^{3 *}}=0.6
  g$.}
\label{div}
\end{center}
\end{figure}

The differential cross section for the reaction $W_L^+ W_L^- \to W_L^+
+ W_L^-$ can be found in figure \ref{dwave} for the reaction involving
the neutral $d$-wave, figure \ref{zprime} for that involving the
$W^{3*}$ spin 1 boson and figure \ref{Hstar} for that involving the
$H^{*}$ scalar. The particles $W^{3*}$ and $H^{*}$ are assumed to
couple, in a first approximation, only to the $W$'s. This allows to
compute their decay rates using standard model formulas. As mentioned
previously, it is not an easy task to predict the mass spectrum of the
model, thus we assumed, for numerical illustration, three different
masses: 350 GeV, 500 GeV and 800 GeV. The coupling constants are
assumed to sizable (see the figures \ref{dwave}, \ref{zprime},
\ref{Hstar} and \ref{div}).  If the cross sections are extrapolated to
very high energies, unitarity is violated.  However, as expected
in any substructure models, it will be restored by bound states
effects.

It is very instructive to plot the ratio of the differential cross
section involving new physics to the standard model differential cross
section. We have done so for the neutral $d$-wave (fig. \ref{div}). It
is obvious from this picture that any deviation from the standard
model, even at high energy will manifest itself already in a
deviation from one for that ratio. Already at an energy which is low
compared to the mass of the new particle, i.e. well bellow the
resonance, one observes a deviation from unity.

Nevertheless the calculation of the full reaction e.g. $e^+ e^- \to
W^+ W^- \nu \bar \nu$ involves the convolution of the cross section of
the reaction $W^+ W^- \to W^+ + W^-$ with functions describing the
radiative emission of the $W$'s from the fermions. When this integral
is performed some sensitivity is lost. Nevertheless the effects are
expected to be so large that they cannot be overlooked.  The reaction
will allow to test a mass range of a few TeV's so that even if the new
particles are too massive to be produced on-shell, their effects will
be noticeable at future colliders.

\section{Conclusions}

We have discussed the production of a neutral $d$-wave $D^3$ at the
LHC or at a linear collider. If the mass of this particle is of the
order of the scale of the theory, i.e. 300 GeV, it can be produced at
these colliders.  We have also shown that this particle as well as
radial excitations of the Higgs boson and $Z$ boson would spoil the
cancellation of the leading powers in $s$ of in the reaction $W^+_L
+W^-_L \to W^+_L +W^-_L$, thus any new particle contributing to that
reaction will have a large impact already at energies well below the
mass of this new particle. This reaction is thus not only of prime
interest if the Higgs boson is heavy but should also be studied if the
Higgs boson was light.
\chapter{The substructure of fermions}
If the duality breaks down entirely at a certain energy scale, it is
conceivable that effects from the substructure of the fermions will
become manifest. We shall discuss a quite generic parametrization of
the contribution of the substructure of a lepton to its anomalous
magnetic moment. Assuming a mixing matrix, we can then consider
radiative lepton decays that are conceivable if leptons have a
substructure. The results presented in this chapter were published in
\cite{Calmet:2001dc,Calmet:2001zq}.

\section{Anomalous magnetic moment}

A new contribution to the magnetic moment of the muon can be described
by adding an effective term ${\cal L}^{eff}$ to the Lagrangian of the
standard model as follows:

\begin{eqnarray} \label{effL1}
  {\cal L}^{eff} &=& \frac{e}{2 \Lambda} \bar \mu
  \left(A + B \gamma_5 \right) \sigma_{\mu \nu}
  \mu F^{\mu \nu}
  \left(1-\frac{4 \alpha}{\pi}\ln \frac{\Lambda}{m_\mu} \right), 
\end{eqnarray}
where $\mu$ is the muon field, $F^{\mu \nu}$ the electromagnetic field
strength, $\Lambda$ the compositeness scale and $A$ is a constant of
order one and $B$ is probably much smaller since it parametrizes
$CP$-violation. We have taken the QED one loop correction into account
\cite{Degrassi:1998es}. The leading order contribution has been
considered in \cite{Calmet:2001dc}. We have included a $\gamma_5$-term
in view of a possible $CP$ violation of the confining interaction.

The constants in ${\cal L}^{eff}$ depend on dynamical details of the
underlying composite structure. If the latter is analogous to QCD,
where such a term is induced by the hadronic dynamics, the constant
$A$ is of the order one. One obtains the following contribution to the
anomalous magnetic moment of the muon:
\begin{eqnarray} \label{amu1CH6}
  \Delta a_\mu &=& \left(\frac{m_\mu}{\Lambda}\right)
  \left(1-\frac{4 \alpha}{\pi}\ln \frac{\Lambda}{m_\mu} \right).
\end{eqnarray}
The $\gamma_5$-term does not contribute to the anomalous magnetic
moment.
  
The magnetic moment term (\ref{effL1}) has the same chiral structure
as the lepton mass term. Thus one expects that the same mechanism
which leads to the small lepton masses ($m_\mu \ll \Lambda$), e.g. a
chiral symmetry, leads to a corresponding suppression of the magnetic
moment \cite{Brodsky:1980zm}. In this case the effective Lagrangian
should be written as follows:

\begin{eqnarray} \label{effL2}
  {\cal L}^{eff} &=& \frac{e}{2 \Lambda} \frac{m_\mu}{ \Lambda}
        \bar \mu \left(A + B \gamma_5 \right) \sigma_{\mu \nu}
  \mu F^{\mu \nu} 
  \left(1-\frac{4 \alpha}{\pi}\ln \frac{\Lambda}{m_\mu} \right).  
\end{eqnarray}
The contribution of the compositeness to the magnetic moment is in
this case given by
\begin{eqnarray} \label{amu2CH6}
  \Delta a_\mu &=&  \left(\frac{m_\mu}{\Lambda}\right)^2
  \left(1-\frac{4 \alpha}{\pi}\ln \frac{\Lambda}{m_\mu} \right).
\end{eqnarray}

\section{Radiative lepton decays}

If the leptons have a composite structure, the question arises whether
effects which are absent in the standard model, in particular
flavor-changing transitions, e.g. the decays $\mu \to e \gamma$ or
$\tau \to \mu \gamma$ arise.

We shall study flavor changing magnetic-moment type transitions which
indeed lead to radiative decays of the charged leptons on a level
accessible to experiments in the near future.

We start by considering the limit $m_e=m_\mu=0$, i.e. only the third
lepton $\tau$ remains massive. Neutrino masses are not considered. In
this limit the mass matrix for the charged leptons has the structure
$m_{l^-}= m_\tau \mbox{diag} (0,0,1)$ and exhibits a ``democratic
symmetry'' \cite{Fritzsch:1999rd,Fritzsch:1994yx}. Furthermore there
exists a chiral symmetry $SU(2)_L \times SU(2)_R$ acting on the first
two lepton flavors. The magnetic moment term induced by compositeness,
being of a similar chiral nature as the mass term itself, must respect
this symmetry. We obtain
\begin{eqnarray} \label{effL3}
  {\cal L}^{eff} &=& \frac{e}{2 \Lambda} \frac{m_\tau}{ \Lambda}
        \bar \psi \tilde{M}\left(A + B \gamma_5 \right)  \sigma_{\mu \nu}
  \psi F^{\mu \nu}
\left(1-\frac{4 \alpha}{\pi}\ln \frac{\Lambda}{m_\psi} \right)
  .
\end{eqnarray}
Here $\psi$ denotes the vector $(e, \mu, \tau)$ and $\tilde{M}$ is given by
$\tilde{M}=\mbox{diag}(0,0,1)$. 

Once the chiral symmetry is broken, the mass matrix receives non-zero
entries, and after diagonalization by suitable transformations in the
space of the lepton flavors it takes the form $M=\mbox{diag}(m_e,
m_\mu, m_\tau)$. If after symmetry breaking the mass matrix $M$ and
the magnetic moment matrix $\tilde{M}$ were identical, the same
diagonalization procedure which leads to a diagonalized mass matrix
would lead to a diagonalized magnetic moment matrix. However there is
no reason why $\tilde{M}$ and $M$ should be proportional to each other
after symmetry breaking. The matrix elements of the magnetic moment
operator depend on details of the internal structure in a different
way than the matrix elements of the mass density operator. Thus in
general the magnetic moment operator will not be diagonal, once the
mass matrix is diagonalized and vice versa. Thus there exist
flavor-non-diagonal terms (for a discussion of analogous effects for
the quarks see \cite{Fritzsch:1999rd}), e.g. terms proportional
to $\bar e \ \sigma_{\mu \nu} \left ( A+ B \gamma_5 \right)\mu$. These
flavor-non-diagonal terms must obey the constraints imposed by the
chiral symmetry, i.e. they must disappear once the masses of the light
leptons involved are turned off. For example, the $e-\mu$ transition
term must vanish for $m_e \to 0$. Furthermore the flavor changing
terms arise due to a mismatch between the mass density and the
magnetic moment operators due to the internal substructure. If the
substructure were turned off ($\Lambda \to \infty$), the effects should
not be present. The simplest Ansatz for the transition terms between
the leptons flavors $i$ and $j$ is $const. \sqrt{m_i m_j}/\Lambda$. It
obeys the constraints mentioned above: it vanishes once the mass of
one of the leptons is turned off, it is symmetric between $i$ and $j$
and it vanishes for $\Lambda \to \infty$. In this case the magnetic
moment operator has the general form:

\begin{eqnarray} \label{effL4}
  \! \! \! \!  \! \! \! \!
  {\cal L}^{eff} &=& \frac{e}{2 \Lambda} \frac{m_\tau}{ \Lambda}  \bar \psi
\left( \begin{array}{ccc} \frac{m_e}{m_\tau}& C_{e \mu}
    \frac{\sqrt{m_e m_\mu}}{\Lambda} &
  C_{e \tau}
    \frac{\sqrt{m_e m_\tau}}{\Lambda}
    \\
  C_{e \mu}
    \frac{\sqrt{m_e m_\mu}}{\Lambda}
    &
\frac{m_\mu}{m_\tau}
    &
 C_{\mu \tau}
    \frac{\sqrt{m_\mu m_\tau}}{\Lambda}
    \\
C_{e \tau}
    \frac{\sqrt{m_e m_\tau}}{\Lambda}
    &
C_{\mu \tau}
    \frac{\sqrt{m_\mu m_\tau}}{\Lambda}
    &
1
    
\end{array} \right ) 
\psi \left(A + B \gamma_5 \right)  \sigma_{\mu \nu}  F^{\mu \nu}
\nonumber \\ && \times
  \left(1-\frac{4 \alpha}{\pi}\ln \frac{\Lambda}{m_\psi} \right).
\end{eqnarray}
Here $C_{ij}$ are constants of the order one. In general one may
introduce two different matrices (with different constants $C_{ij}$)
both for the 1-term and for the $\gamma_5$-term, but we shall limit
ourselves to the simpler structure given above.

Based on the flavor-changing transition terms given in eq.
(\ref{effL4}), we can calculate the decay rates for the decays
$\mu \to e \gamma$, $\tau \to \mu \gamma$ and $\tau \to e \gamma$.
We find:
\begin{eqnarray}
  \Gamma(\mu \to e \gamma)&=& e^2  \frac{m_\mu}{8 \pi} \left(
    \frac{\sqrt{m_\mu m_e}}{\Lambda} \right)^2
    \left (\frac{m_\mu}{\Lambda}\right)^2
    \left (\frac{m_\tau}{\Lambda}\right)^2
    \left( |A|^2+|B|^2\right)
    \nonumber \\ && \times
    \left(1-\frac{8 \alpha}{\pi}\ln \frac{\Lambda}{m_\mu} \right),
    \\
  \Gamma(\tau \to \mu \gamma)&=& e^2 \frac{m_\tau}{8 \pi} \left(
    \frac{\sqrt{m_\tau m_\mu}}{\Lambda} \right)^2
    \left (\frac{m_\tau}{\Lambda}\right)^2
    \left (\frac{m_\tau}{\Lambda}\right)^2
    \left( |A|^2+|B|^2\right)\nonumber \\
&& 
 \times
    \left(1-\frac{8 \alpha}{\pi}\ln \frac{\Lambda}{m_\tau} \right)
  ,
\\
  \Gamma(\tau \to e \gamma)&=& e^2 \frac{m_\tau}{8 \pi} \left(
    \frac{\sqrt{m_\tau m_e}}{\Lambda} \right)^2
    \left (\frac{m_\tau}{\Lambda}\right)^2
    \left (\frac{m_\tau}{\Lambda}\right)^2 
    \left( |A|^2+|B|^2\right)\nonumber \\
&& 
 \times
    \left(1-\frac{8 \alpha}{\pi}\ln \frac{\Lambda}{m_\tau} \right)
  .
\end{eqnarray}
In the following we take $|A|=1$. The parameter $|B|$ can be
constrained using the limits for the electron EDM. This limit gives
the most stringent constraint on this parameter.

Recently an indication was found that the anomalous magnetic moment of
the muon $\mu^+$ is slightly larger than expected within the standard
model \cite{Brown:2001mg}. The deviation is of the order of $10^{-9}$:
\begin{eqnarray}
  \Delta a_\mu &=& a_\mu(exp)-a_\mu(SM) =  (4.3\pm 1.6) \times 10^{-9}. 
\end{eqnarray}
For a review of the contribution of the standard model to the
anomalous magnetic moment of the muon see Ref.
\cite{Czarnecki:2001pv}. The observed effect (2.6 $\sigma$ excess)
does not necessarily imply a conflict with the standard model, in view
of the systematic uncertainties in the theoretical calculations due to
the hadronic corrections. If this result is confirmed by further
experimental data and theoretical work, it might be interpreted as the
first signal towards an internal structure of the leptons
\cite{Lane:2001ta}, although other interpretations (vertex corrections
due to new particles or non-minimal couplings due to a more complex
space-time structure \cite{Calmet:2001si}) are also possible.

The BNL result would give: $\Lambda \approx 2 \times 10^{9}$ GeV using
eq. (\ref{amu1CH6}).  Using eq. (\ref{amu2CH6}) and the central value
of $\Delta a_\mu$, one obtains: $\Lambda\approx 1.54$ TeV, i.e.
$\Lambda$ is much smaller due to the chiral symmetry argument
\cite{Brodsky:1980zm}. The $95 \%$ confidence level range for
$\Lambda$ is
\begin{eqnarray} \label{rangeL}
 1.16\ \mbox{TeV} < \Lambda <  3.04 \ \mbox{TeV}.
\end{eqnarray}
We can use this experimental input to illustrate the contribution of
the fermion substructure to its anomalous magnetic moment and to
compute the

The Lagrangian (\ref{effL4}) yields the following EDM for the
electron:
\begin{eqnarray}
   d_e &=& \frac{e}{ \Lambda} \frac{m_e}{ \Lambda}|B|
\left(1-\frac{4 \alpha}{\pi}\ln \frac{\Lambda}{m_e} \right)
=3.7\times 10^{-24} |B|
   \mbox{\ e-cm},
\end{eqnarray}
which has to be compared to the experimental limit
${d_e}^{\mbox{exp}}<(0.18 \pm 0.12 \pm 0.10) \times 10^{-26}
e-\mbox{cm}$ \cite{Groom:2000in}, we thus see that $|B|$ must be much
smaller than $|A|$. We set $|B|=0$ in the following. The corresponding
branching ratios are:
\begin{eqnarray}
  \mbox{Br}(\mu \to e \gamma) &\approx& 1.5 \times 10^{-10},
  \\
   \mbox{Br}(\tau \to \mu \gamma) &\approx& 3.5 \times 10^{-10},
   \\
   \mbox{Br}(\tau \to e \gamma) &\approx& 1.7 \times 10^{-12},
\end{eqnarray}
using the central value of $\Delta a_\mu$ to evaluate $\Lambda$.  One
obtains the following ranges for the branching ratios
\begin{eqnarray} \label{range}
 8.3\times 10^{-10}  &>\mbox{Br}(\mu \to e \gamma)&> 2.5 \times 10^{-12}, 
 \ \ \ \ \\
 1.9 \times 10^{-9} &> \mbox{Br}(\tau \to \mu \gamma) &> 5.8 \times 10^{-12},
   \\
 9.3\times 10^{-12}  &> \mbox{Br}(\tau \to e \gamma) &> 2.8 \times 10^{-14},
\end{eqnarray}
using the $95 \%$ confidence level range for $\Lambda$ (\ref{rangeL}).
  
These ranges are based on the assumption that the constant $A$ of
order one is fixed to one. The upper part of the range for the $\mu
\to e \gamma$ decay given in (\ref{range}) is excluded by the present
experimental limit: $\mbox{Br}(\mu \to e \gamma)<1.2 \times 10^{-11}$
\cite{Groom:2000in}. Our estimates of the branching ratio should be
viewed as order of magnitude estimates. In general we can say that the
branching ratio for the $\mu \to e \gamma$ decay should lie between
$10^{-13}$ and the present limit.
 
The decay $\tau \to \mu \gamma$ processes at a level which cannot be
observed, at least not in the foreseeable future. The decay $\tau \to
e \gamma$ is, as expected, much suppressed compared to $\tau \to \mu
\gamma$ decay and cannot be seen experimentally.
 
Numerically, the effect of the QED one loop correction is small
compared to the ``tree level'' calculation \cite{Calmet:2001dc}
because there is a cancellation between two effects: the extracted
composite scale is larger but the decay rates are suppressed by the
factor $\left(1-\frac{8 \alpha}{\pi}\ln \frac{\Lambda}{m_f}\right)$,
where $m_f$ is the mass of the decaying lepton.

Note added: the QCD uncertainties finally settled down
\cite{Knecht:2001qg,Knecht:2001qf,Blokland:2001pb,Hayakawa:2001bb}.
The deviation is only of the order of 1.6 $\sigma$ which allows to put
a limit of 2 TeV for the compositeness scale of the muon. This scale
corresponds to the following branching ratios
\begin{eqnarray}
  \mbox{Br}(\mu \to e \gamma) &\approx& 3.1 \times 10^{-11},
  \\
   \mbox{Br}(\tau \to \mu \gamma) &\approx& 7.1 \times 10^{-11},
   \\
   \mbox{Br}(\tau \to e \gamma) &\approx& 3.5 \times 10^{-13},
\end{eqnarray}
for the radiative lepton decays.

\chapter{Conclusions}

We have presented a duality between the standard model and a model
based on the same gauge group but where $SU(2)_L$ is confining its
charges instead of being broken by means of the Higgs mechanism.  This
duality allows a calculation of the electroweak mixing angle and of
the mass of the Higgs boson.

If the duality is unbroken, we do not expect any physics beyond the
standard model, as both phases are identical. But, both the
confinement phase and the Higgs phase are necessary to extracted all
the informations present in the theory.  Left-handed particles, the
electroweak bosons and the Higgs boson have a point like and a bound
state like character. The duality allows a calculation of the
electroweak mixing angle and of the Higgs boson mass.

We have considered a supersymmetric extension of the duality, and
shown that our ideas are compatible with a supersymmetric extension.

Albeit the author does not expect it, this duality might only be a low
energy phenomenon. If the standard model breaks down in the Yukawa
sector and if Nature is described by the confinement phase, the decay
modes of the Higgs boson can be dramatically affected. In particular
it might not couple to $b$-quarks. In that case the decay channels of
the Higgs boson would differ strongly from the standard model
expectations. The strategy for the Higgs boson searches would differ
from the standard one. Instead of searching for decays of the Higgs
boson to $b$-quark which is the dominant decay channel for a light
standard model Higgs boson, one should rather search for a Higgs boson
decaying to gluons.  This would be an example of a low energy failure
of the duality.

The absence of a phase transition between the confinement phase and
the Higgs phase implies that there is the same number of degrees of
freedom in both phases. But, if the duality breaks down, new
particles, like excitations of the electroweak bosons and of the Higgs
boson, will appear and will make sizable contributions to standard
model processes. Of particular interest are the spin 2 excitations of
the electroweak bosons which should make a sizable contribution to the
electroweak boson scattering. We have shown that, due to the neutral
$d$-wave, the cross section of the reaction $W^+_L +W^-_L \to W^+_L
+W^-_L$ would strongly differ from the standard model expectations
already at energies well bellow the mass scale of that new particle.
This would be an example of a high energy break down of the duality.

In the case of a total breakdown of the duality, effects of the
fermion substructure could appear and lead to sizable effects in low
energy observables like the anomalous magnetic moment of the muon.

Finally, the best test of the duality will be to find a Higgs boson
with a mass around $130$ GeV. This does not only represent a test of
the duality, but also of the standard model which has this duality
property.  This mass can therefore be seen as a prediction of the
standard model, which might have a problem if the Higgs boson mass is
much different from $130$ GeV.

We shall like to conclude by emphasizing that the model in the
confinement phase we have presented is basically different from
composite models that can be found in the literature.  The first
difference is the weak coupling confinement.  Secondly we are
considering bound states that are point like in space time but have an
extension in momentum space.  Those are the reasons why this model is
dual to the standard model.

\newpage
\newpage
\end{document}